\documentclass[cernpreprint,  texlive=2023, USenglish, texmf, orcidlogo]{atlasdoc}
 
\usepackage{atlaspackage}
\usepackage{atlasbiblatex}
 
\usepackage{atlasphysics}
\usepackage{atlasbsm}

\graphicspath{{logos/}{figures/}}
 
\usepackage{ANA-SUSY-2019-14-PAPER-defs}

\AtlasTitle{Search for displaced photons produced in exotic decays of the Higgs boson using 13~TeV $pp$ collisions with the ATLAS detector}

\AtlasAbstract{A search is performed for delayed and nonpointing photons originating from the displaced decay of a neutral long-lived particle (LLP).
The analysis uses the full Run~2 data set of proton--proton collisions
delivered by the LHC at a center-of-mass energy of $\sqrt{s}=13$~TeV between 2015 and 2018 and recorded by the ATLAS detector,
corresponding to an integrated luminosity of 139 fb$^{-1}$.
The capabilities of the ATLAS electromagnetic calorimeter are exploited to precisely measure the arrival times and trajectories of photons. The results are
interpreted in a scenario where the LLPs are
pair-produced in exotic decays of the 125 GeV Higgs boson, and
each LLP subsequently decays into a photon and a particle that
escapes direct detection, giving rise to missing transverse
momentum. No significant excess is observed above the expectation due to
Standard Model background processes. The
results are used to set upper limits on the branching ratio of the
exotic decay of the Higgs boson.
A model-independent
limit is also set on the production of photons with large values of
displacement and time delay.
 
}

\AtlasRefCode{SUSY-2019-14}

\PreprintIdNumber{CERN-EP-2022-096}

\AtlasJournalRef{Phys. Rev. D 108 (2023) 032016}
\AtlasDOI{10.1103/PhysRevD.108.032016}
 
\AtlasJournal{Phys. Rev. D}

\hypersetup{pdftitle={ATLAS document},pdfauthor={The ATLAS Collaboration}}
 
\begin{document}
 
\maketitle
 
\tableofcontents

\pagebreak
\section{Introduction}
\label{sec:intro}

 
The Standard Model (SM)
provides an accurate description of the fundamental particles and interactions at energy scales up to a few hundred \GeV.
However, the SM fails to explain several key phenomena, such as gravity, dark matter, and the matter--antimatter asymmetry
in the universe.
Such shortcomings suggest the presence of new physics, motivating
the search for beyond-the-SM (BSM) physics.
 
Supersymmetry (SUSY)~\cite{Golfand:1971iw,Neveu:1971rx,Neveu:1971iv,Ramond:1971gb,Volkov:1973ix,Wess:1973kz,Wess:1974tw} is a theoretically well-motivated extension of the SM that proposes the
existence of new SUSY partners (sparticles) of the SM particles.
The sparticle has identical quantum numbers to its SM partner, differing only by half a unit of spin.
A new quantum number, dubbed R-parity, distinguishes between supersymmetric and SM particles.
In R-parity-conserving SUSY
models~\cite{Fayet:1976et,Fayet:1977yc,Farrar:1978xj,Fayet:1979sa,Dimopoulos:1981zb}, the lightest SUSY particle~(LSP) is stable, and
SUSY production in proton--proton ($pp$) collisions at the LHC would produce sparticles in pairs. The sparticles would then decay in cascades
involving other sparticles and SM particles until the LSP is
produced. A neutral, weakly interacting LSP would usually escape the detector without interacting,
so LHC searches for SUSY often require a large amount of missing
transverse momentum (\MET) in the event.
 
In gauge-mediated SUSY breaking~(GMSB) models~\cite{Dine:1981gu,AlvarezGaume:1981wy,Nappi:1982,Dine:1993yw, Dine:1994vc,Dine:1995ag}, the gravitino (\gravitino) is the LSP for typical model parameter values. GMSB phenomenology is largely determined by the properties of the
next-to-lightest supersymmetric particle~(NLSP), since the decay
chains of the sparticles with higher
mass would typically lead to the NLSP, which would subsequently
decay to produce the LSP.  The
weak coupling of the NLSP to the gravitino LSP could generate a non-neglible lifetime of the NSLP, leading to
displaced NLSP decays~\cite{Dine:1994vc}.

An electrically neutral long-lived NLSP would avoid direct detection as
it traveled partway
(or entirely) through the ATLAS detector before decaying, with
a prominent decay mode being to a photon plus the LSP. The photons resulting from
displaced NLSP decays would be produced with some delay compared
to prompt photons, due mostly to the time-of-flight of the
massive NLSP before its decay. In addition, due to the opening angle between the photon and LSP produced in the NLSP decay, the photons
would tend to be
``nonpointing'', meaning that their flight path would not
be consistent with originating from
the primary vertex (PV) of the hard scatter of the event.
The capabilities of the ATLAS liquid-argon (LAr) electromagnetic (EM)
calorimeter to make precise ``pointing'' measurements of the flight direction
and the arrival time of photons (\tg) can be exploited to
search for such photons. Given the size of the ATLAS detector, requiring LAr calorimeter
measurements of the delayed photons
restricts the sensitivity of the analysis to NLSP
lifetimes of $\mathcal{O}$(ns).
 
Previous ATLAS analyses searched for nonpointing and delayed
photons produced in
long-lived NLSP decays in the data sets
of $pp$ collisions collected at center-of-mass
energies of 7~\TeV~\cite{SUSY-2012-15} and 8~\TeV~\cite{SUSY-2013-17} during Run~1 of the LHC.
Neither search found an excess above the SM background expectation, and set limits in the context of a particular set of GMSB SUSY models. 
A recent Run~2 CMS search for long-lived particles in such models
also found that their data agreed with the SM prediction~\cite{CMS-EXO-19-005}.
 
The discovery in 2012 of a particle with a mass near 125~\GeV\ that is consistent with expectations for
the SM Higgs boson
opens new avenues for searches with BSM sensitivity.
Specifically, the coupling of the Higgs boson to mass leads to search prospects for pairs of sparticles
produced in Higgs boson decays, provided the sparticles are sufficiently light.
Additional motivation for this production mechanism arises
from current uncertainties in the branching ratios of the Higgs boson.
Current experimental constraints from the Higgs boson property measurements allow the branching ratio for Higgs boson decays into invisible BSM states to be as large as 13\% at 95\% CL~\cite{HIGG-2021-23}.

\begin{figure}[hbpt]
\begin{centering}
\includegraphics[width=0.5\textwidth]{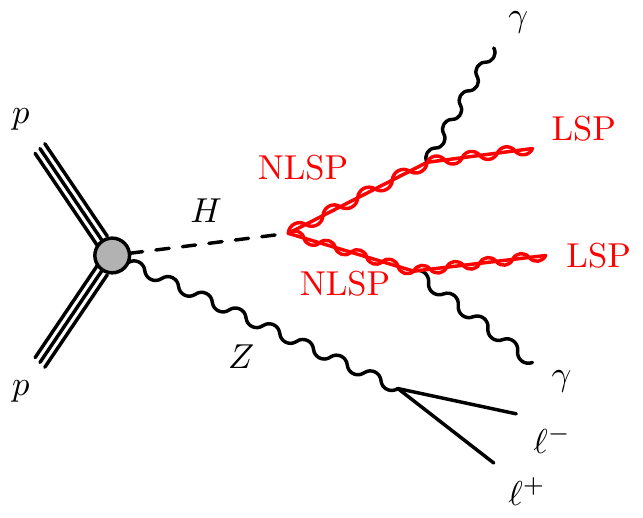}
\caption{Example Feynman diagram of the signal model, with a Higgs boson produced in association with a $Z$ boson. The Higgs
boson decays into a pair of electrically neutral
NLSP particles which are long-lived, each of which
then decays into a photon and an LSP. The $Z$ boson decays into a charged-lepton pair (electrons or muons).}
\label{fig:introduction_feynmanDiagrams}
\end{centering}
\end{figure}

This analysis utilizes the full Run~2 ATLAS data set of 13~\TeV\ $pp$
collisions,
and searches for evidence of Higgs boson decays into a pair of long-lived
electrically neutral
NLSP particles, each of which subsequently decays into an LSP,
which escapes the detector, and a photon.
Given the value of the Higgs boson mass, the
energies of the final-state photons,
as well as the value of \MET, are not
sufficiently high to provide a trigger for the events. Instead,
the analysis considers SM Higgs boson production
in association with either a $Z$ boson, $W$ boson or $t\bar{t}$ system. Leptonic decays of the associated $W/Z$ boson or $t\bar t$ system are exploited to provide a single-lepton (electron or muon) trigger.
An example leading-order Feynman diagram, for the case of $H + Z$ production, is shown in Figure~\ref{fig:introduction_feynmanDiagrams}.
To reduce the model-dependence
of the results, the analysis considers a simplified model where
the masses of the NLSP and LSP and the lifetime of the NLSP
are treated as independent parameters.
 
In addition to the requirement of the lepton trigger,
selected
events must have at least one photon candidate.
Events are divided into two
orthogonal final states, with the $1\gamma$ final state requiring exactly one photon and the $\geq2\gamma$
final state requiring two or more photons.
As detailed in Section~\ref{sec:selection},
the analysis strategy exploits measures of photon pointing
and timing. The pointing information is used to divide the
selected events into five nonoverlapping categories with different
signal-to-background ratios, while the timing information
provides the distribution for a likelihood fit used to search for the possible presence
of signal events.
 
The analysis exploits the expectation of
significant \MET \ in signal events, due to the
escaping LSPs, as well as any neutrinos produced in the leptonic
decays of the associated system. In addition, the signal would produce delayed photons,
which would have a positive value for their time of arrival at the EM calorimeter, whereas prompt backgrounds
have time distributions centered at zero.
The analysis therefore performs the signal search utilizing
a signal region (SR) defined by requiring
high \MET \ and $\tg > 0$.
The data-driven background estimate is determined using
a control region (CR) with low values of \MET. The background estimate
is validated using two different
validation regions (VRs): the first requires intermediate values of \MET and is denoted by VR(\MET),
while the second, denoted by VR($t$), imposes the same \MET \ requirement as the SR but reverses the
timing requirement by requiring $\tg < 0$. By construction, the various regions are mutually
exclusive, and any signal contamination of the CR, VR(\MET) and VR($t$) data samples
can be safely neglected.
The development of the background modeling using the CR, and its
validation using the VRs, were finalized
before the data in the SR was unblinded and the final results obtained.


\section{ATLAS detector}
\label{sec:detector}

 
\newcommand{\AtlasCoordFootnote}{
ATLAS uses a right-handed coordinate system with its origin at the nominal interaction point (IP)
in the center of the detector and the \(z\)-axis along the beam pipe.
The \(x\)-axis points from the IP to the center of the LHC ring,
and the \(y\)-axis points upwards.
Cylindrical coordinates \((r,\phi)\) are used in the transverse plane,
\(\phi\) being the azimuthal angle around the \(z\)-axis.
The pseudorapidity is defined in terms of the polar angle \(\theta\) as \(\eta = -\ln \tan(\theta/2)\).
Angular distance is measured in units of \(\Delta R \equiv \sqrt{(\Delta\eta)^{2} + (\Delta\phi)^{2}}\).}

The ATLAS detector~\cite{PERF-2007-01} at the LHC covers nearly the entire solid angle around the collision point.\footnote{\AtlasCoordFootnote}
It consists of an inner tracking detector surrounded by a thin superconducting solenoid, electromagnetic and hadron calorimeters,
and a muon spectrometer incorporating three large superconducting air-core toroidal magnets.
 
The inner-detector system (ID) is immersed in a \SI{2}{\tesla} axial magnetic field
and provides charged-particle tracking in the range \(|\eta| < 2.5\).
The high-granularity silicon pixel detector covers the vertex region and typically provides four measurements per track,
the first hit normally being in the insertable B-layer installed before Run~2~\cite{ATLAS-TDR-19,PIX-2018-001}.
It is followed by the silicon microstrip tracker, which usually provides eight measurements per track.
These silicon detectors are complemented by the transition radiation tracker (TRT),
which enables radially extended track reconstruction up to \(|\eta| = 2.0\).
The TRT also provides electron identification information
based on the fraction of hits (typically 30 in total) above a higher energy-deposit threshold corresponding to transition radiation.
 
The calorimeter system covers the pseudorapidity range \(|\eta| < 4.9\).
Within the region \(|\eta|< 3.2\), electromagnetic (EM) calorimetry is provided by barrel (EMB) and
endcap (EMEC)
high-granularity lead/liquid-argon (LAr) calorimeters,
with an additional thin LAr presampler covering \(|\eta| < 1.8\)
to correct for energy loss in material upstream of the calorimeters.
Hadron calorimetry is provided by the steel/scintillator-tile calorimeter,
segmented into three barrel structures within \(|\eta| < 1.7\), and two copper/LAr hadron endcap calorimeters.
The solid angle coverage is completed with forward copper/LAr and tungsten/LAr calorimeter modules,
optimized for electromagnetic and hadronic energy measurements respectively.
The abilities of the LAr EM calorimeter to measure the flight direction and the arrival time of photons, which are key to this analysis, are described in detail in Section~\ref{sec:variables}.
 
The muon spectrometer (MS) comprises separate trigger and
high-precision tracking chambers measuring the deflection of muons in a magnetic field generated by the superconducting air-core toroidal magnets.
The field integral of the toroids ranges between \num{2.0} and \SI{6.0}{\tesla\metre}
across most of the detector.
Three layers of precision chambers, each consisting of layers of monitored drift tubes, covers the region \(|\eta| < 2.7\),
complemented by cathode-strip chambers in the forward region, where the background is highest.
The muon trigger system covers the range \(|\eta| < 2.4\) with resistive-plate chambers in the barrel, and thin-gap chambers in the endcap regions.
 
Events of interest are selected by the first-level trigger system implemented in custom hardware,
followed by selections made by algorithms implemented in software in the high-level trigger~\cite{TRIG-2016-01}.
The first-level trigger accepts events from the \SI{40}{\MHz} bunch crossings at a rate below \SI{100}{\kHz},
which the high-level trigger further reduces in order to record events to disk at about \SI{1}{\kHz}.
 
An extensive software suite~\cite{ATL-SOFT-PUB-2021-001} is used in the reconstruction and analysis of real
and simulated data, in detector operations, and in the trigger and data acquisition systems of the experiment.


\section{Data and Monte Carlo samples}
\label{sec:samples}

 
This analysis uses a data set of $\sqrt{s} = 13$~\TeV \
$pp$ collisions recorded by the ATLAS detector from 2015 to 2018, corresponding to the full Run~2 period of the LHC.
Signal events with a Higgs boson produced in association with a $Z$ boson were collected using the unprescaled single-electron and single-muon triggers~\cite{TRIG-2018-01,TRIG-2018-05} with the lowest transverse momentum threshold for the electron or muon.
After data quality requirements~\cite{DAPR-2018-01} are applied to ensure that all detector components
were in good working condition, the data set corresponds to an integrated luminosity of
$139.0 \pm 2.4$~\ifb~\cite{DAPR-2013-01,Avoni_2018}.
The mean number of interactions per bunch crossing  (``pileup'') during Run~2 was $\langle\mu\rangle = 33.7$.
 
No SM background Monte Carlo (MC) simulation
samples are needed, since the analysis uses a fully data-driven background estimation procedure.
To study the signal, MC samples for the $ZH$ production process were generated at leading order (LO) using
\MGNLO[2.6.2]~\cite{Alwall:2011uj}, with a Higgs boson mass of 125~\GeV, and normalized to the cross sections reported in Ref.~\cite{deFlorian:2016spz}.
The generated events were interfaced to
\PYTHIA[8]~\cite{Sjostrand:2014zea} to simulate subsequent decays of the Higgs boson and supersymmetric particles in the signal model and also the parton showering. The A14 set of
tuned parameters~\cite{ATL-PHYS-PUB-2014-021} and the \NNPDF[2.3lo] PDF set~\cite{Ball:2014uwa}  were used in the event generation.
The signal model MC
events were passed through a \GEANT~\cite{Agostinelli:2002hh}
simulation of the ATLAS detector~\cite{SOFT-2010-01} and reconstructed with the same
software~\cite{ATL-SOFT-PUB-2021-001} as used for the data. The generation of the
simulated event samples includes the effect of pileup, both within
the same bunch crossing as well as in bunch crossings before or after the one containing the hard interaction.
Events in the simulation are weighted in order to reproduce
the amount of pileup observed in the Run~2 data-taking period.
 
The
final state and kinematics of the Higgs boson decay depend on the
masses of the NLSP and LSP, in addition to the lifetime of the NLSP. The signal generation was implemented
with the \ninotwo as
the NLSP and the \ninoone as the LSP, but the NLSP and LSP
masses and NLSP lifetime were treated as
free parameters. Therefore, GMSB signal models were generated in a three-dimensional grid.
The grid of MC signal models was constructed with
NLSP masses  from 30~\GeV \ up to 60~\GeV, near the kinematic limit enforced by the Higgs boson mass.
For each value of NLSP mass, LSP masses from 0.5~\GeV \
up to a value of $\mNLSP - 10$~\GeV\ were considered. Given the typical step size of 10~\GeV \
used for both the NLSP and LSP masses, a total of 18 different (\mNLSP, \mLSP) pairs were simulated
for a given value of NLSP lifetime.
For each set of NLSP and LSP mass
values, at least two different NLSP lifetimes, typically 2~ns and 10~ns, were simulated. Results
can be generalized to other lifetime values by appropriately
reweighting the samples of simulated events.
 
Some cross-checks of detector performance
utilize MC samples of \Zee \ events  and events with radiative
$Z \rightarrow \ell\ell\gamma$ decays, with $\ell\ell$
designating an opposite-sign same-flavor lepton
(electron or muon) pair.
The photon time resolution is not modeled accurately in the
MC simulation, which underestimates the time resolution
observed in data. Therefore, additional smearing is applied to the photon times as simulated in
the MC signal events. In addition to a contribution applied independently
to each photon, the additional smearing includes a correlated
event-level contribution to account
for the impact of the spread of the actual $pp$
collision times, which results from the longitudinal profiles of the colliding proton bunches along the
LHC beamline. The combined smearing contributions are tuned to match the
timing performance observed for electrons in data because,
due to their similar EM shower
development, electrons and photons have similar timing performance.
The correlated and uncorrelated contributions
to the time measurement
are deconvolved by studying the times of
electron--positron pairs in $Z \ra ee$ events.


\section{Photon pointing and timing measurements}
\label{sec:variables}

 
Photons produced in the decays of long-lived NLSPs can be nonpointing, delayed, or both.
This analysis is made possible by the capability of the
LAr EM calorimeters to perform precise measurements of the pointing  and timing of EM
objects, and in particular photons, as discussed in more detail below.
The measurements of pointing and timing are almost completely
uncorrelated for prompt backgrounds.
The lack of correlation
results from the fact that, as described below,
the pointing measurement uses the spread
of the EM shower to precisely measure its centroids in
the first two layers in the EM calorimeter, while the
timing measurement
uses the time reconstructed from the pulse-shape of only the
second-layer cell with the maximum energy deposit.
In contrast, these measurements would be correlated for a photon that is produced in the decay of a heavy NLSP.
Using both variables therefore provides a powerful way to separate
signal from background.
 
\subsection{Pointing measurements}
 
For $\abseta < 2.5$,
the LAr EM calorimeter is segmented into three layers in depth that can be
used to measure the longitudinal profile of the shower. The first layer uses highly
granular ``strips'' segmented in the $\eta$ direction,
with a typical transverse segmentation of
$\Delta\eta \times \Delta\phi = 0.003 \times 0.1$,
designed to allow efficient
discrimination between single-photon showers
and two overlapping showers from the decay of a $\pi^0$ meson. The second layer has a typical transverse
segmentation of $\Delta\eta \times \Delta\phi = 0.025 \times 0.025$, and
collects most of the energy deposited in the calorimeter by EM showers
initiated by electrons or photons.
Very high energy EM showers can leave significant energy deposits in the third layer, which serves as a ``tail catcher'' and which can also
be used to correct for energy leakage beyond the EM calorimeter.
 
By precisely measuring the centroids of the EM shower in the first and second EM calorimeter layers,
the flight direction of photons can be determined, from which one can calculate the value of
\zorigin, defined as the $z$-coordinate of the extrapolation of the
line through these centroids in the $z$--$r$ plane back to the beamline ($x = y = 0$).
The angular resolution of the EM calorimeter's measurement of the flight direction of prompt photons is
about 60~mrad/$\sqrt{E}$, where $E$ is the photon energy in \GeV.
This angular resolution
corresponds to a resolution in \zorigin \ of
about 15~mm for prompt photons with
energies in the range of 50--100~\GeV\ in the EM barrel calorimeter.
Given the geometry, the
\zorigin resolution is worse for photons reconstructed in the endcap calorimeters.

This analysis  uses the measurement of the
photon flight direction to search for photons that do not point
back to the PV. The pointing variable used in the
analysis is $\zDCA = \zorigin - \zpv$, with
\zpv \ being the $z$-coordinate of the PV of the event, selected as
described in Section~\ref{sec:eventsel}.
Given that \zpv \ is measured with high precision using
the tracker, the \zdca
resolution is essentially equivalent to the resolution in \zorigin.
 
While the geometry of the EM calorimeter is optimized
for detecting
particles that point back to near
the nominal interaction point at the center of the detector
(i.e.\ $x = y = z = 0$), the fine segmentation allows good
pointing performance to be achieved over a  wide range of photon impact angles.  Using a sample of electrons from \Zee \ events,
the calorimeter pointing performance is
validated in data by using the finite spread of the $pp$ collision
region along the $z$-axis.
The pointing resolution degrades gradually
with increasing \abszdca (see Figure~1 in
Ref.~\cite{SUSY-2013-17}). The \GEANT-based
simulation of the ATLAS detector shows good agreement
with the resolution observed in data over the
range of \abszdca values (up to about 200~mm) that can
be accessed in data. This gives confidence in the ability of the MC
simulation to estimate the resolution at the larger \abszdca\ values characteristic
of signal photons. The resolution
remains much smaller than \abszdca\ in the full
region where the
signal is expected. The studies also show similar
pointing performance for photons and electrons, as expected given their similar EM shower development.

\subsection{Timing measurements}
\label{sec:timing}
 
Photons from long-lived NLSP decays would reach the LAr calorimeter with
a slight delay compared to prompt photons produced directly in the hard scattering. This
delay results
mostly from the flight time of the heavy NLSP, which would have  a  distribution
of relativistic speed ($\beta = v/c$) that typically peaks near 0.9 and
has a tail to much lower values.  In addition,
the opening angle in the NLSP decay, which causes the photon to be nonpointing,
results in the geometrical path to the calorimeter being longer than that for a prompt photon from the PV.
 
The EM calorimeter, with its novel ``accordion'' design, and
its readout, which incorporates fast shaping, has excellent time
resolution. Test-beam measurements~\cite{2008NIMPA.597..178A}
of individual EMB calorimeter modules demonstrated a time resolution
of $\approx100$~ps in response to high-energy electrons. In situ
calibration tests in the ATLAS cavern show a time
resolution of $\approx 70$~ps~\cite{Abreu:2010zzc}, limited not by the
calorimeter but by the jitter of the calibration pulse injection system.
To avoid that the electronic readout significantly degrades the intrinsic
timing performance of the calorimeter,
quality-control tests during production of
the electronics required the clock jitter on the LAr readout
boards to be less than 20~ps,
with typical values of 10~ps~\cite{Buchanan:2008zza}.
 
During the Run~2 period from 2015 to 2018, the hardware configuration
of the readout of the various LAr channels was adjusted such that the
channels were synchronized online within a uniformity of order
1~ns.  While this level of precision is sufficient for the trigger
system to correctly identify the bunch crossing corresponding to
a particular event of interest, a dedicated calibration
for each channel is applied offline to improve the time resolution significantly for EM clusters.  The time calibration is based on a large sample of \Wenu \  events and is performed with zero defined as the expected value for a prompt electron from the PV of the hard collision.
The time calibration includes corrections for the dependence of the time measurement on energy, crosstalk, flight path from the PV, collision conditions, and the behavior of individual channels.

The LAr energy and time for each calorimeter cell
are reconstructed by applying the
optimal filtering algorithm~\cite{cleland} to the set of four samples of the signal shape
read out for each calorimeter channel,
with successive samples of the waveform separated by 25~ns.  More specifically,
the deposited energy per cell and the time of the deposition are calculated using
appropriately weighted linear combinations of the set of samples of the waveform:
 
\begin{equation*}
\label{eq:energy}
E  = \sum_{i=0}^{3} a_i S_i  \ \ \ \   \mathrm{and} \ \ \ \
t  = \frac{1}{E} \sum_{i=0}^{3} b_i S_i  \ \     ,
\end{equation*}
 
\noindent where $S_i$
denotes the four samples of the signal waveform. The parameters
$a_i$ and $b_i$ are the optimal filter coefficients (OFC), the values of which are
calculated, knowing the pulse shape and noise autocorrelation matrix, to deliver the
best energy and time resolutions.
 
For this analysis, the arrival time of an EM object
is measured using the second-layer EM calorimeter cell
with the largest energy deposit among cells in the associated EM cluster (\Ecell).
For the EM shower of an electron or photon with an energy
in the range of interest, this cell typically
contains about 20\%--50\% of the total energy deposited in the EM shower.

To cover the full dynamic range of physics signals of interest, the
ATLAS LAr calorimeter readout boards~\cite{Buchanan:2008zza} employ three
overlapping linear gain scales, dubbed High, Medium and Low, where
the relative gain is reduced by a factor of about ten for each successive scale.  For a given event,
any individual LAr readout channel is digitized using the gain scale that provides
optimal energy resolution, given the energy deposited in that calorimeter cell.
The calibration of the time is determined separately for High and
Medium gain for each channel. The number of electron candidates from
the \Wenu \ sample digitized using Low gain, typically those with
energies above about 200~\GeV, is insufficient to obtain
statistically precise results for the calibration constants. Therefore, the analysis
requires that selected photons be digitized using either High or Medium gain, which has a negligible impact on
the signal efficiency for this analysis.

An independent  sample
of \Zee \ events is used to validate the time calibration and determine the resolution that is obtained by performing Gaussian fits to the time distributions in bins of cell energy. Figure~\ref{fig:timingresn} shows the time resolution for High and Medium gain cells with $|\eta| < 0.4$, as a function of \Ecell,
the energy in the second-layer calorimeter cell used to calculate the time for the sample of \Zee \ events.  Similar results are obtained over the full coverage of the EM calorimeter.
 
The time resolution, $\sigma(t)$, is expected to follow
the form $\sigma(t) = p_0/\Ecell \oplus p_1$, where \Ecell \ is the cell energy,
$\oplus$ denotes addition in quadrature, and the fit parameters
$p_0$ and $p_1$ are the coefficients of the so-called noise term
and constant term, respectively.
Superimposed on Figure~\ref{fig:timingresn} are the results of fits to
this expected form of the time resolution function. The fits yield
values of $p_1$, which gives the time
resolution in the limit of large energy deposits, of
208~ps (217~ps)  for
High (Medium) gain.
The slightly poorer resolution for Medium
gain is due to limited event yields in
the \Wenu \ sample used to determine the time calibration constants.
By examining the timing distributions of the two electrons in \Zee \ candidate events,
the time measurements are observed to include a correlated contribution of $\approx190$~ps,
which, for a given PV position, agrees well with the expected distribution of collision times due to the
longitudinal spread of the proton bunches along the LHC beamline.
Assuming this contribution to the time resolution to be uncorrelated with the LAr contributions implies the LAr contributions are $\sim$~50--150~ps, depending on $|\eta|$, demonstrating that the offline calibration
achieves an in situ time resolution comparable to that achieved in
the more constrained testbeam meaurements~\cite{2008NIMPA.597..178A}.
There is only a modest variation in time resolution over the ranges of photon energies considered since, as described in Section~\ref{sec:strategy} (see Table~\ref{tab:AnalysisCutsBins}), selected photons in the analysis are required to have \Ecell\ $>$ 7 GeV.

\begin{figure}[hbpt]
\centering
\includegraphics[width=0.45\textwidth]{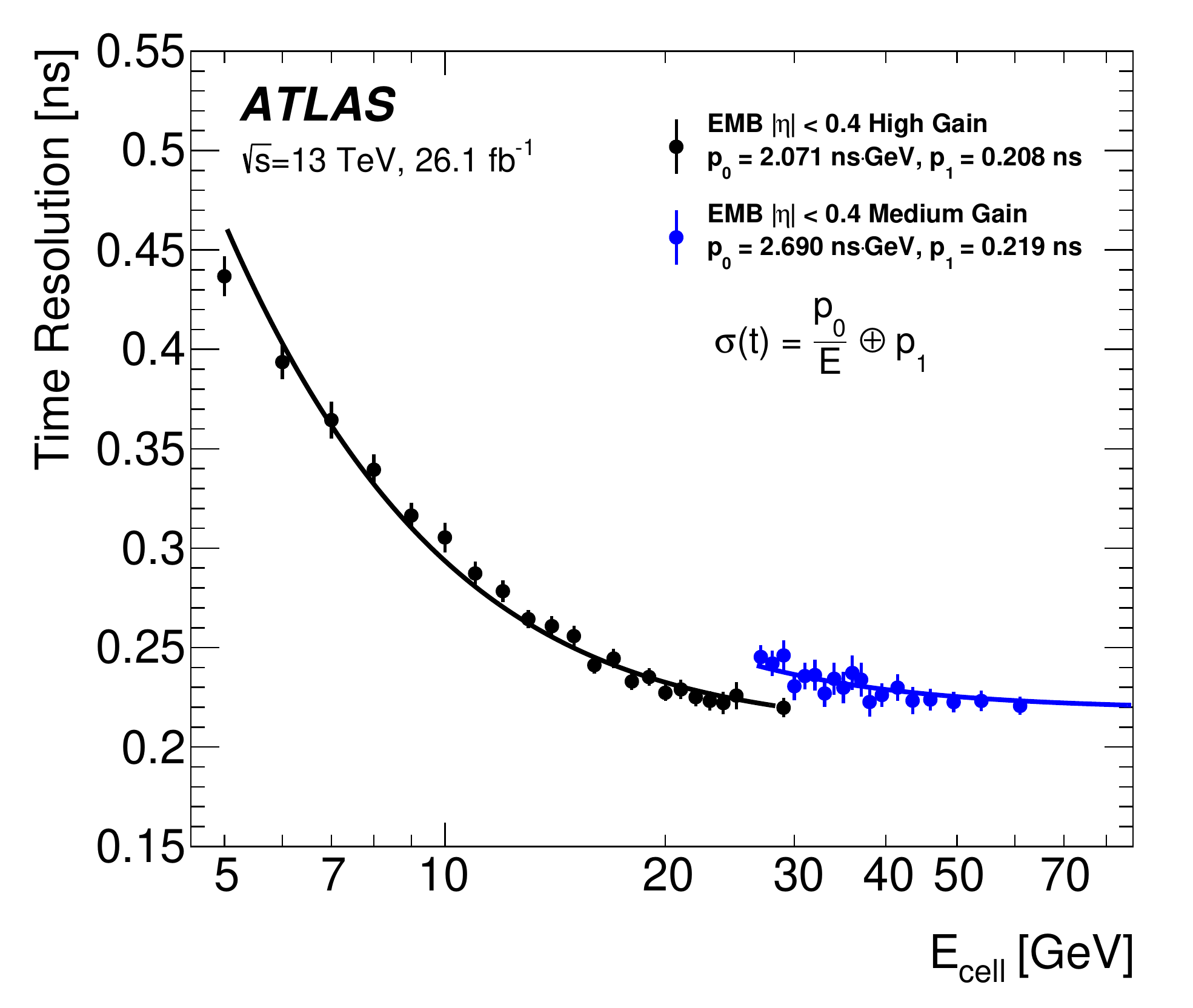}
\caption{Time resolution obtained
with the LAr EM calorimeter as a function of \Ecell, the energy in the second-layer cell with the largest energy deposit. The solid lines show the results of fitting the expected form, namely $\sigma(t) = p_0/\Ecell \oplus p_1$, to the data points.
The results shown are obtained using \Zee \ events recorded in 2018, for electrons in the EMB calorimeter with $|\eta| < 0.4$, and for both the High and Medium gain readout scales. Similar results are obtained over the full Run~2 data set and for the full coverage of the EM calorimeter.
}
\label{fig:timingresn}
\end{figure}
 
The signal timing templates are constructed using the timing variable of the simulated signal samples.
Additional resolution uncertainties are assigned to cover the observed discrepancy between radiative $Z \rightarrow \ell \ell \gamma$ events in data and those from MC simulation. The photon timing resolution uncertainty is implemented in the likelihood function (see Section~\ref{sec:systematics} for details) such that the effects in the timing distribution are correlated between pointing categories. Uncertainties accounting for fluctuations in the statistically limited signal MC samples are also included in the likelihood function.


\section{Event selection }
\label{sec:selection}

The requirements applied in the selection of various physics
objects in the event are described below, followed by the selection of events and the optimization of these selections.

\subsection{Object selection}
\label{sec:object}
 
The reconstruction and identification of electrons and
photons are described in Ref.~\cite{EGAM-2018-01}. Shape variables computed
from the lateral and longitudinal energy profiles of
the EM showers in the calorimeter are used to identify
photons and electrons and to distinguish them from backgrounds.
 
Photons are required to satisfy $\pt > 10$~\GeV\ and
$|\eta|<2.37$, excluding the region $1.37 < |\eta| < 1.52$, which corresponds to the transition
region between the EMB and EMEC calorimeters.
In addition, to reduce the
probability of jets being misidentified as photons, photons
must satisfy both calorimeter-based and track-based isolation requirements.
The calorimeter isolation variable is defined as the energy of EM calorimeter-cell clusters
in a cone of size $\Delta R=0.2$ around the photon candidate, excluding a fixed window
containing most of the photon shower. Corrections accounting for photon energy leakage
from this window into the surrounding cone, for pileup, and for the
underlying event are applied~\cite{EGAM-2018-01}.
The calorimeter isolation is required to be less than 6.5\% of the photon's transverse energy.
The track-based isolation variable is defined as the scalar sum of the \pt of all tracks
with $\pt>1$~\GeV\ within $\Delta R = 0.2$ of the photon candidate.
The track isolation variable is required to be less than 5\% of the photon's transverse energy.
 
A set of photon selection criteria, designed for high efficiency and modest background rejection,
defines the so-called Loose photon identification used in this analysis. The
Loose photon identification does not impose requirements on the EM shower shape in the first (strips)
layer of the EM calorimeter, and uses only variables that describe the
shower shape in the second layer of the EM calorimeter and leakage into the hadronic
calorimeter. The Loose photon identification thereby minimizes the bias against identification
of nonpointing photons. Figure~\ref{fig:zdcaeffic} shows the Loose
photon identification efficiency as a function of \abszdca, with
the simulated results for the signal photons from this analysis
superimposed on data and MC results
for SM $Z \rightarrow \ell \ell \gamma$  events. For
the $Z \rightarrow \ell \ell \gamma$  events, \abszdca is
measured from the center of the detector ($z = 0$),
exploiting the spread of $pp$ collisions along the beamline.
The radiative $Z \rightarrow \ell \ell \gamma$ sample is selected by requiring
events to have a photon plus two same-flavor opposite-sign leptons
(electrons or muons), with $m(\ell\ell\gamma)$ being within 10~\GeV\ of the
known $Z$ boson mass, and the dilepton invariant
mass being less than 83~\GeV. To reduce the impact of electrons
being misidentified as photons, candidate events in the $Z \rightarrow ee\gamma$ sample are vetoed if the invariant mass of either $e\gamma$
pair is consistent with the $Z$ boson mass.
The $Z \rightarrow \ell \ell \gamma$ selection results in a very pure sample of photons, from which the efficiency of applying the Loose photon identification requirement is measured after removing the remaining background component using a template fit to the $\ell \ell \gamma$ invariant mass distribution~\cite{PERF-2017-02}.
As seen in Figure~\ref{fig:zdcaeffic}, the Loose photon identification
efficiency is very high and quite constant
within the beam spot (\abszdca \ less
than about 200~mm), and the results for the various samples are in
good agreement, providing confidence in the use of MC simulation to
estimate the identification efficiency for signal photons at higher \abszdca \ values.
As shown, the efficiency
decreases gradually for photons produced further from the center of
the detector. To cover potential mis-modeling of the efficiency drop in the high pointing region,  photon identification uncertainty is introduced and discussed in Section~\ref{sec:systematics:signalshape}
 
\begin{figure}[hbpt]
\centering
\includegraphics[width=0.45\textwidth]{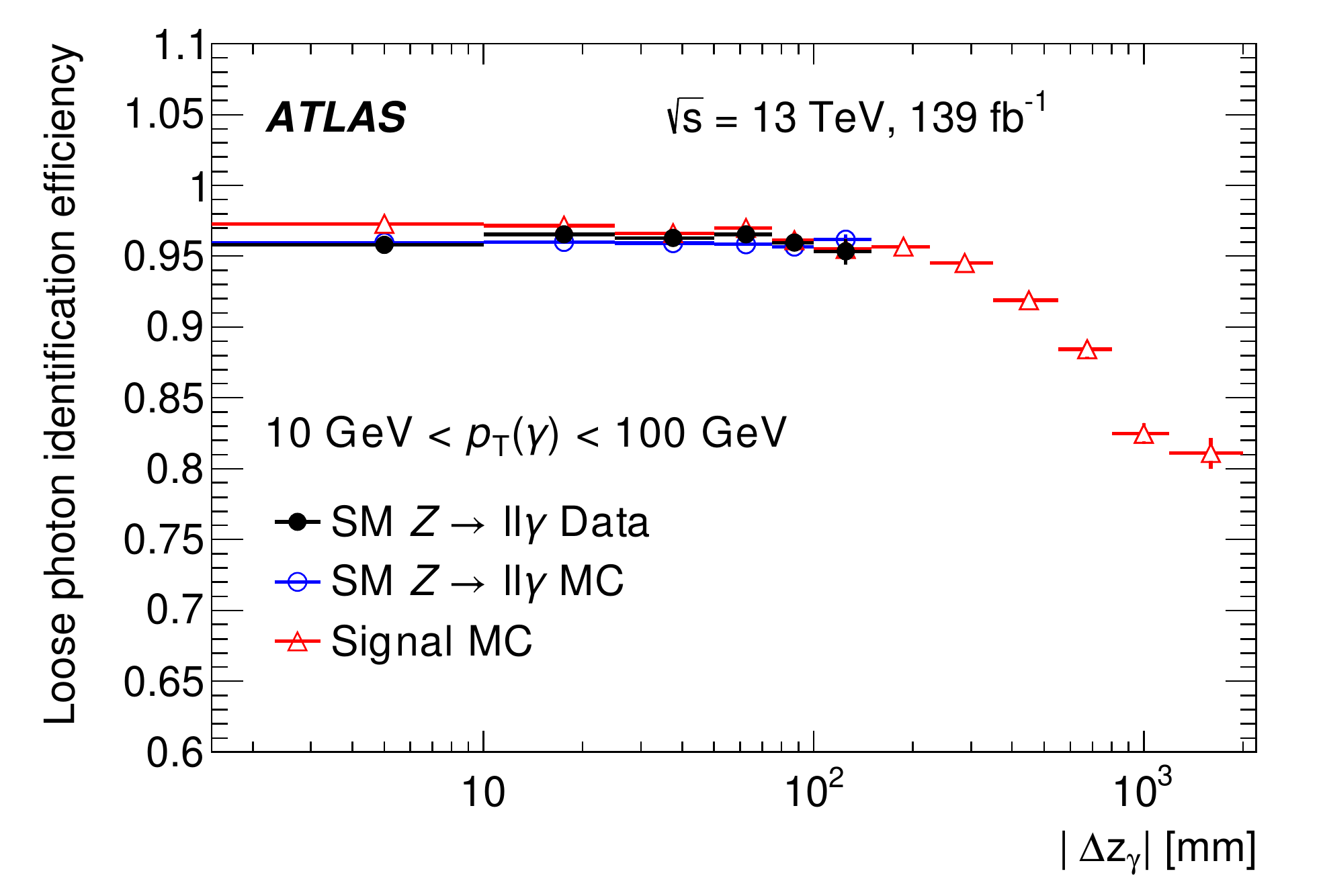}
\caption{The Loose photon identification efficiency, shown as a function of \abszdca, for photons with transverse momentum
in the range 10--100~\GeV\ and with $\Ecell > 7$~\GeV. The
simulated results for signal photons from long-lived particle
decays in this analysis are overlaid with data and MC results
for SM $Z \rightarrow \ell \ell \gamma$  events. For
the $Z \rightarrow \ell \ell \gamma$  events, \abszdca is
measured from the center of the detector ($z = 0$),
exploiting the spread of $pp$ collisions along the beamline.
}
\label{fig:zdcaeffic}
\end{figure}
 
Some studies in this analysis
use the Medium photon working point, which adds a
requirement on one shower shape variable in the strips layer in order
to increase the photon purity at the cost of some decrease
in efficiency.
Detailed descriptions of both the Loose and Medium photon identification criteria
can be found in Ref.~\cite{EGAM-2018-01}.
 
Electrons are reconstructed by matching tracks in the ID to topological clusters of calorimeter cells
formed using the same dynamical, topological cell-clustering algorithm as
in the photon reconstruction~\cite{EGAM-2018-01}.
Electron candidates are required to have $\pt>10$~\GeV\ and $|\eta|<2.47$, and must satisfy a Medium
electron identification selection based on a likelihood discriminant using
calorimeter shower shapes and track parameters \cite{EGAM-2018-01}.
Isolation criteria using calorimeter- and track-based information are applied to electrons.
The reconstructed track matched to the electron candidate must be consistent with the PV
position: its longitudinal impact parameter $z_0$ relative to the PV is required to satisfy
$|z_0 \sin\theta|<0.5$~mm. In addition, the electron track's transverse impact parameter with respect to the
beam axis divided by its uncertainty, $|d_0|/\sigma(d_0)$, must be less than 5.
 
Muons are reconstructed by matching tracks from the MS and ID
subsystems. Muons in the range $2.5<|\eta|<2.7$ without an ID track but whose MS track is
compatible with the interaction point are also considered. Muon candidates are required to have $\pt>10$~\GeV\ and $|\eta|<2.7$, and must satisfy the Medium muon
identification requirements~\cite{MUON-2018-03}.
Muons are required to satisfy calorimeter- and track-based isolation requirements~\cite{MUON-2018-03} that are
95\%--97\% efficient for muons with $\pt\in[10,60]$~\GeV\ and 99\% efficient for $\pt>60$~\GeV.
Muon tracks must satisfy $|z_0 \sin\theta|<0.5$~mm and $|d_0|/\sigma(d_0)<3$.
Finally, $\tau$ leptons are required to have at least 20 GeV of \pt and to lie within $|\eta|$ < 1.37 or 1.52 < $|\eta|$ < 2.5
 
Jets are used only indirectly in the analysis, through an overlap removal procedure and corrections
to the calculation of \met. Jets are reconstructed by applying a particle-flow algorithm~\cite{PERF-2015-09} to noise-suppressed positive-energy topological clusters~\cite{PERF-2014-07} of calorimeter cells from the anti-$k_t$ algorithm~\cite{Cacciari:2008gp,Fastjet} with  radius parameter $R=0.4$.
Jets must have $\pt>25~$\GeV\ and $|y|<4.4$.
A jet-vertex-tagger (JVT) multivariate discriminant~\cite{PERF-2014-03} is applied to jets with $\pt < 60$~\GeV\ and $|\eta|<2.4$ to suppress jets from pileup.
 
An overlap removal procedure is performed in order to avoid double-counting of objects.
First, electrons overlapping with photons within $\Delta R<0.4$
are removed.
Jets overlapping with photons ($\Delta R<0.4$) and electrons ($\Delta R<0.2$) are removed. Electrons
overlapping with the remaining jets ($\Delta R<0.4$) are removed, to match the requirements imposed when measuring isolated electron efficiencies.
Finally, muons overlapping with photons or jets ($\Delta R<0.4$) are removed.
 
The value of \MET\ is defined as the magnitude of the negative vector sum of the
transverse momenta of the selected photons, electrons, muons and jets, as well as the
transverse momenta of remaining low-\pt particles, estimated using tracks matched to the PV but not to any of the selected objects~\cite{PERF-2016-07}.

\subsection{Event selection}
\label{sec:eventsel}
 
Selected events are required to have a candidate PV for the
hard scatter, reconstructed
from at least two charged tracks, each with $\pT > 500$~\mev.
If multiple vertices are reconstructed, the PV is selected as the one with
the largest sum of the $\pT^2$ values of the associated tracks.
 
Events are
required to pass the lowest unprescaled single-lepton
(electron or muon) trigger. For data taken in 2015, the electron
(muon) trigger required
the lepton to satisfy $\pt > 24 \ (20)$~\GeV. For 2016--2018, these kinematic
requirements on the trigger leptons were both increased to $\pt > 26$~\GeV.
The offline event selection required at least one electron or muon, with
$\pt > 27$~\GeV, satisfying Medium identification, and
matching the
trigger lepton.
 
Triggered events are subject to a common set of preselection
requirements. In addition to the lepton requirement above, selected
events are required to have at least one photon candidate with
$\pt > 10$~\GeV \ and satisfying Loose photon identification.
 
To reject events
with an electron misreconstructed as
a photon, events with an electron
are rejected if the invariant mass of the electron--photon pair
satisfies $|m_{e\gamma} - m_Z| < 15$~\GeV.
 
To suppress photons from earlier or later bunch crossings,
photons are required to have measured arrival times at the LAr
calorimeter of $|t_\gamma| < 12$~ns.

\subsection{Analysis strategy and optimization}
\label{sec:strategy}
 
Due to their different signal-to-background ratios,
the $1\gamma$ and  $\ge2\gamma$ final states are treated
separately in the analysis. The $\ge2\gamma$ final state has
lower background, and higher sensitivity for smaller values
of the NLSP
lifetime, for which there is a significant probability
for both NLSPs to decay before reaching the EM
calorimeter.
The use of the $1\gamma$ final state recovers sensitivity for
events where one photon is missed, or where one NSLP decays
beyond the boundaries of the ID.
 
For all selected events, the
final statistical analysis uses the pointing and timing measurements
of only one photon. Since the \zdca \ distribution should be symmetric for
both signal and background, the pointing distribution is
folded by taking $|\zdca|$ as the variable of interest instead
of \zdca \ itself. The inputs to the statistical analysis are therefore the values of $|\zdca|$ and \tg \ measured for the
photon selected in each event.
Given the fact that,
due to geometry, the pointing resolution along
the $z$-axis is better for EMB photons than for EMEC photons,
the photon selected for the statistical analysis
must be detected in the EMB. Thus, $1\gamma$ events must
have their single photon in the EMB. For $\ge2\gamma$ events,
at least one photon must be in the EMB; for events with more
than one EMB photon, the highest-\ET \ EMB
photon is selected for the statistical analysis.

A study was performed in order to optimize the SR definition
across the two-dimensional signal grid of NLSP and LSP masses. The kinematic distributions of the signal have a
significant dependence on the value of the mass splitting, $\Delta m = \mNLSP - \mLSP$.
Higher values of $\Delta m$ typically lead to production of higher-energy photons and to more modest
values of \MET. In contrast, lower $\Delta m$ leads to lower
photon energies and higher values of \MET. The optimization
study revealed that close-to-optimal results can be achieved
across the entire signal grid by the
definition of two different SR selections:
the \lowdm \ SR selection is applied to signal models with
$\Delta m = 10$~\GeV\
 
and
the \highdm \ SR selection is applied to signal models with
$\Delta m > 10$~\GeV.
The definitions of the two SR selections, which include different requirements
on \Ecell \ and \MET,  are  summarized in Table~\ref{tab:AnalysisCutsBins}, along with
definitions of the corresponding CRs and VRs.
 
While exploring the optimal SR selections,
the optimization procedure also examined how to best choose
the number of bins and locations of the corresponding bin edges
for both the photon pointing and timing variables. Considerations
included maximizing the sensitivity while avoiding any empty bins
in the required background templates constructed from CR data.
The binning optimization was performed separately for the
$1\gamma$ and $\ge2\gamma$ final states. The results led
to the choice of a common pointing binning, with a total
of five pointing categories.
Optimization of the timing binning led to seven (six) bins
for the $1\gamma$ ($\ge2\gamma$) final state, the
difference being mostly due to the lower CR event yields
for the $\ge2\gamma$ channel. The bin definitions are included
in Table~\ref{tab:AnalysisCutsBins}.

\begin{table*}[hbpt]
\centering
\caption{Optimized requirements defining the \lowdm \ and \highdm \ SR selections, as well as the corresponding
CRs and VRs. Also included are the optimized binnings for photon pointing and timing
for both the $1\gamma$ and $\ge2\gamma$ channels.}
\label{tab:AnalysisCutsBins}
\begin{tabular}{|c|cccccccc|}
\hline
& \multicolumn{4}{c|}{Low-$\Delta m$ selection} & \multicolumn{4}{c|}{High-$\Delta m$ selection} \\ 
& \multicolumn{4}{c|}{($\Delta m = 10$~\GeV)} & \multicolumn{4}{c|}{($\Delta m > 10$~\GeV)} \\ \cline{2-9}
Parameter & \multicolumn{1}{c|}{\ \  \ CR \ \ \ } & \multicolumn{1}{l|}{VR(\MET)} & \multicolumn{1}{l|}{VR($t$)} & \multicolumn{1}{l|}{\ \  \ SR \ \ \ }
& \multicolumn{1}{c|}{\ \  \ CR \ \ \ } &
\multicolumn{1}{l|}{VR(\MET)} & \multicolumn{1}{l|}{VR($t$)} & \multicolumn{1}{l|}{\ \ \ SR \ \ \ } \\ \hline \hline
\Ecell \ [\GeV] & \multicolumn{4}{c|}{> 7} & \multicolumn{4}{c|}{> 10} \\ \hline
\MET \ [\GeV] & \multicolumn{1}{c|}{< 30} & \multicolumn{1}{c|}{30--50} & \multicolumn{1}{c|}{> 80} & \multicolumn{1}{c|}{> 80} & \multicolumn{1}{c|}{< 30} & \multicolumn{1}{c|}{30--50} & \multicolumn{1}{c|}{> 50} & > 50 \\ \hline
\tg \ [ns] & \multicolumn{1}{c|}{~~> 0} & \multicolumn{1}{c|}{~~~~~> 0} & \multicolumn{1}{c|}{~~< 0} & \multicolumn{1}{c|}{~~> 0} & \multicolumn{1}{c|}{~~> 0} & \multicolumn{1}{c|}{~~~~~> 0} & \multicolumn{1}{c|}{~~< 0} & ~~> 0 \\ \hline
\ \abszdca \ bins [mm] \ & \multicolumn{8}{c|}{[0, 50, 100, 200, 300, 2000]} \\ \hline
$|\tg|$ \ bins [ns] & \multicolumn{8}{c|}{ } \\
\multicolumn{1}{|r|}{$1\gamma$ channel} & \multicolumn{8}{c|}{[0, 0.2, 0.4, 0.6, 0.8, 1.0, 1.5, 12.0]} \\
\multicolumn{1}{|r|}{$\ge2\gamma$ channel} & \multicolumn{8}{c|}{[0, 0.2, 0.4, 0.6, 0.8, 1.0, 12.0]~~~~~~~~} \\ \hline
\end{tabular}
\end{table*}
 
The signal acceptance is about 22\% (10\%) for signal models with an
NLSP lifetime of 2~ns (10~ns), where the acceptance is defined as the fraction of simulated signal events passing the trigger
and having at least one lepton and at least one photon reconstructed
in the final state. The lower acceptance for higher NLSP lifetimes
is expected, as more NLSP decays occur outside the ATLAS ID and their
decay photons are not captured by the EM calorimeter.
The signal efficiency is around 40\% for all the signal models,
defined as the fraction of simulated signal events
within the signal acceptance that pass the SR selection requirements.


\section{Background estimation}
\label{sec:modeling}

 
The background is expected to be completely dominated
by $pp$ collision events, with possible backgrounds
due to cosmic rays, beam-halo events, or other noncollision
processes being negligible. The source of the
Loose photons in background events contributing to the
selected sample is expected to be either a prompt photon,
an electron misidentified as a photon, or a jet
misidentified
as a photon. In each case, the object providing the
Loose photon signature is prompt and
originates from the PV.
 
The pointing and timing distributions expected for the
background sources are determined entirely using
control samples in data.
This avoids a reliance on the
precise MC simulation of the pointing and timing performance
for the backgrounds, and especially of the
tails of their $|\zdca|$ \ and $t_\gamma$ distributions.
Additionally, using
data samples naturally accounts for the influence of pileup, the
possibility of selecting the wrong PV, and any instrumental
or other effects that might influence the background
measurements.
 
It is known~\cite{SUSY-2013-17}
that the measured $|\zdca|$ and \tg \ distributions of
genuine photons are narrower than those of
other physics objects (particularly jets) that
are misreconstructed as
photons. The $|\zdca|$  distribution is particularly sensitive
to the background composition, since it depends on details
of the EM shower development. Consequently, the statistical
analysis (see Section~\ref{sec:stats}) uses the $|\zdca|$
value only
to separate the events into mutually independent pointing
categories, and implements
the normalization of the background in each pointing category
as an independent, unconstrained nuisance parameter. Therefore,
the fitting procedure eliminates the need to model
the pointing distribution of the background, and reduces the
background modeling issue to one of determining the shape of the
background timing template.
 
Variations of the shape of the background \tg \ distribution are
explored by dividing the CR data into a
genuine-photon enriched sample and a fake-photon enriched sample.
The actual purities of the individual samples do not need
to be known; they just need to be different from each other, so
that the total background timing template can then be
constructed as a linear combination of the timing
templates from these two samples.
 
The sample of events from the  CR is divided
into mutually exclusive subsamples
according to whether or not the photon passes an identification requirement (typically either the Medium or Tight definitions
given in {Ref.~\cite{EGAM-2018-01}}) that
is more stringent than the Loose identification used to
define the overall CR.
The photons that fail the more stringent identification
are used to form the fake-photon enriched sample, while those
that pass it are put in the genuine-photon enriched sample.
To increase the event yield, the genuine-photon
enriched control
sample also includes photons from the
radiative $Z \rightarrow \ell\ell\gamma$
decay sample.
 
As discussed earlier, the time resolution depends on \Ecell,
the energy deposited in the second-layer cell of the EM
calorimeter that is used to measure the arrival time of the photon.
To mitigate any potential discrepancy between the CR and SR timing shapes that arises from kinematic differences, a
reweighting procedure is applied to the control samples so that the photon \Ecell \ distribution in the CR after reweighting
is the same as that in the SR.
 
Due to imperfections in calibrations, limited event yields, or other effects, the photon timing distributions in the data can have small nonzero mean values, at the level of up to a few tens of
picoseconds. Before constructing the final background timing templates, the means are determined by fitting
the timing distributions, which are then shifted to eliminate any small residual nonzero mean values.
To avoid any bias being introduced by possible signal contributions, the fits are restricted to the core of the timing
distributions.
 
Figure~\ref{fig:templates} shows the unit-normalized shapes of
the pointing and timing distributions for the inclusive
genuine-photon enriched and fake-photon enriched samples.
Superimposed are the distributions for selected signal
models.
 
\begin{figure}[hbpt]
\centering
\includegraphics[width=.48\textwidth]{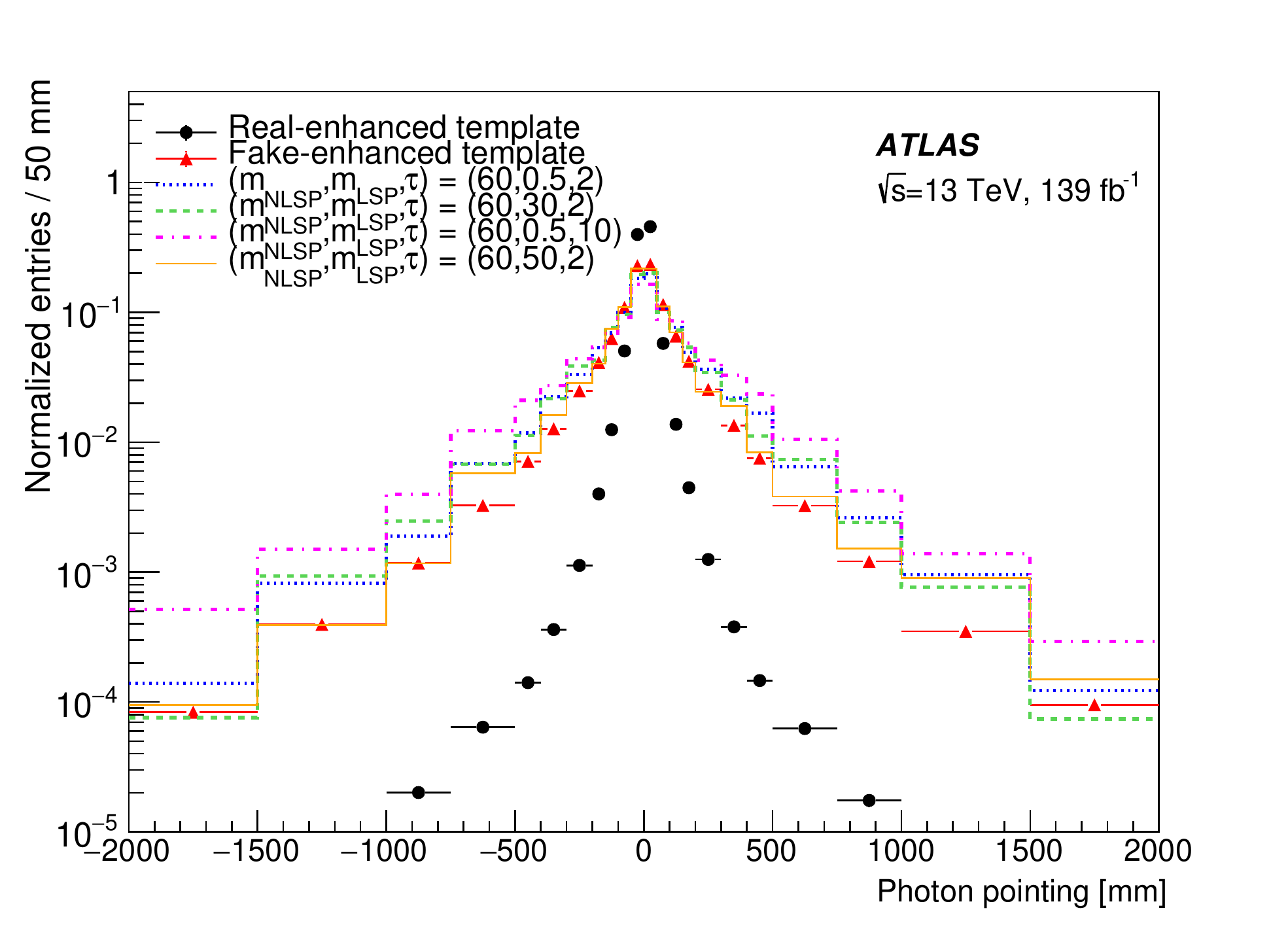}
\includegraphics[width=.48\textwidth]{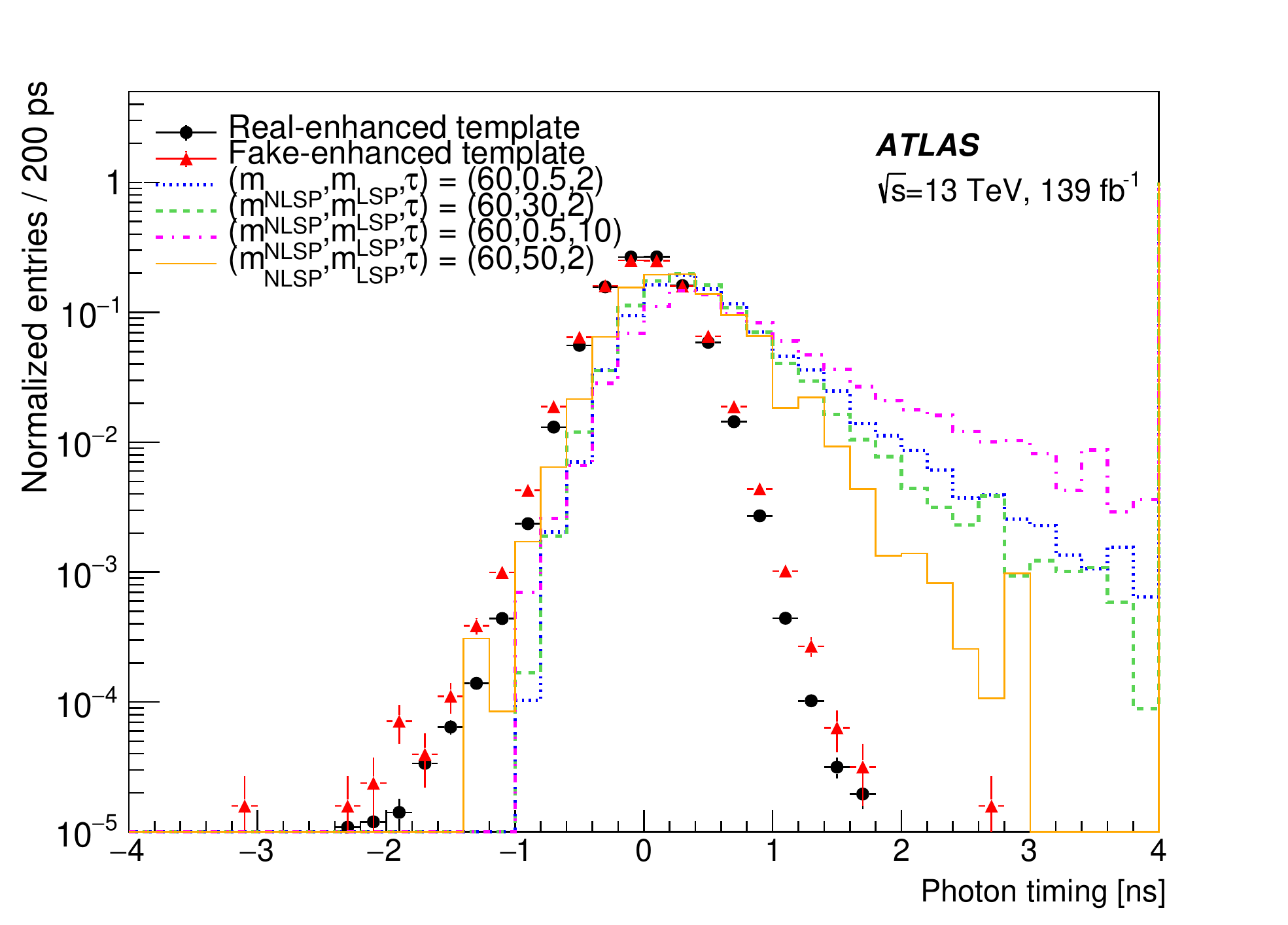}
\caption{Photon pointing (\zdca) and timing (\tg) distributions
for the genuine-photon enriched and fake-photon enriched
samples. Superimposed are the expected distributions for
selected signal models.
The signal
models are labeled by the values of their NLSP and LSP masses
($m_{\mathrm{NLSP}}$ and $m_{\mathrm{LSP}}$ in \GeV) and NLSP lifetime ($\tau$ in ns).}
\label{fig:templates}
\end{figure}
 
The timing templates are derived separately in each of the pointing categories. In general, the fraction of the genuine-photon enriched
template in the linear combination is a free nuisance parameter in the likelihood function and can be determined when fitting to data. In two of the pointing categories, the shapes
of the timing distributions for the genuine-photon enriched and fake-photon enriched
samples are found to be statistically consistent according to a Kolmogorov--Smirnov test,
so the fit cannot obtain a well-determined value for the nuisance parameter.
For these two pointing categories, the genuine-photon
enriched and fake-photon enriched template fractions are each
set to a constant value of 0.5, to exploit the
statistical power obtained by combining the two samples.


\section{Systematic uncertainties}
\label{sec:systematics}

 
Systematic uncertainties affecting the search are primarily experimental. Theoretical uncertainties are also relevant,
when the search result is interpreted to yield a constraint on the decay of the Higgs boson into long-lived particles.
 
In the statistical analysis,  the background
normalization for each \abszdca \ category
is determined using an independent nuisance parameter.
Therefore, it is not necessary to include systematic uncertainties for
the background's normalization or its shape in the variable \abszdca.
The various systematic
uncertainties relevant for this analysis can be divided into two categories: so-called
``flat'' uncertainties are not a function of \abszdca \ or \tg \ and affect only the overall
signal yield, while ``shape'' uncertainties are those that are related to the shapes of the
unit-normalized \abszdca \ and \tg distributions for signal or to the shape
of the background \tg template.
 
The various systematic uncertainties are outlined below.
However, their impact on the limits is very minor,
and the results of the analysis are dominated by the statistical
uncertainties. Implementing the full set of systematic
uncertainties degrades the sensitivity, as measured by
the expected 95\% CL limits on the branching ratio $\mathcal{B}$(\Hnlspnlsp), by only
about 5\% compared to the results obtained with only
statistical uncertainties considered.

\subsection{Signal normalization uncertainties}
 
Theoretical uncertainties of the signal model arise from the fact that the Higgs boson production processes are calculated at LO in QCD, from uncertainties in the value of the strong coupling constant \alphas, and from the choice of parton distribution functions. While the cross sections of the Higgs boson production processes are normalized to the best available theoretical calculations~\cite{deFlorian:2016spz}, the missing higher order QCD calculations in the event generation may lead to a potential mis-modeling of signal efficiency. Such uncertainties are estimated by varying
the factorization and renormalization scales used in the event generation separately  by factors of 0.5 and  2.0 relative to their nominal values, and the uncertainty is taken as the maximum difference between the yields obtained with the nominal scales and the different variations. In addition, the variation in signal selection efficiency due to the choice of  PDF sets in the event generation is also estimated.
The total combined theoretical uncertainty in the
signal yield reaches about 17\%, and
is included as a
single nuisance parameter in the fit.
 
Examples of instrumental systematic uncertainties which affect the overall signal yield include
uncertainties in photon and lepton reconstruction and identification efficiencies, reconstruction of \MET, and the integrated luminosity measurement.
The uncertainty in the integrated luminosity is 1.7\%,
evaluated using the methodology described in Ref.~\cite{DAPR-2013-01}.
The photon identification efficiency is studied using a
sample of photons from radiative
$Z \rightarrow \ell\ell\gamma$ decays,
selected as described in Section~\ref{sec:modeling}.
The efficiency of the Loose identification is around 95\%
for $\pt(\gamma)\approx 20$~\GeV, increasing gradually
to about 99\%
for $\pt(\gamma) > 60$~\GeV. The measured efficiency in
data is well reproduced in MC simulations.
 
Summing the instrumental uncertainties
in quadrature yields a small total uncertainty in the
signal yield, up to a few percent.
These uncertainties are incorporated into the likelihood model as a single nuisance parameter scaling the signal yields uniformly across the ten
pointing categories, where the magnitude
of the uncertainty is calculated separately for
the $1\gamma$ and $\ge2\gamma$ channels.

\subsection{Signal shape systematic uncertainties}
\label{sec:systematics:signalshape}
Uncertainties in photon timing and pointing measurements, the identification efficiency for photons with large pointing values, as well as the timing templates,  affect the distribution of
relative event yields between timing bins in different pointing categories. These uncertainties therefore affect the
shapes of the signal distributions, and are incorporated
into the likelihood model such that varying their corresponding nuisance parameters introduces correlated shape effects in the timing and pointing distributions.
For example, as shown in Figure~\ref{fig:zdcaeffic},
the Loose photon identification requirements provide
very high efficiency for identifying photons,
but that efficiency has
a modest dependence on \abszdca, dropping by about
15\% (5\%) for the highest \abszdca \ bin for low
(high) $\pT(\gamma)$ values. To obtain a conservative estimate of the uncertainty
in identifying nonprompt photons, the ratio of
efficiency in each \abszdca \ category
above 150~mm to the
efficiency measured in data for $\abszdca =100$--150~mm is calculated.
The deviation of these
ratio values from unity provides a binned relative
uncertainty in the photon identification efficiency that increases as \abszdca \ increases.
 
\subsection{Background shape systematic uncertainties}
 
Several types of systematic uncertainties are considered for
the shapes of the background timing distributions.
These include the bin-by-bin statistical uncertainties for
both the genuine-photon enriched and fake-photon enriched templates to account for fluctuations in data control samples.
 
In addition, a shape uncertainty correlated between timing bins within the same pointing category is introduced to account for uncertainties arising from the choice of CR samples for
determining the genuine-photon enriched and
fake-photon enriched timing distributions.
The uncertainties are derived by comparing different choices for
the division of events into those two subsamples, namely
by requiring photons to pass the Medium or the Tight
identification requirements in order to be included in the genuine-photon enriched sample.

Finally, a background-only fit of the likelihood function to the VR($t$) data is performed.
The VR($t$) data has the same requirements as the SR but with $\tg < 0$.
This selection ensures negligible signal contamination, and can therefore be
used to test the robustness of the background model.  The residual difference between data in the VR($t$) region and the fitted background is considered as a nonclosure uncertainty for the background templates.  While the nonclosure uncertainties
are zero for most of the time bins, a few have
values up to about 40\% of the background uncertainty in
that bin.


\section{Statistical analysis}
\label{sec:stats}

 
Binned likelihood fits to data in the SR are performed to test the background-only and signal-plus-background hypotheses.
The likelihood function is constructed by simultaneously fitting the ten timing distributions  to their pdfs.
The systematic uncertainties described in Section~\ref{sec:systematics} are incorporated into the likelihood function as nuisance parameters with Gaussian constraint pdfs.
The background normalization and the background template admixture
in each pointing category are floating nuisance parameters constrained by data.
 
A profile likelihood ratio is used to perform frequentist hypothesis tests~\cite{Cowan:2010js}. The $p_0$-value, defined as the probability of statistical fluctuations making the background distributions appear to contain at least as much signal as the data distributions, is calculated for different signal models.

Before applying the statistical analysis to the SR data, the
method and background model were validated by fitting to the
VR(\MET) data. Since the background estimation method essentially models the photon object property of timing and does not depend on the event selection, the fits to the VR(\MET) photon sample, which is statistically independent from the SR, provide a validation of the method.
As shown in Figure~\ref{fig:vrmet}, the VR(\MET)
data are well described by a background-only fit, for both the
\highdm \ and \lowdm \ selections.  Further validation of
the statistical analysis included signal injection tests, in
which varying signal models and strengths were injected
into the VR(\MET) data; in all cases, the fit
correctly extracted the strength of the injected signal.

\begin{figure*}[hbpt]
\centering
\includegraphics[width=0.48\textwidth]{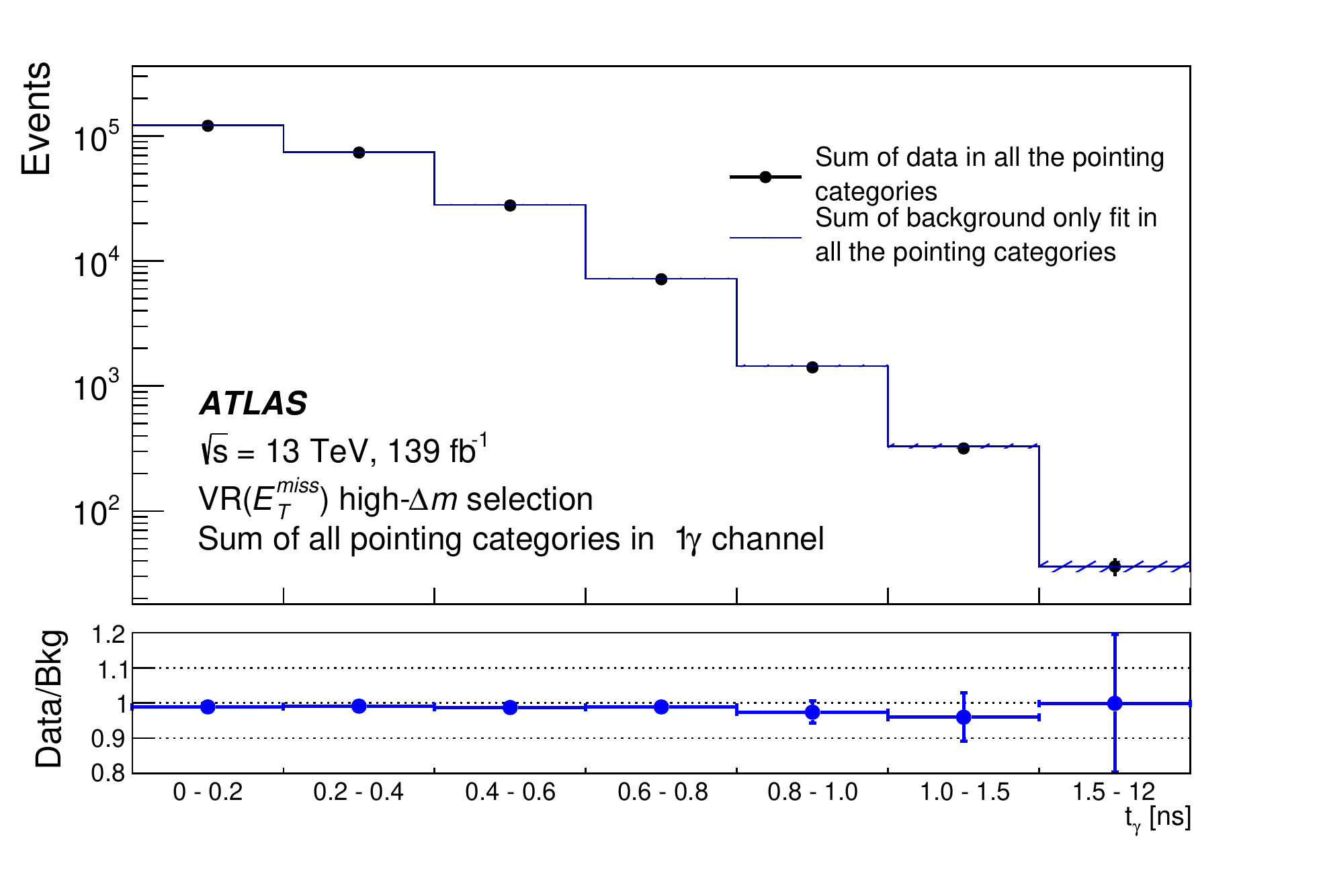}
\includegraphics[width=0.48\textwidth]{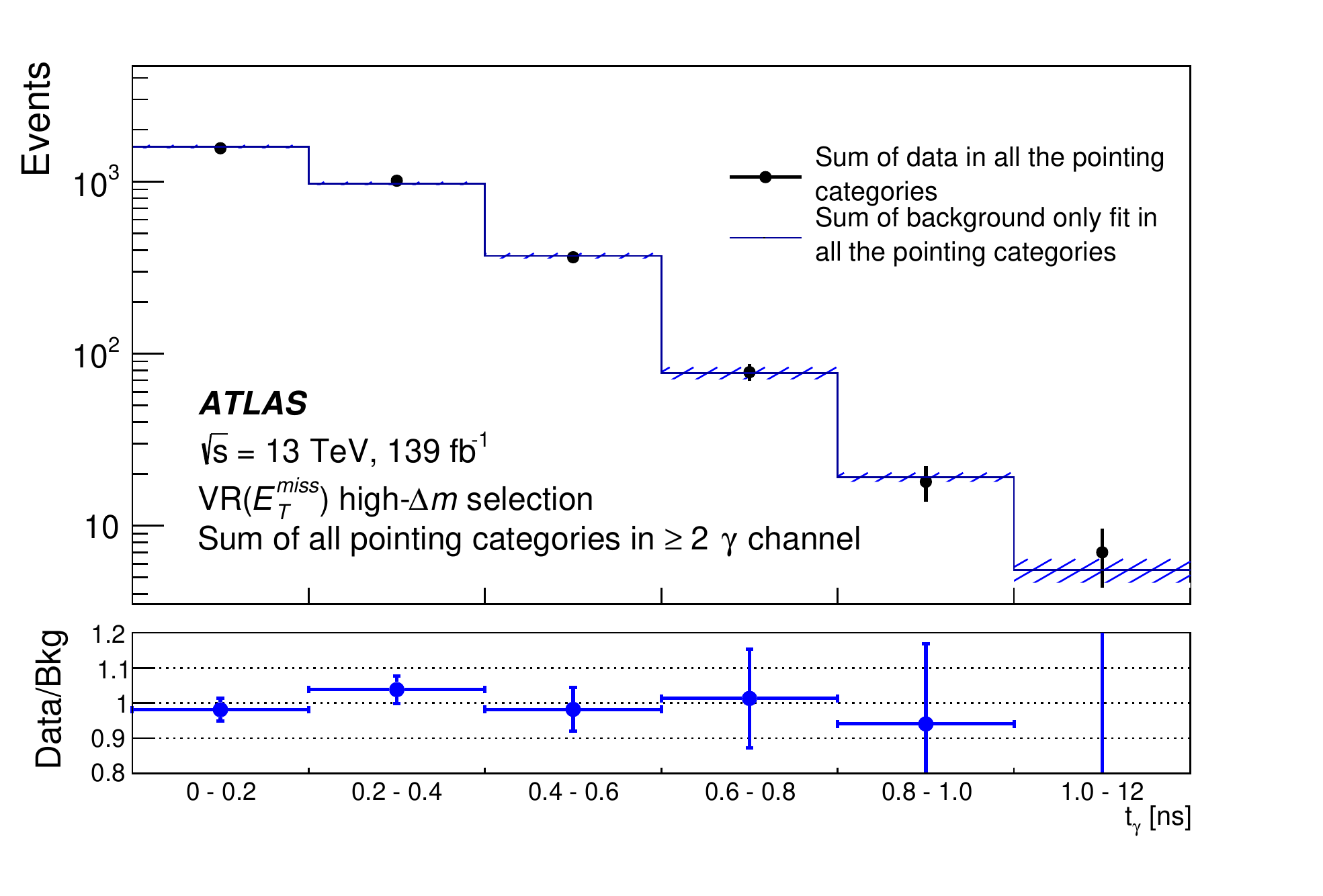}
\includegraphics[width=0.48\textwidth]{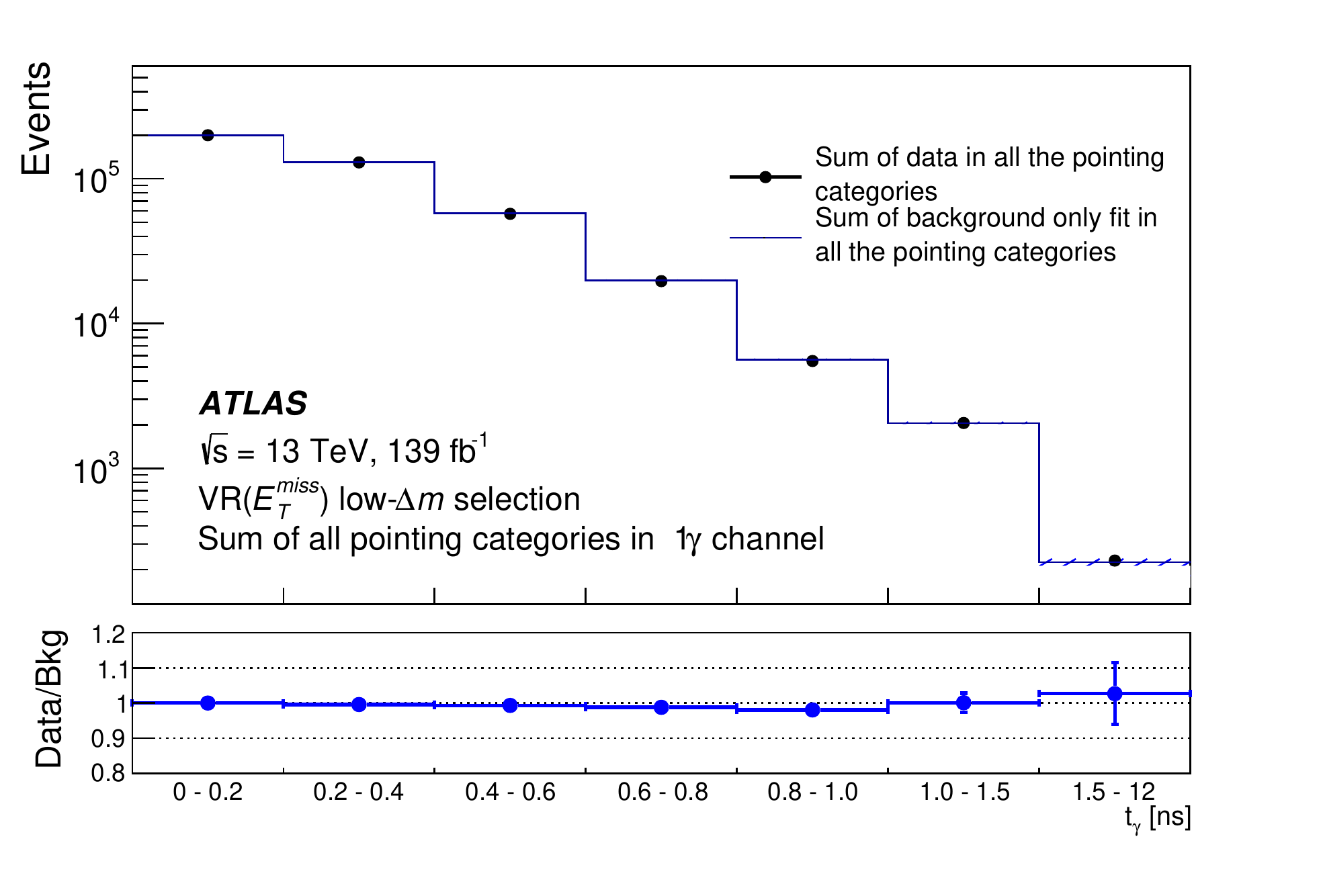}
\includegraphics[width=0.48\textwidth]{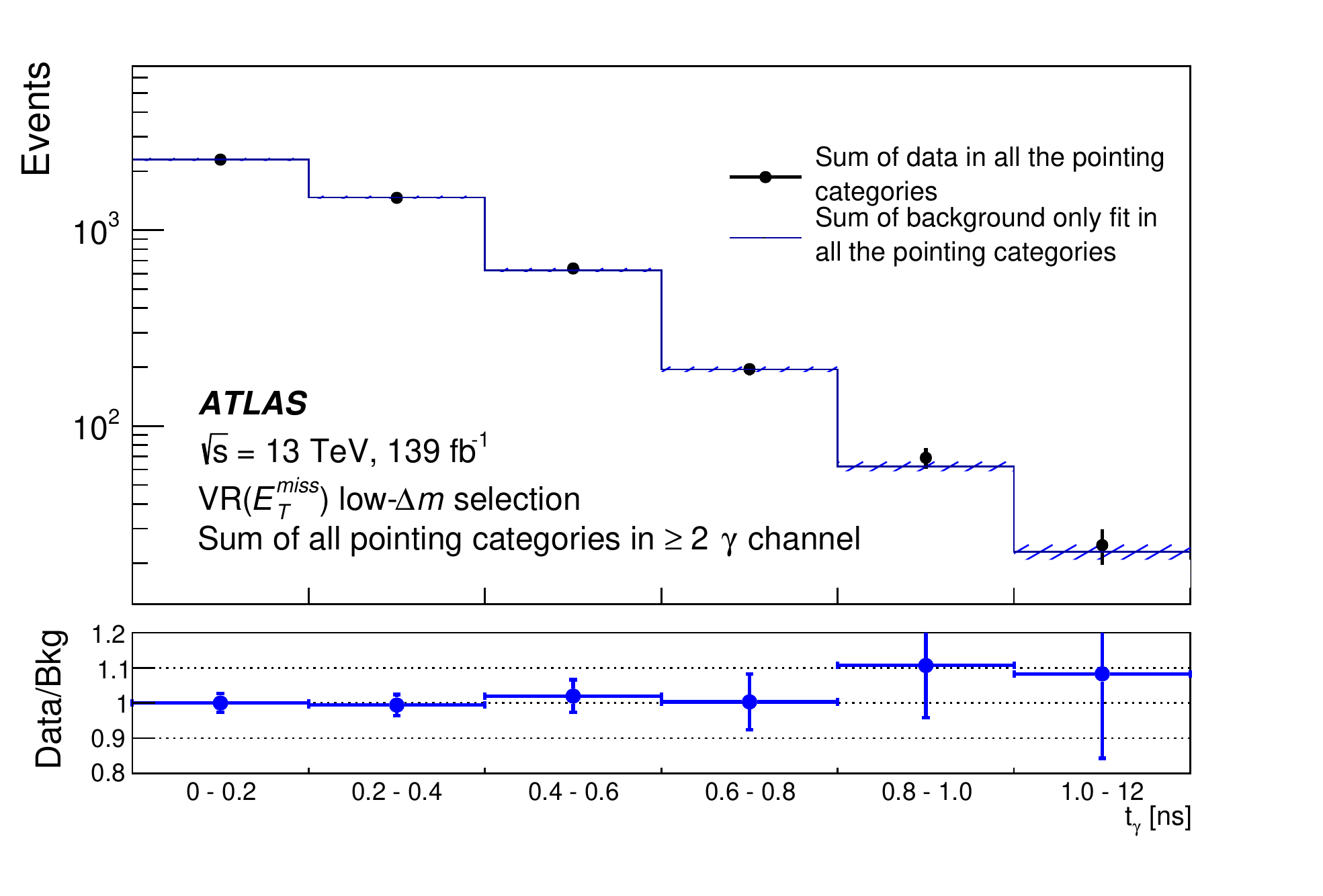}
\caption{Timing distributions for VR(\MET) data and the estimated background, as determined by a background-only fit. The
upper (lower) row shows the results for the \highdm \ (\lowdm) selection. The fit for each
selection is performed simultaneously across ten timing
distributions, namely
the five pointing categories for each of the two final states
($1\gamma$ and $\geq2\gamma$). For conciseness, the figures
show the sum of the data across all five pointing categories,
compared with the sum of the background-only fits, with the
left (right) figure displaying the results for the
$1\gamma$ ($\geq2\gamma$) final state. \label{fig:vrmet}}
\end{figure*}


\section{Results}
\label{sec:result}

The post-fit plots comparing the timing distributions of the SR data and the estimated background
in each pointing category are shown in Figures~\ref{fig:fitplots-srmet50maxe10} and~\ref{fig:fitplots-srmet80maxe7} for the \highdm \
and \lowdm \ selections, respectively. Superimposed on the figures are the expected signal distributions for a
few example signal models.
The data are in good agreement with the fitted background-only hypothesis
in all categories for both the \highdm \
and \lowdm \ selections.
No significant excesses are observed, with the
largest excess for any signal model considered having a
significance of about
1.2~standard deviations.
Therefore,
a modified frequentist CL$_\mathrm{s}$ method~\cite{Read:2002hq} is used to set
95\% CL upper limits on $\mathcal{B}(\Hnlspnlsp)$, assuming that
$\mathcal{B}(\mathrm{NLSP} \rightarrow \mathrm{LSP} + \gamma)$ is equal to unity.
 
\begin{figure*}[!ph]
\centering
\includegraphics[width=6cm]{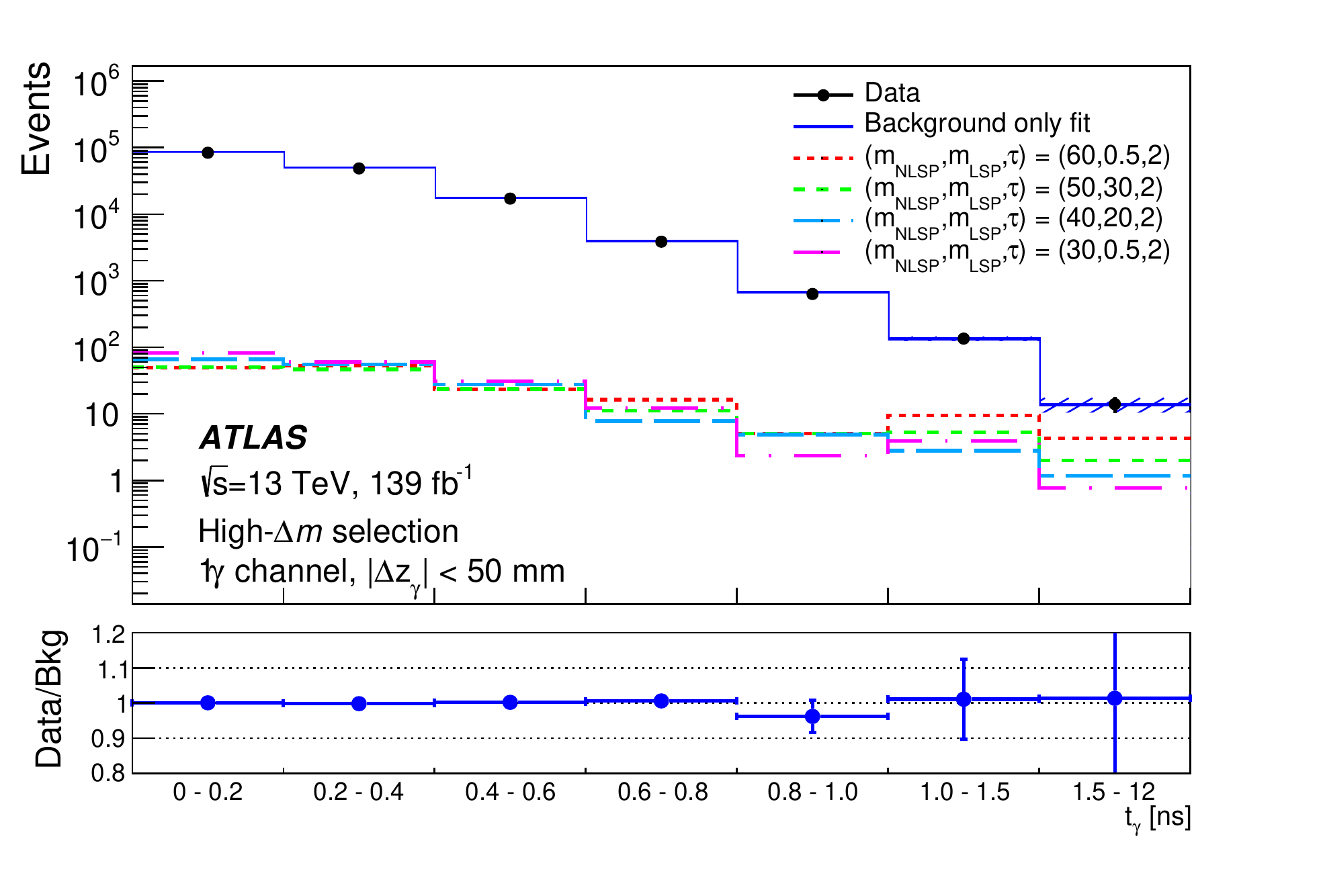}
\includegraphics[width=6cm]{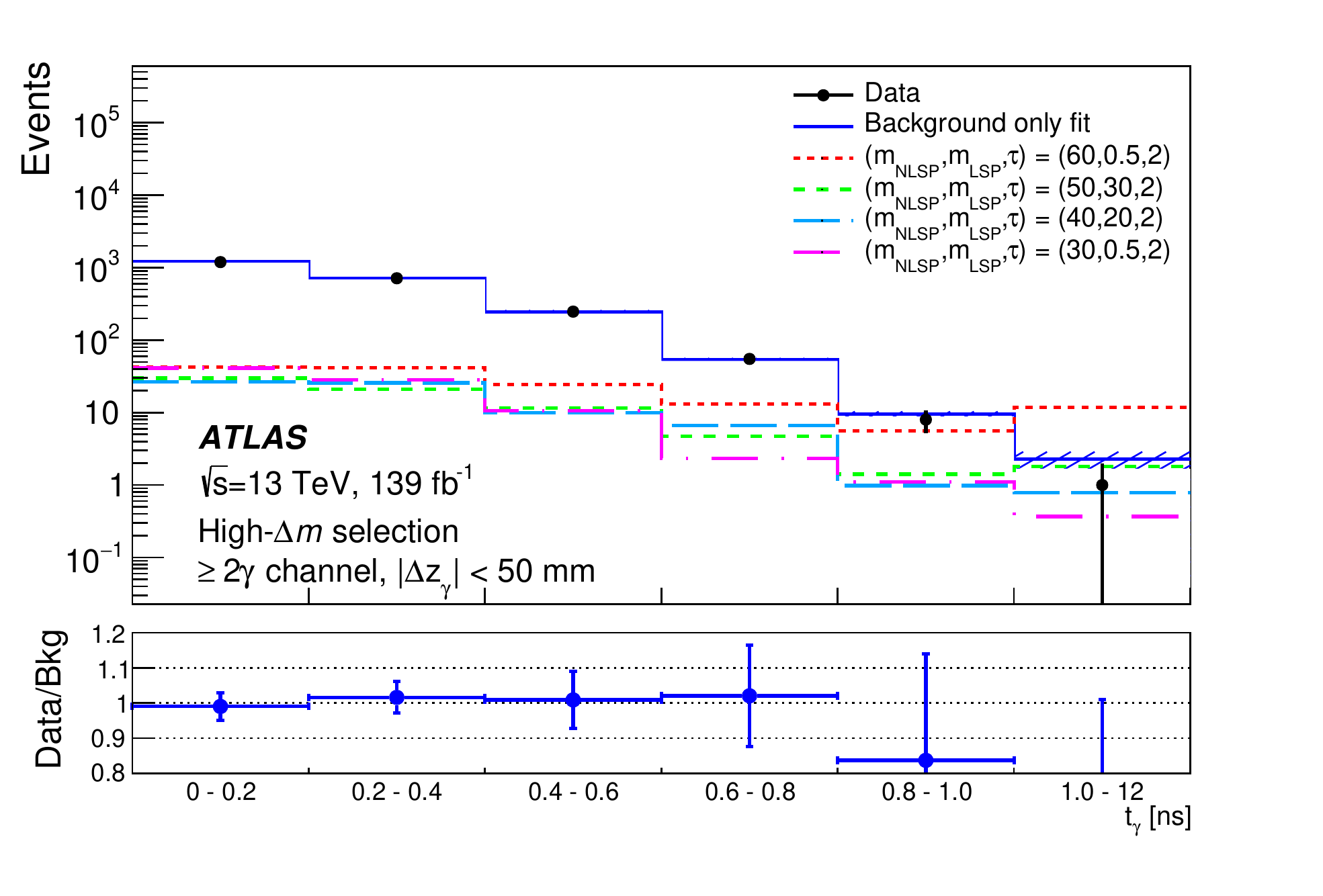}
\includegraphics[width=6cm]{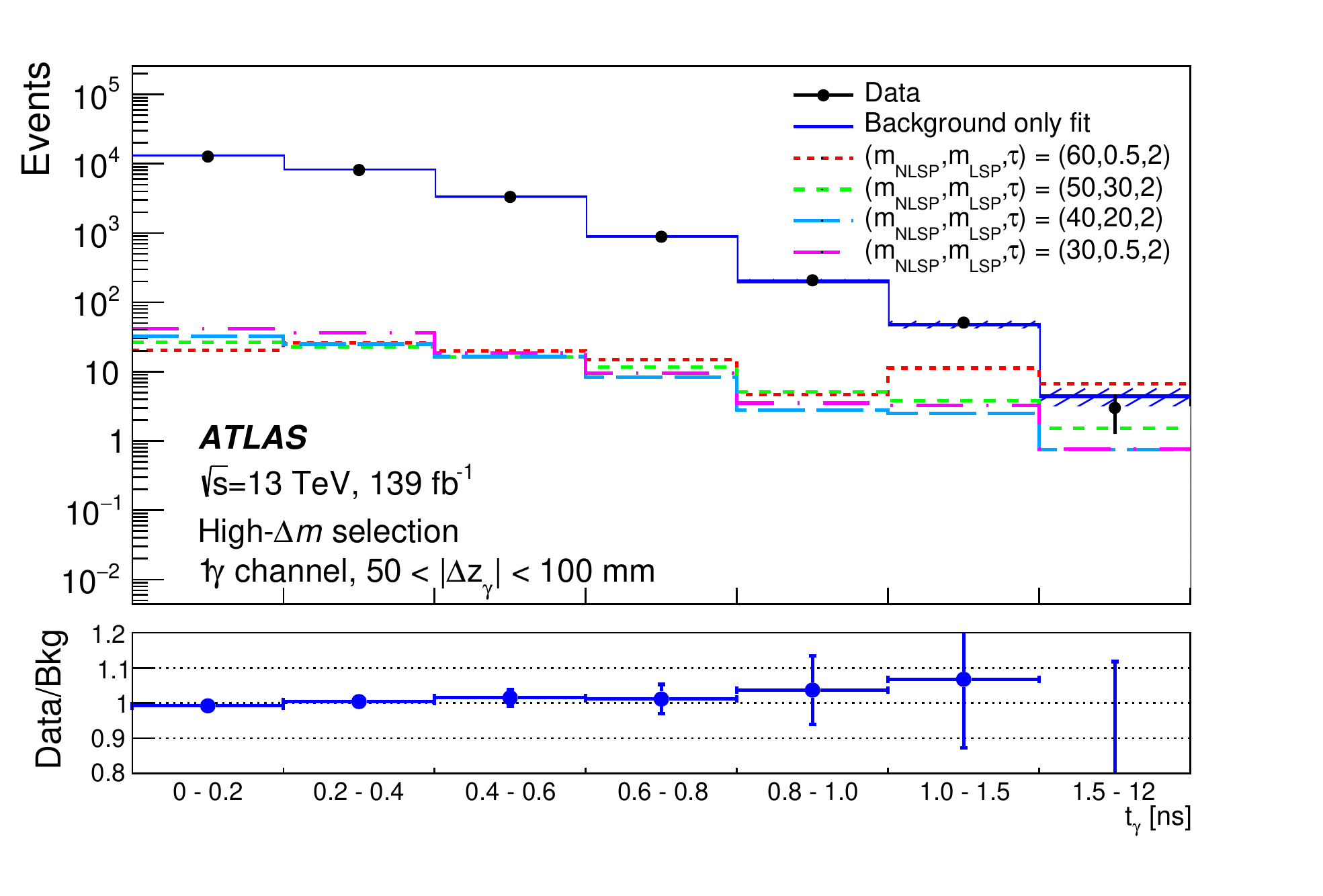}
\includegraphics[width=6cm]{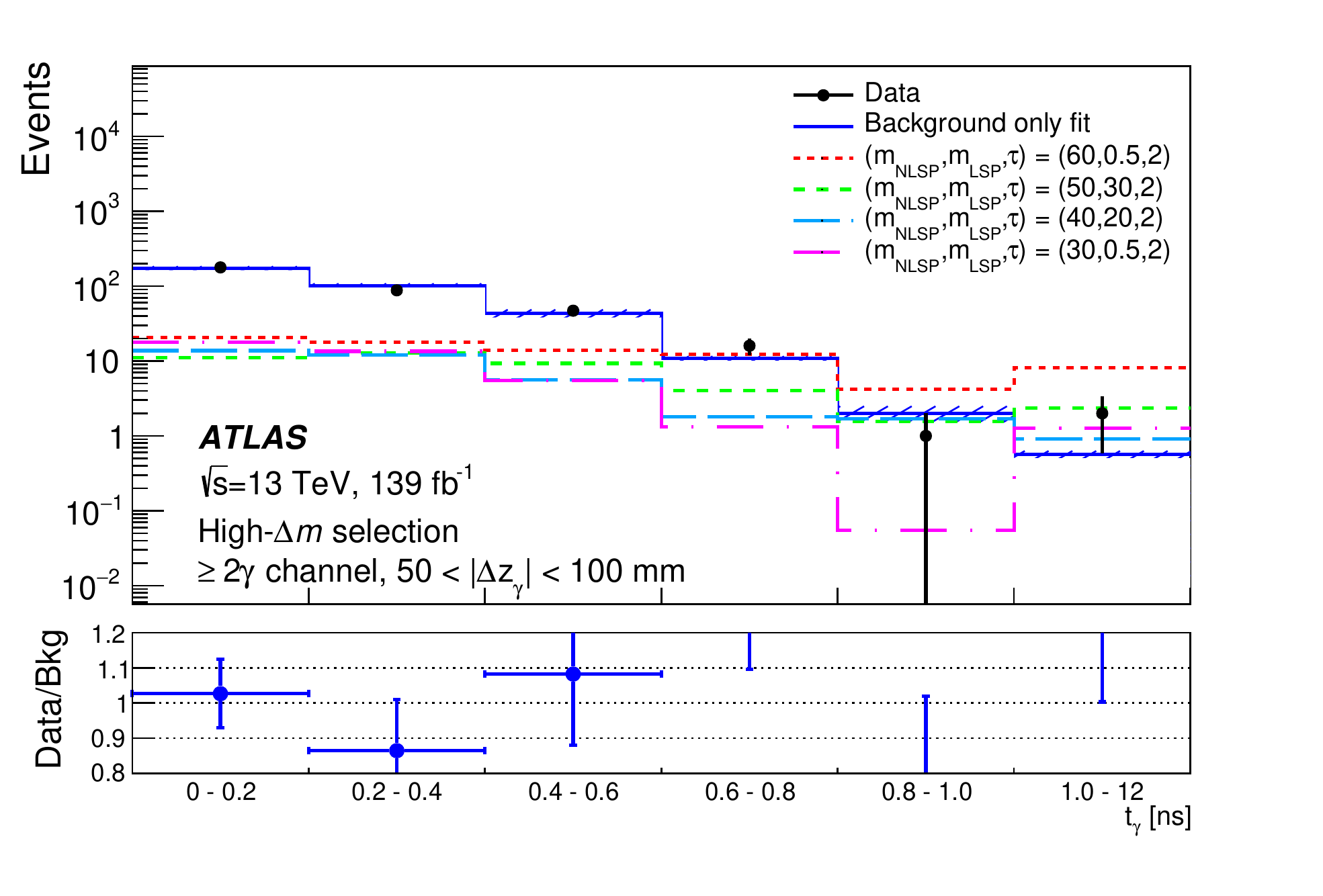}
\includegraphics[width=6cm]{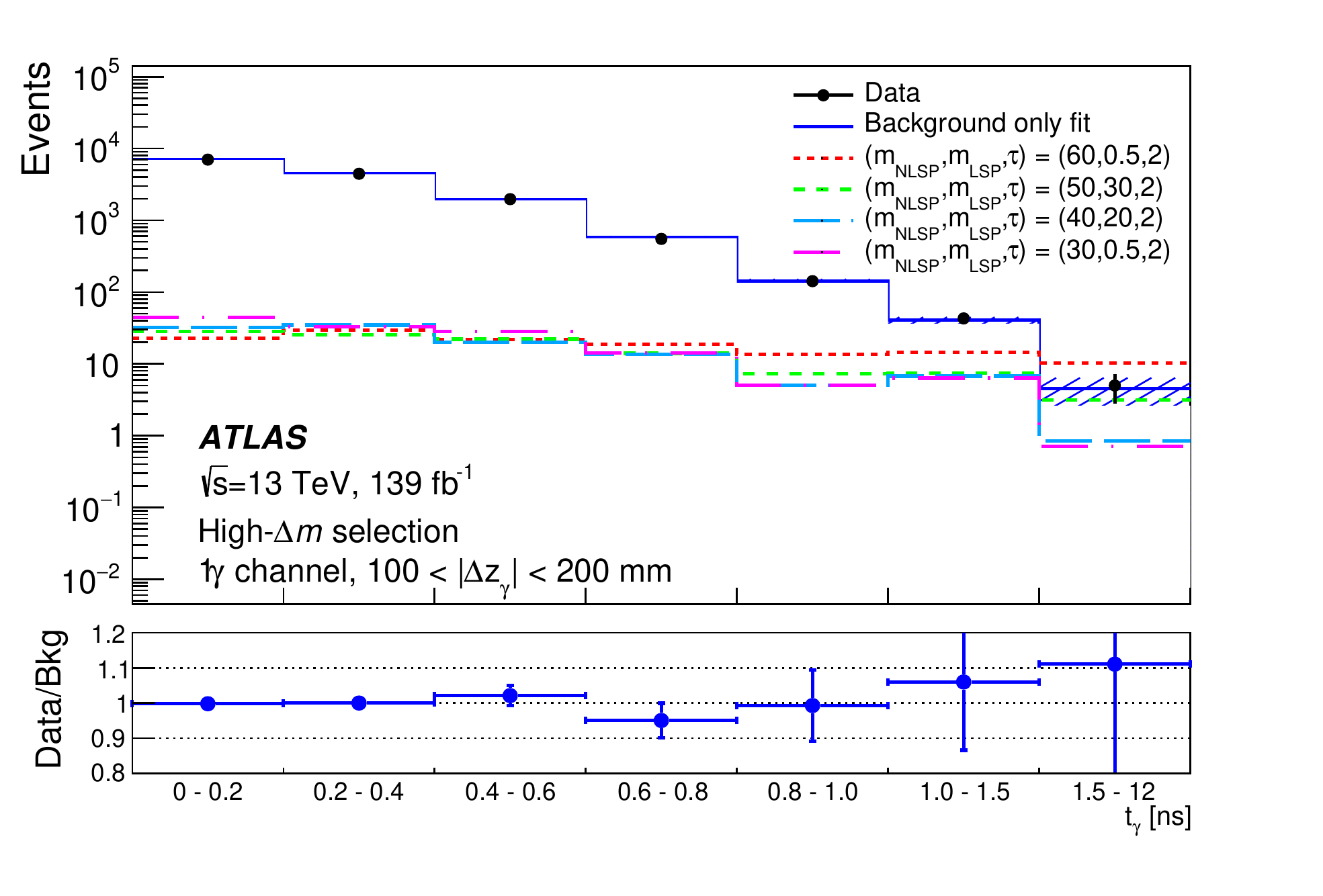}
\includegraphics[width=6cm]{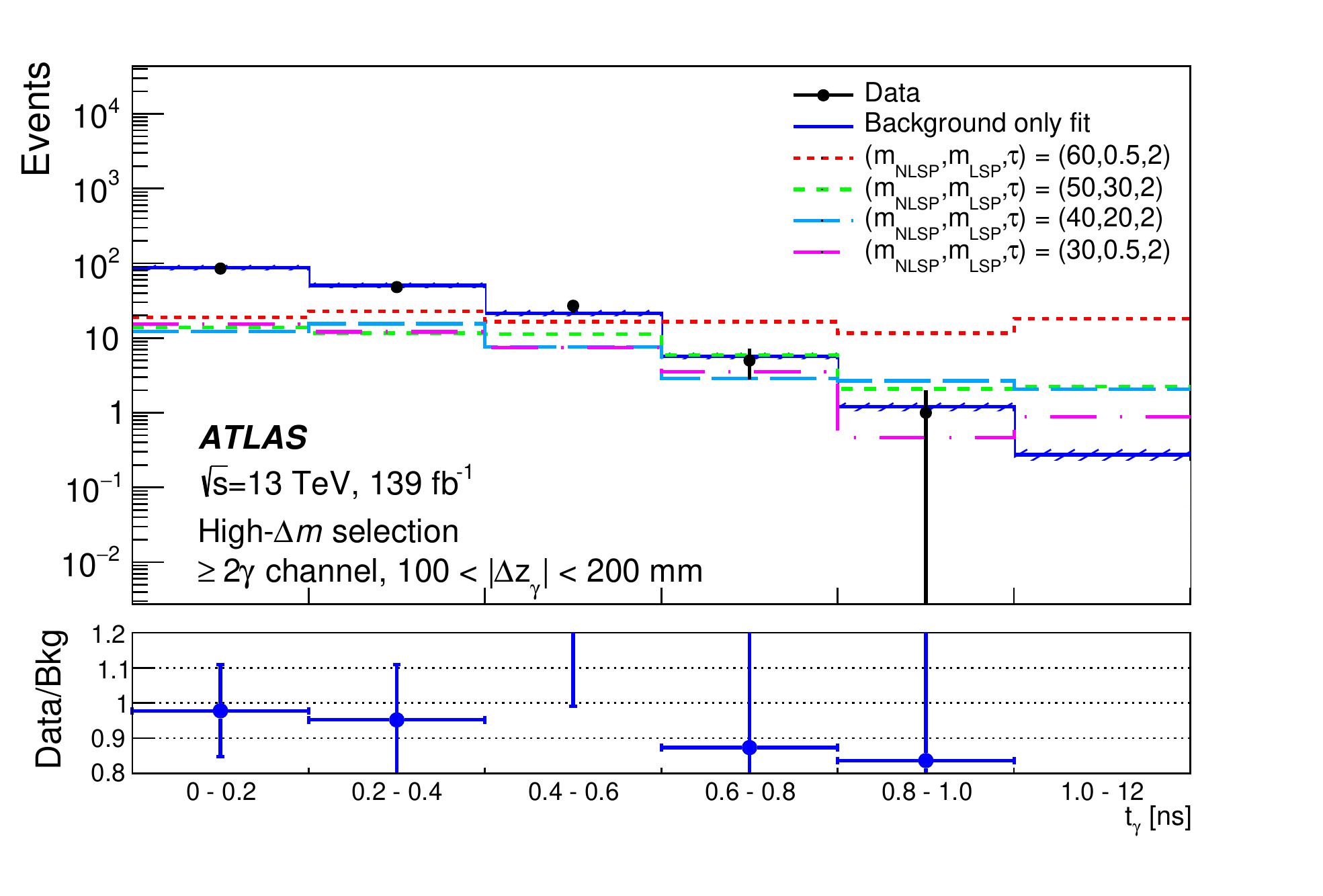}
\includegraphics[width=6cm]{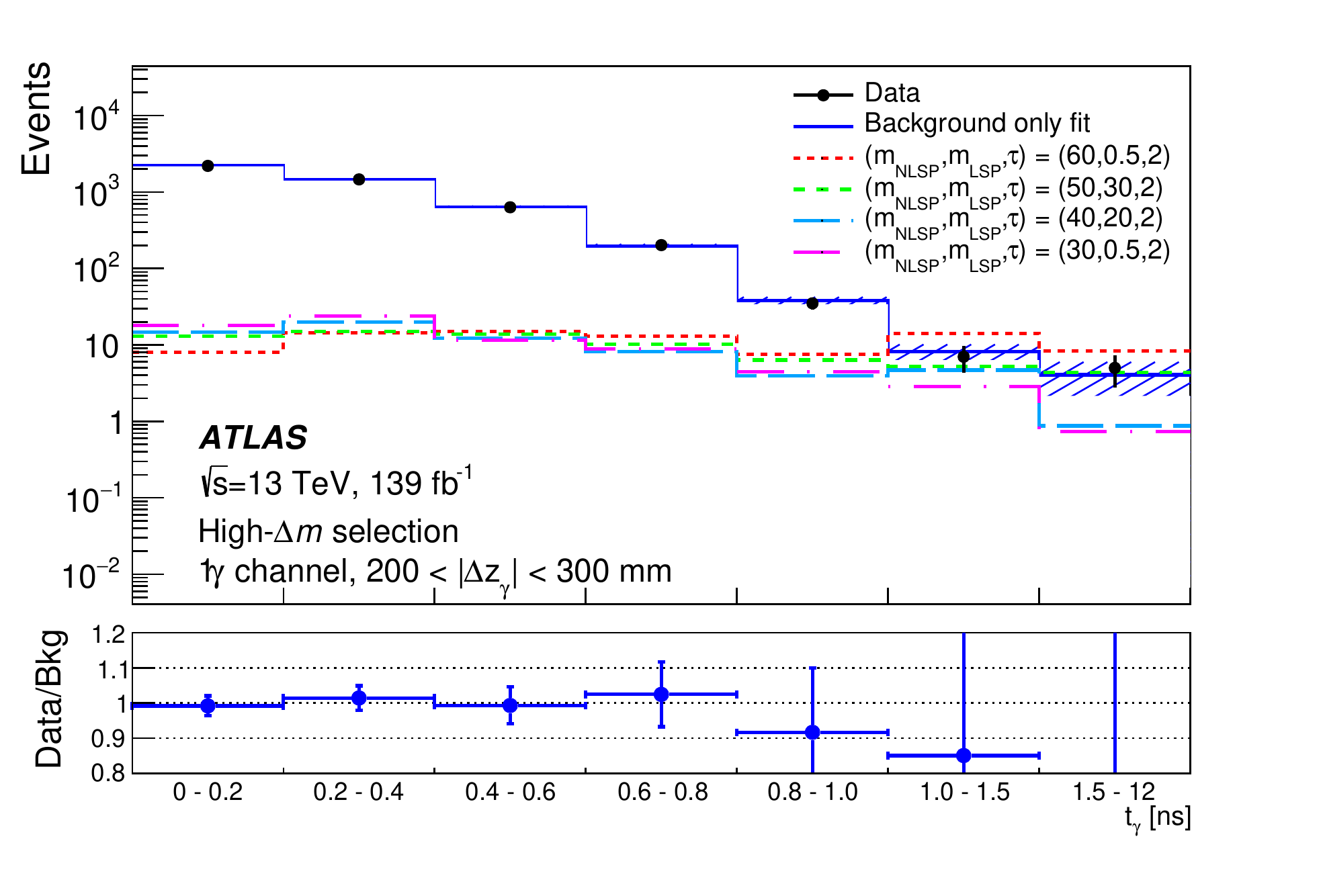}
\includegraphics[width=6cm]{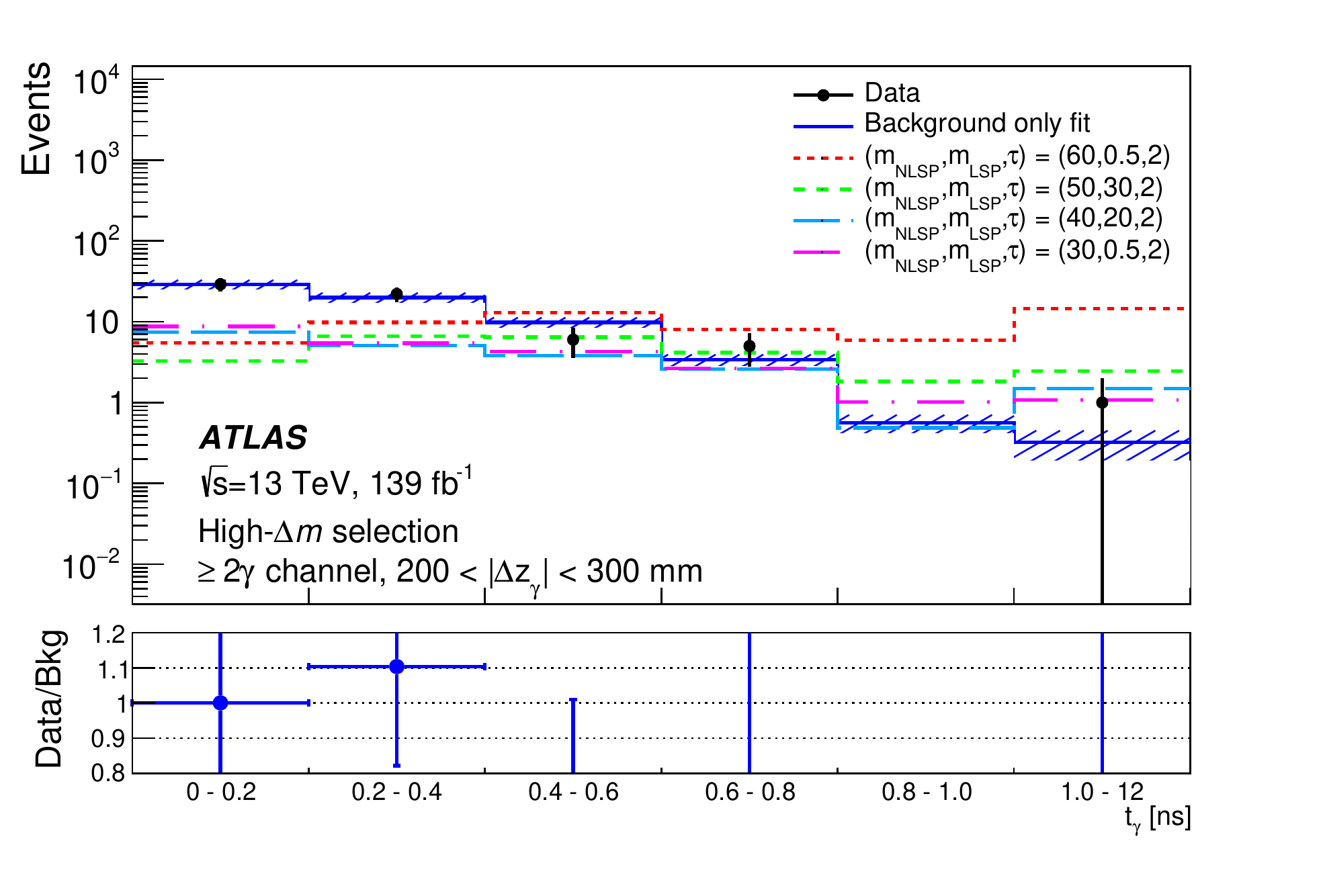}
\includegraphics[width=6cm]{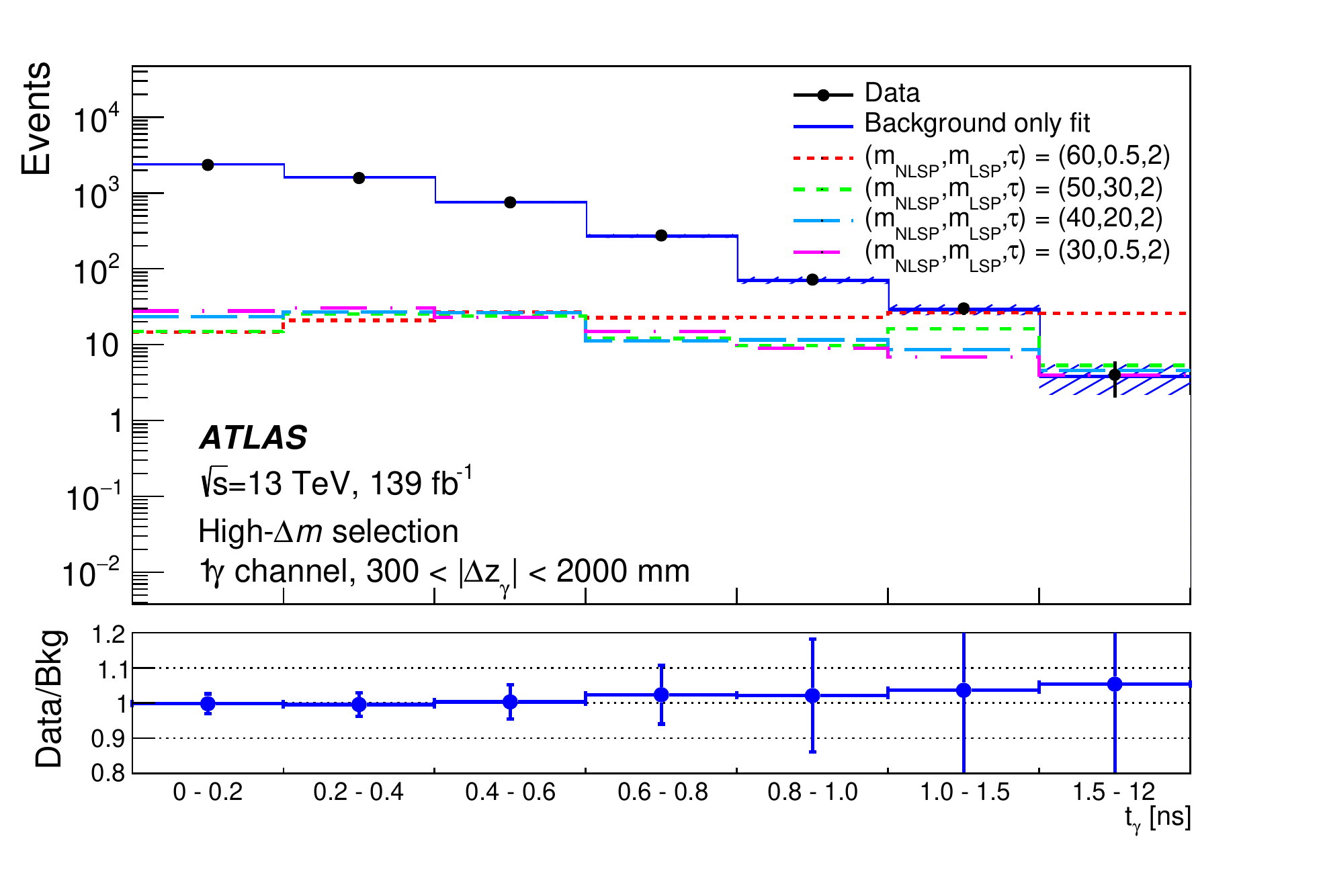}
\includegraphics[width=6cm]{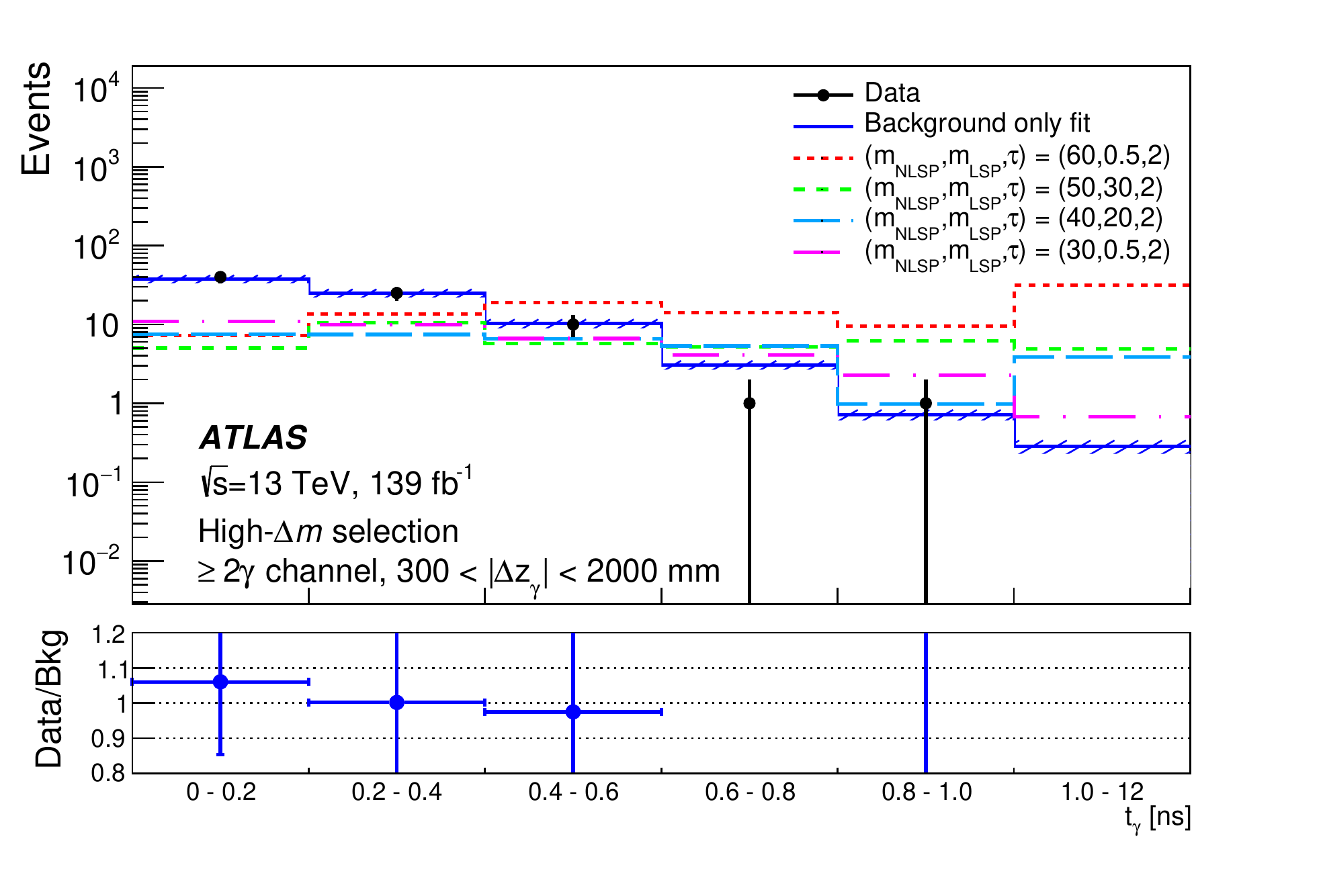}
\caption{Timing distributions for SR data and the estimated background, as determined by the background-only fit, for the \highdm \ selection. The plots
on the left (right) show the time distributions
for each of the five pointing categories for the $1\gamma$ ($\geq2\gamma$) final state.
For comparison, the expected timing shapes for a few different signal points are shown superimposed, with a signal normalization
corresponding to a value of $\mathcal{B}(\Hnlspnlsp) = 20$\%. The signal
models are labeled by their values of the NLSP and LSP masses
(in \GeV) and NLSP lifetime (in ns). \label{fig:fitplots-srmet50maxe10}}
\end{figure*}
 
\begin{figure*}[!ph]
\centering
\includegraphics[width=6cm]{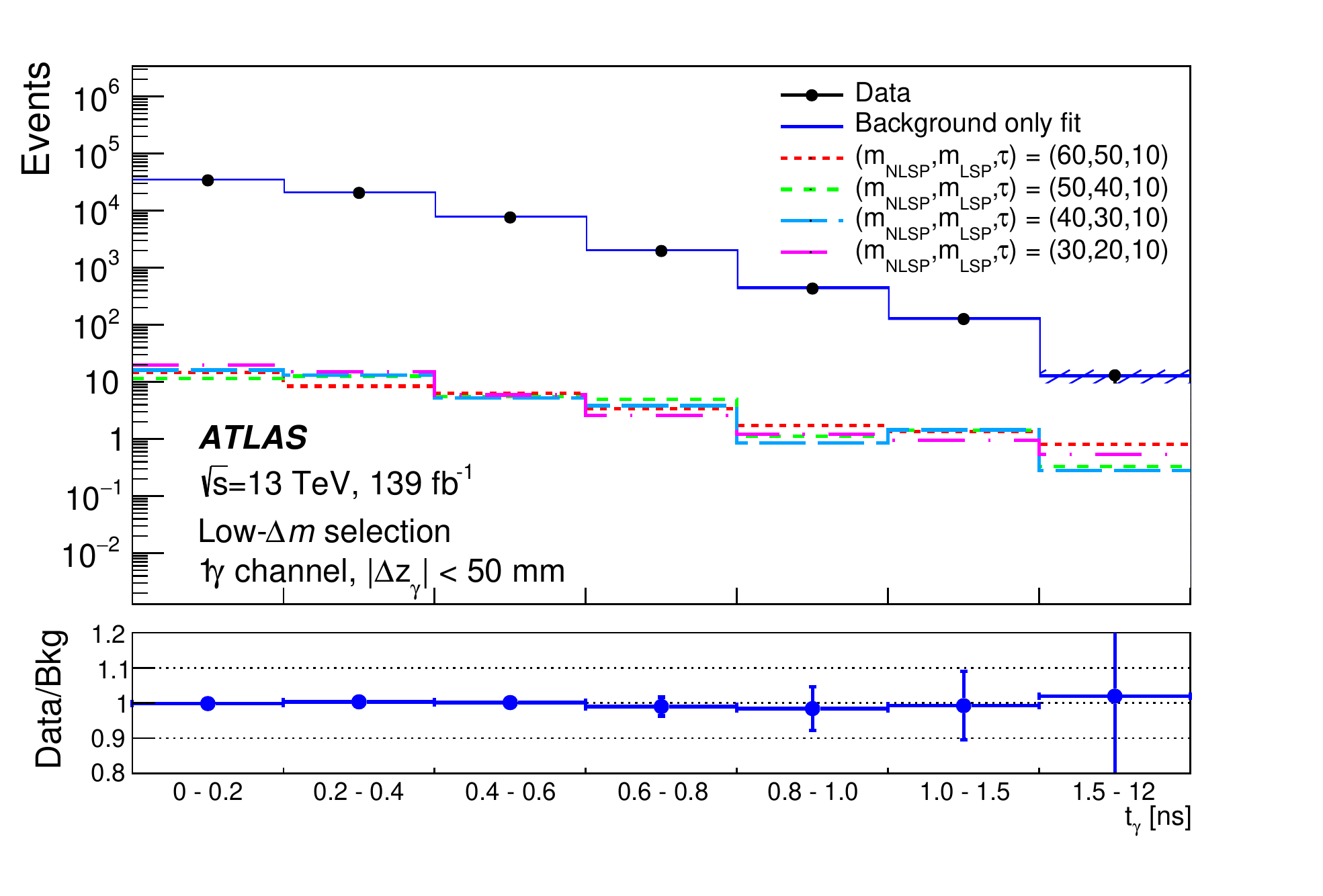}
\includegraphics[width=6cm]{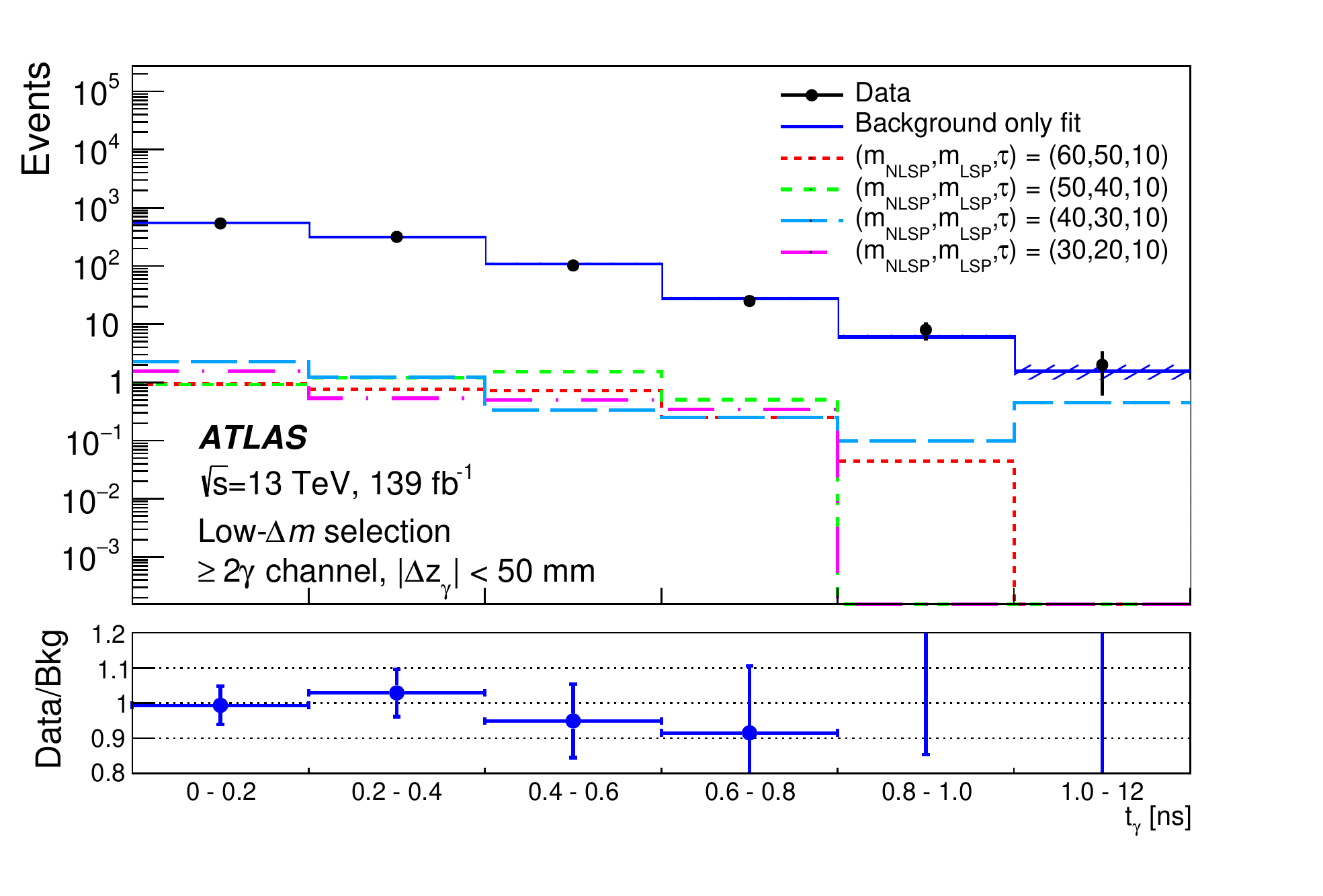}
\includegraphics[width=6cm]{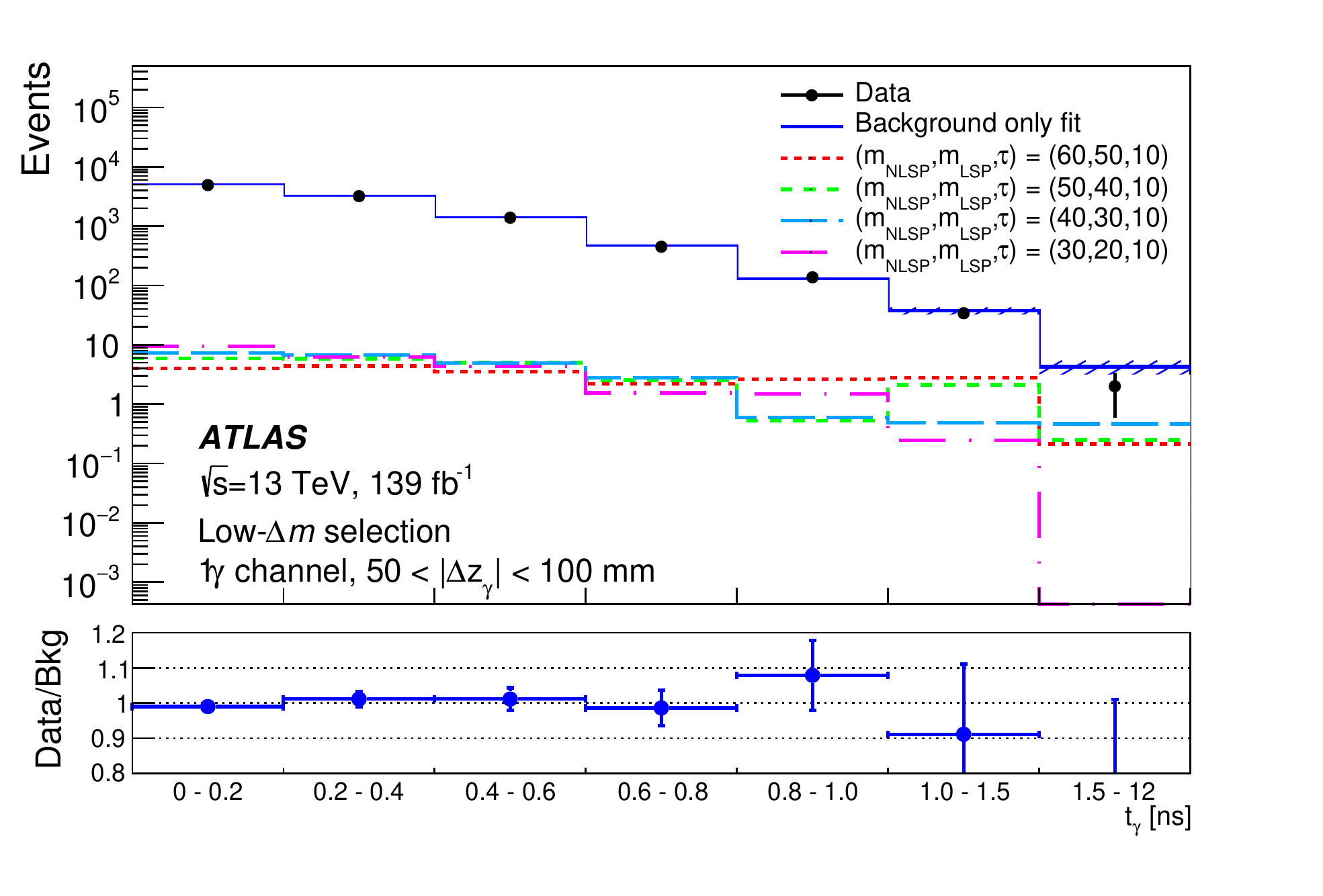}
\includegraphics[width=6cm]{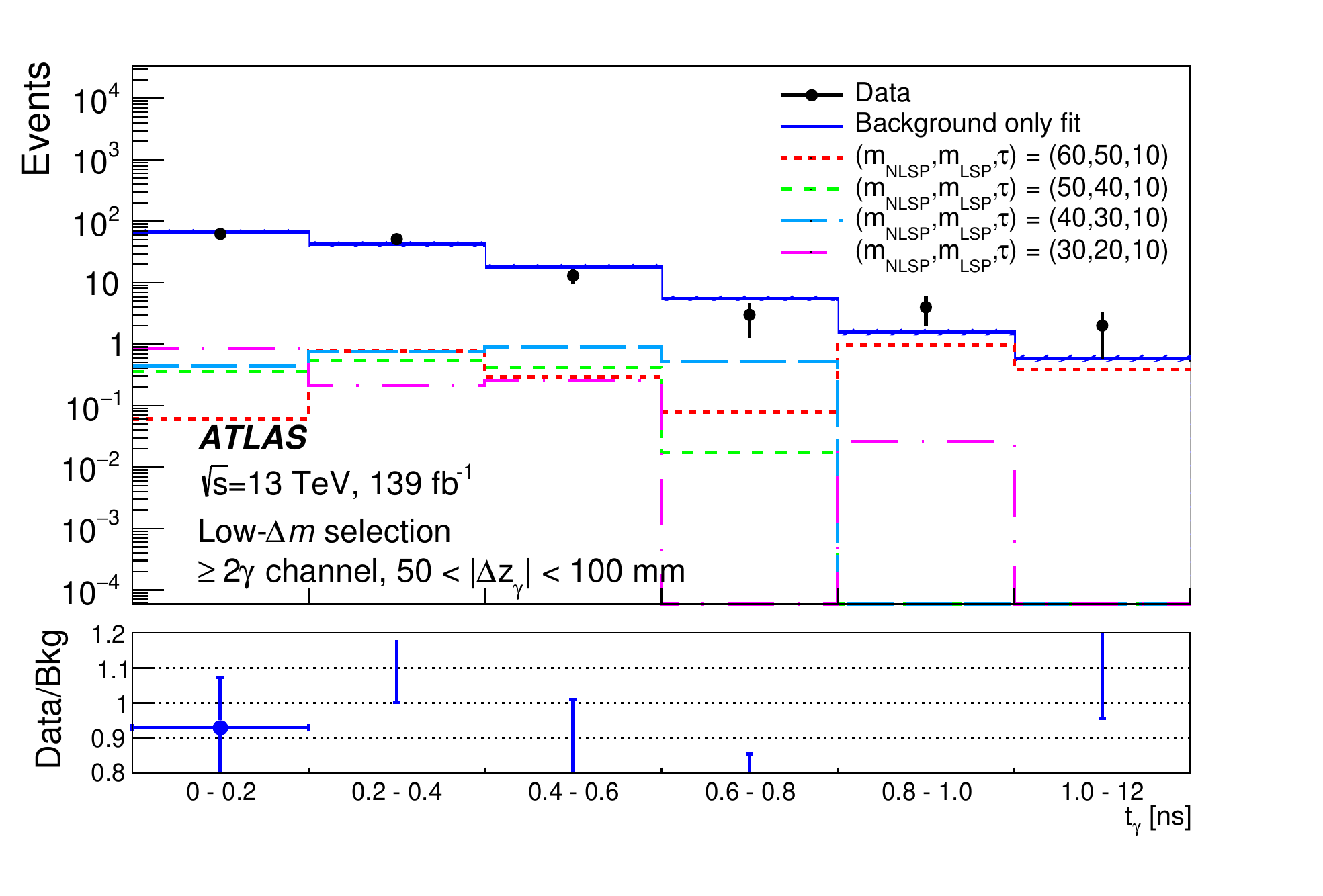}
\includegraphics[width=6cm]{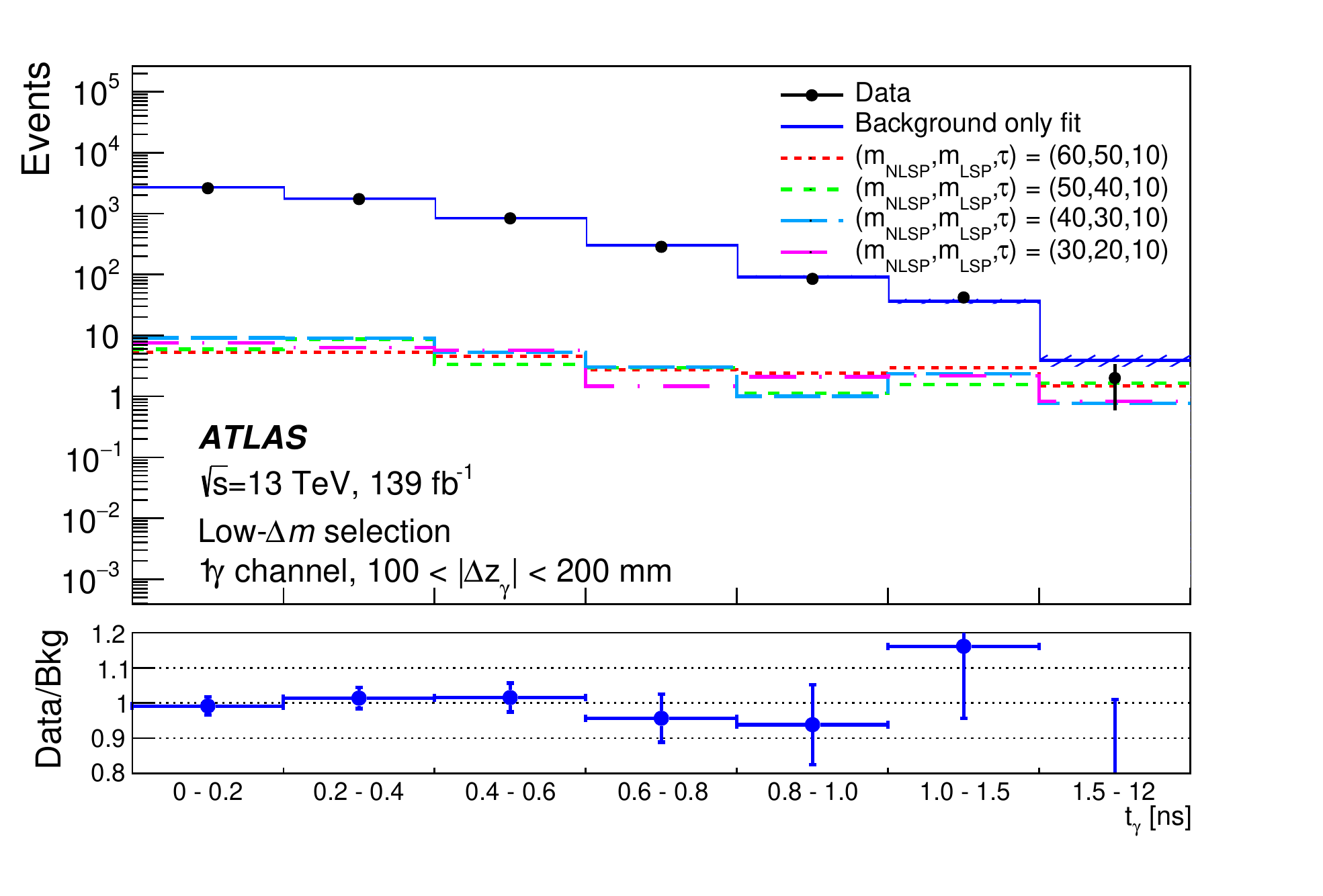}
\includegraphics[width=6cm]{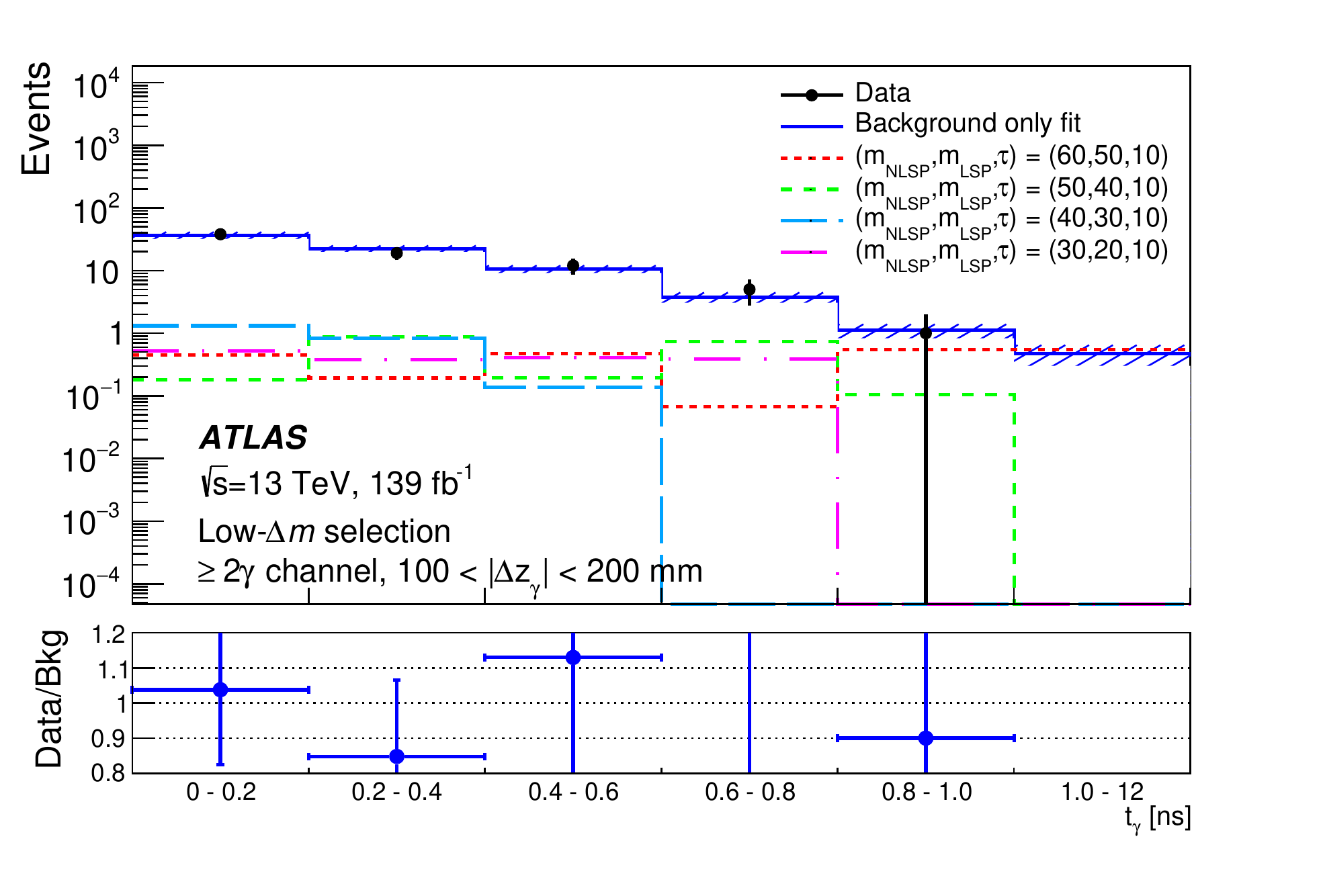}
\includegraphics[width=6cm]{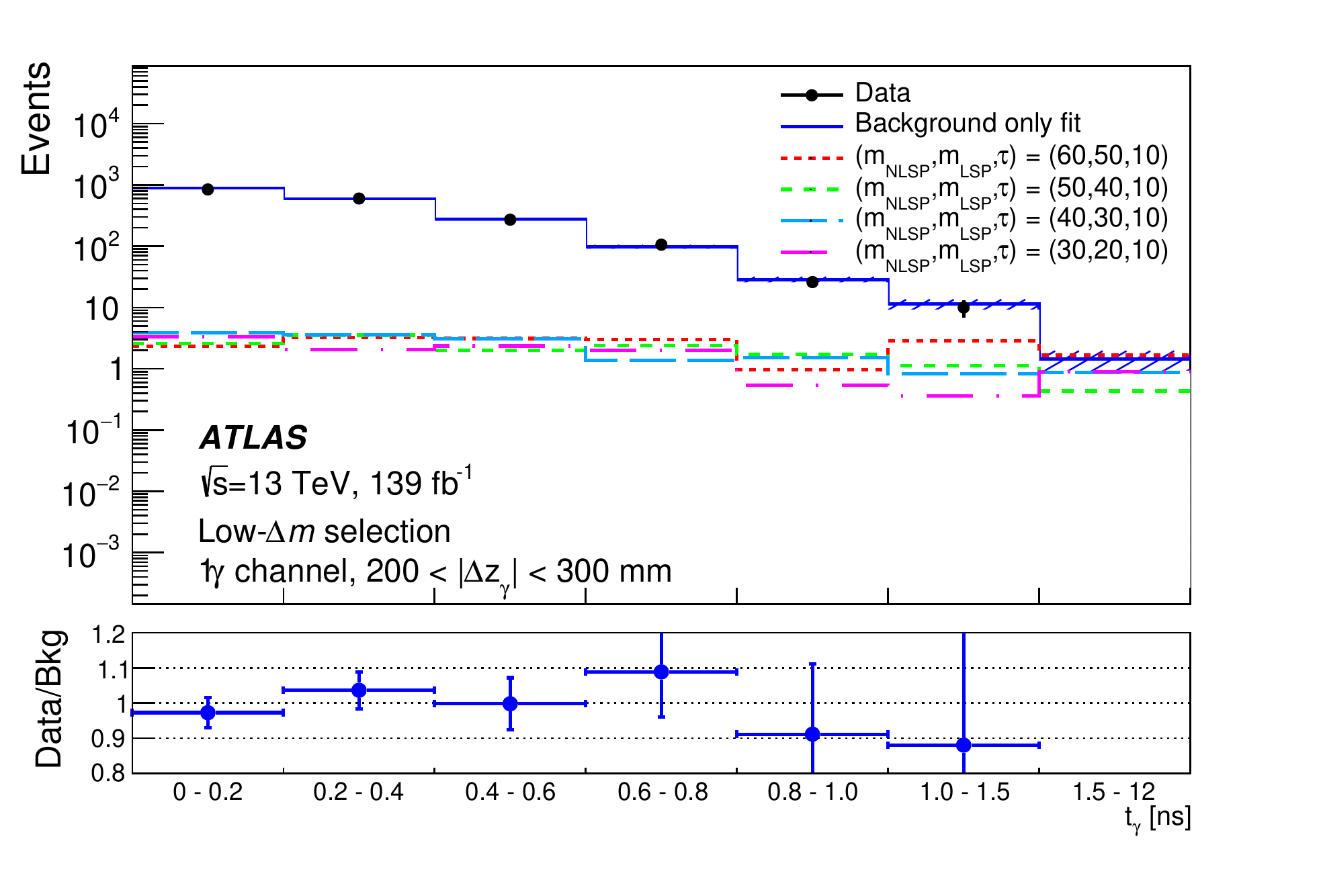}
\includegraphics[width=6cm]{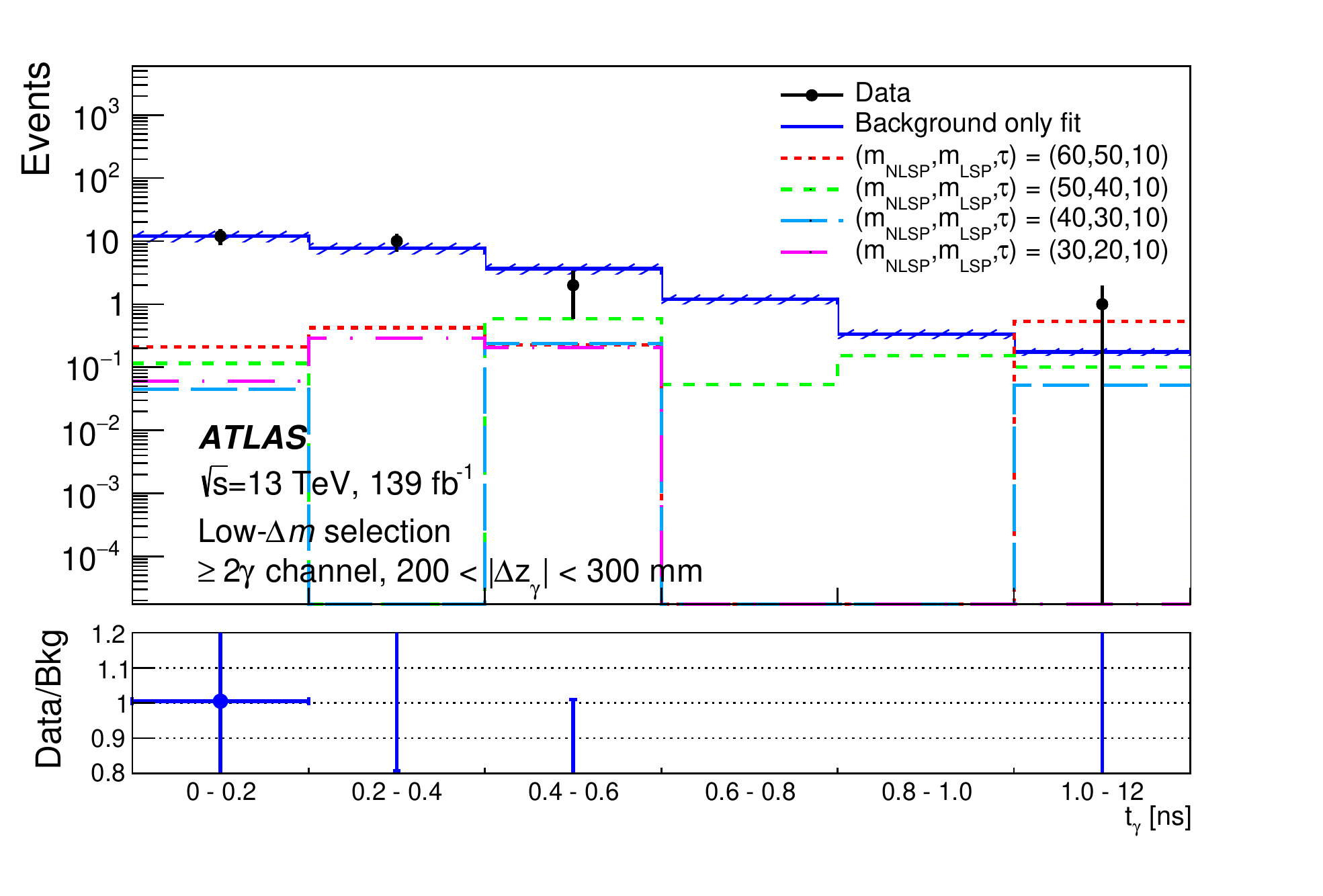}
\includegraphics[width=6cm]{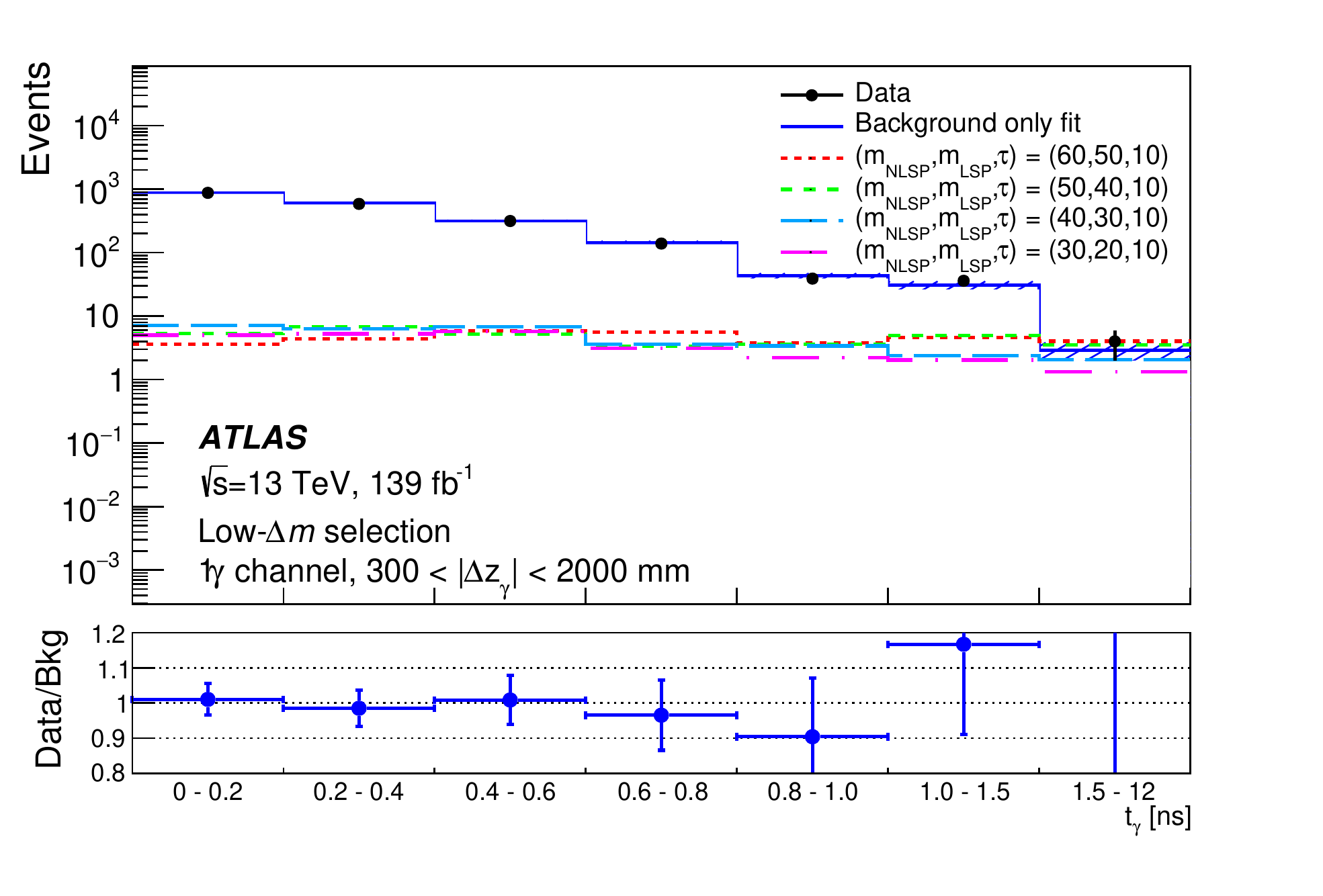}
\includegraphics[width=6cm]{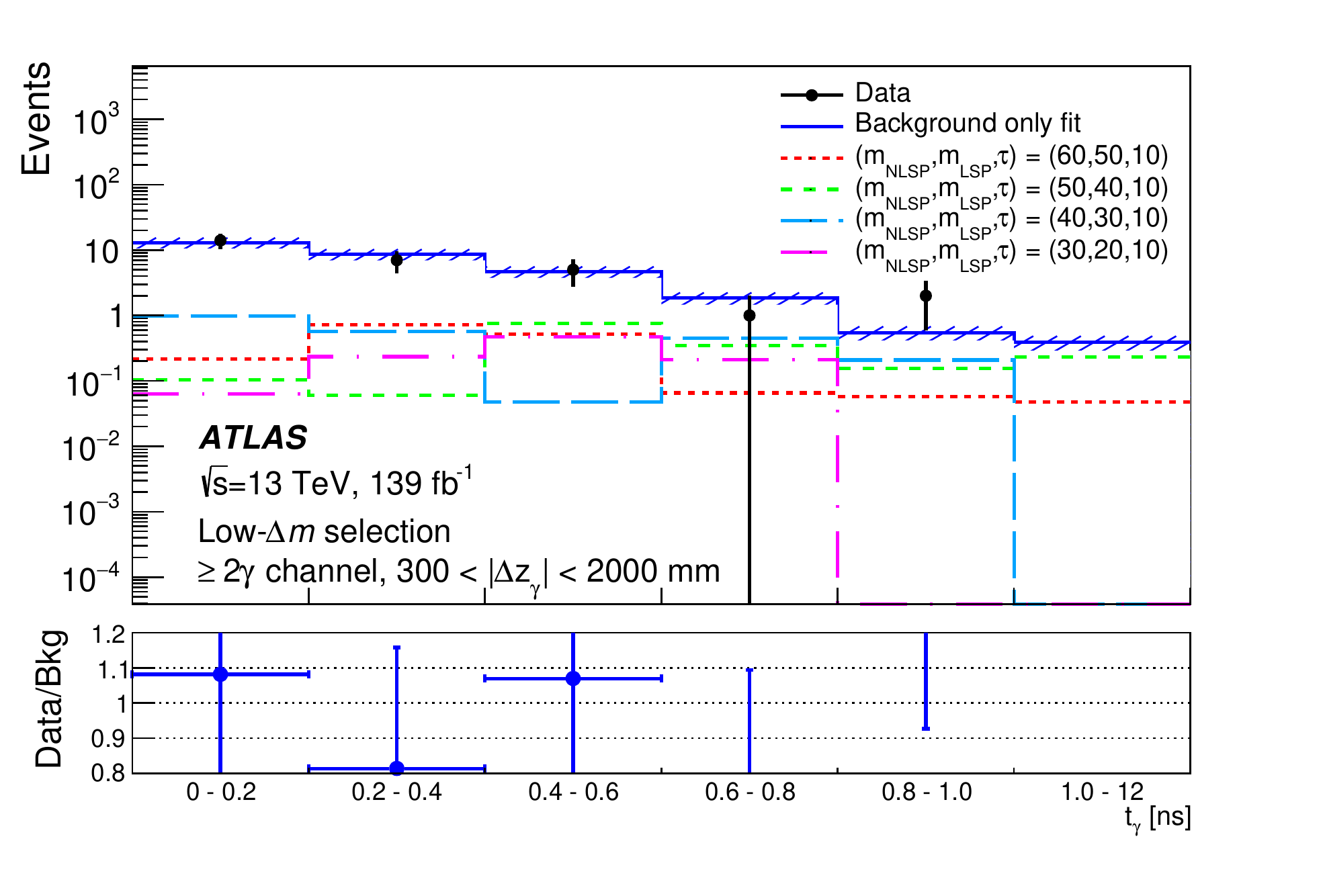}
\caption{Timing distributions for SR data and the estimated background, as determined by the background-only fit, for the \lowdm \ selection. The plots on the left (right) show the time distributions
for each of the five pointing categories for the $1\gamma$ ($\geq2\gamma$) final state.
For comparison, the expected timing shapes for a few different signal points are shown superimposed, with a signal normalization
corresponding to a value of $\mathcal{B}(\Hnlspnlsp) = 20$\%. The signal
models are labeled by their values of the NLSP and LSP masses
(in \GeV) and NLSP lifetime (in ns).\label{fig:fitplots-srmet80maxe7}}
\end{figure*}
\FloatBarrier
 
For a fixed NLSP lifetime, the limits are directly obtained for the 18 simulated signal models, covering a two-dimensional grid of NLSP masses ranging
from 30~\GeV\ to 60~\GeV, and LSP masses from 0.5~\GeV \ to $m_{\mathrm{NSLP}} - 10$~\GeV.
For visualization purposes, a bilinear interpolation between the discrete signal points simulated for
each fixed lifetime is performed in order to obtain the 95\% CL limits across the full available range of NLSP and LSP masses.
In the \highdm \ case, the interpolation is
performed on a two-dimensional grid of \mLSP \ versus \mNLSP.
The results for fixed NLSP lifetimes of 2~ns and 10~ns are shown in Figure~\ref{fig:InterpretationMassHigh}.
A selection of contours corresponding to fixed values of the expected and observed 95\% CL upper limits on $\mathcal{B}(\Hnlspnlsp)$ is superimposed on the $(\mNLSP, \mLSP)$ plane.
 
\begin{figure*}[!ph]
\centering
\includegraphics[width=0.48\columnwidth]{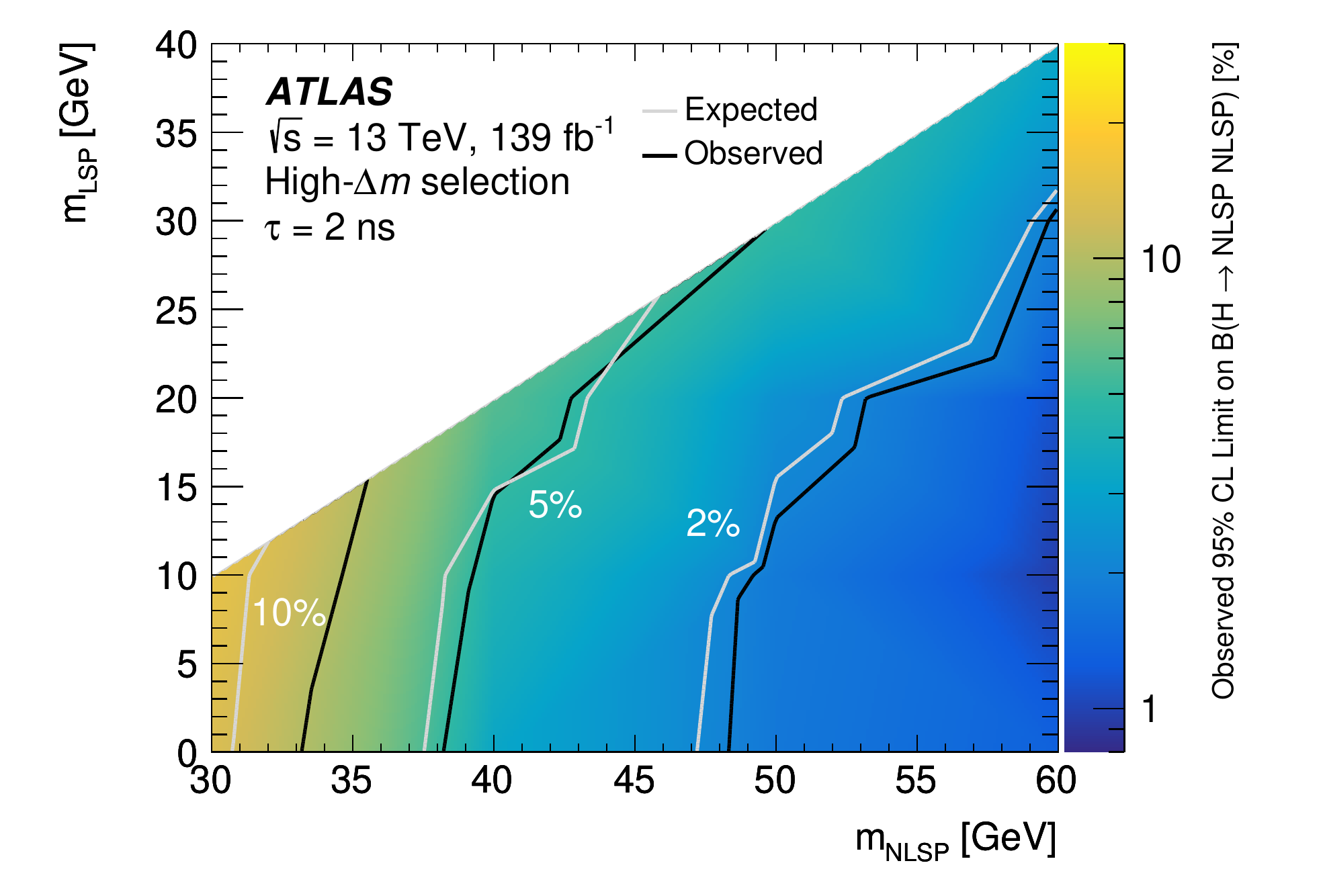}
\includegraphics[width=0.48\columnwidth]{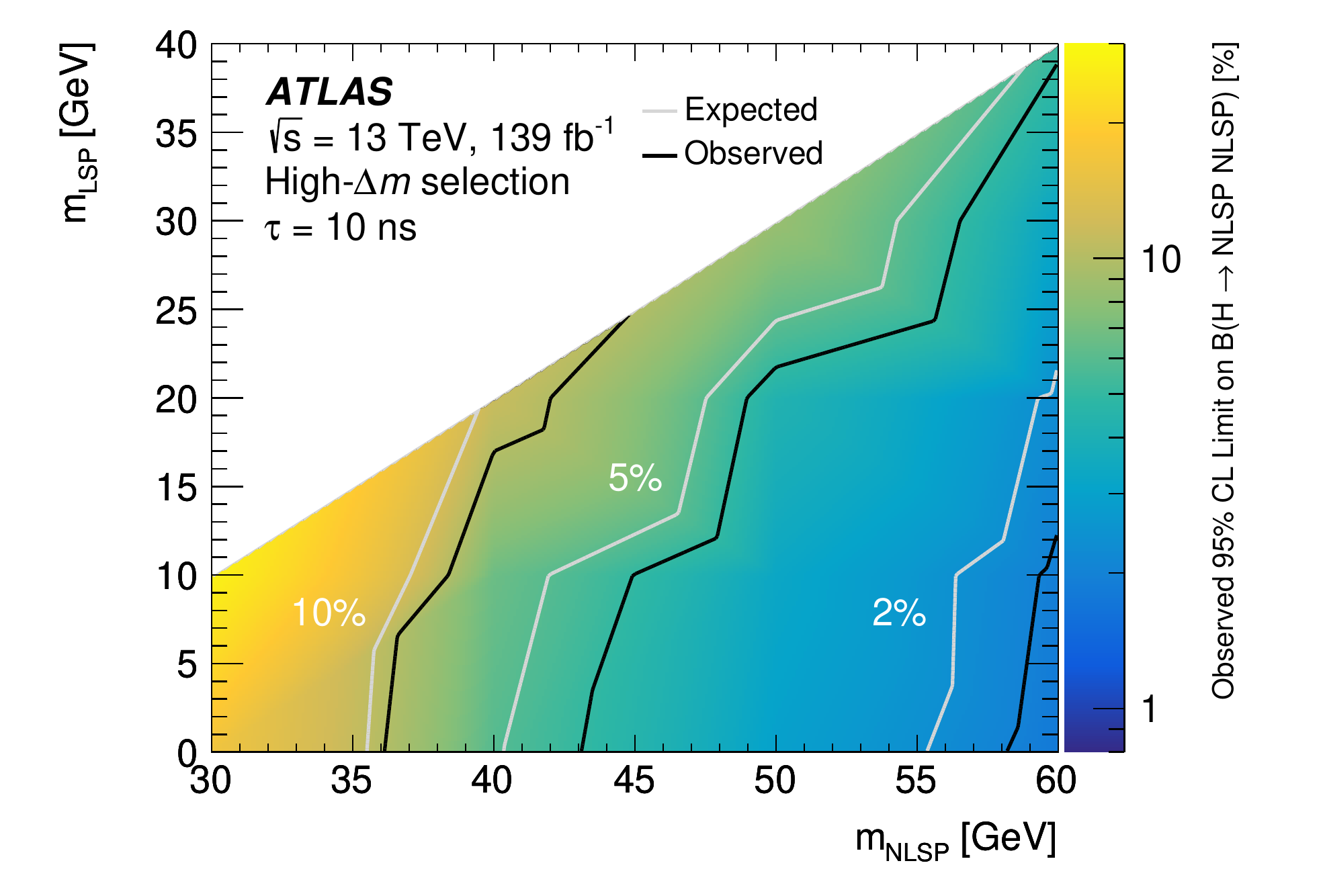}
\caption{Interpolated 95\% CL limits on $\mathcal{B}$(\Hnlspnlsp) in the plane of LSP mass versus NLSP mass for the \highdm \ selection.
The plot on the left (right) corresponds to the
results with the NLSP lifetime equal to 2~ns (10~ns).
A smoothing function is used to continuously fill the
plane, interpolating between the results for the
discrete signal models which
were simulated on a grid with 10~\GeV \ spacing, and with
a minimum value of $\Delta m = 20$~\GeV \ for the signal
models covered by the \highdm \
selection.
For visualization purposes, contours are superimposed indicating where the expected and observed 95\% CL limits
on $\mathcal{B}$(\Hnlspnlsp) reach values of  2\%, 5\%, and 10\%. \label{fig:InterpretationMassHigh} }
\end{figure*}
 
In the \lowdm \ case, the interpolation can only be
performed in one dimension, providing the 95\% CL limit versus the NLSP mass,
since the
LSP mass is fixed according to the relation $\mNLSP - \mLSP = 10$~\GeV.
The observed and expected 95\% CL $\mathcal{B}$(\Hnlspnlsp) upper limit curves for the
\lowdm \ selection
are shown in Figure~\ref{fig:InterpretationMassLow} for NLSP
lifetimes of 2~ns and 10~ns.
 
\begin{figure*}[!ph]
\centering
\includegraphics[width=0.478\columnwidth]{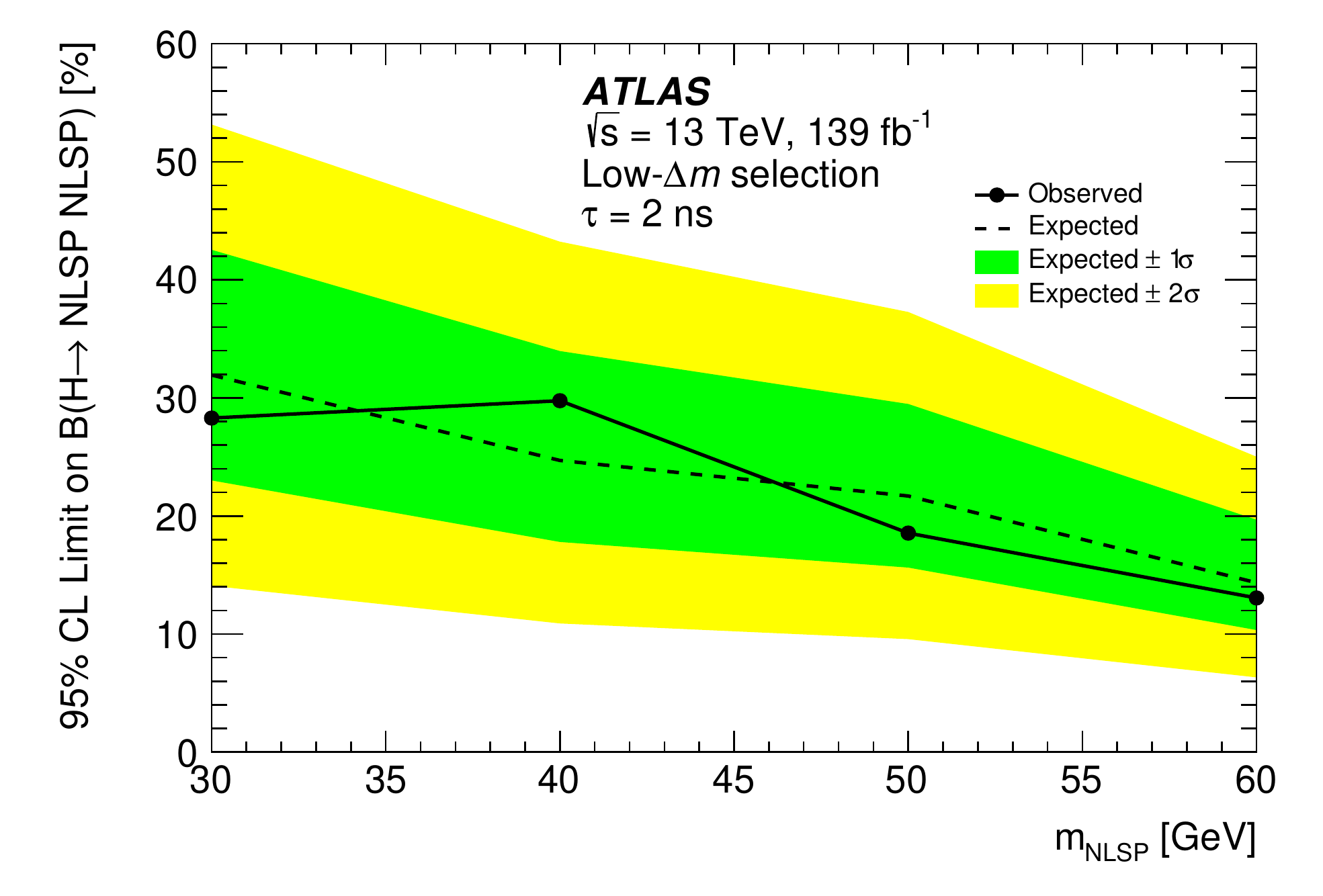}
\includegraphics[width=0.478\columnwidth]{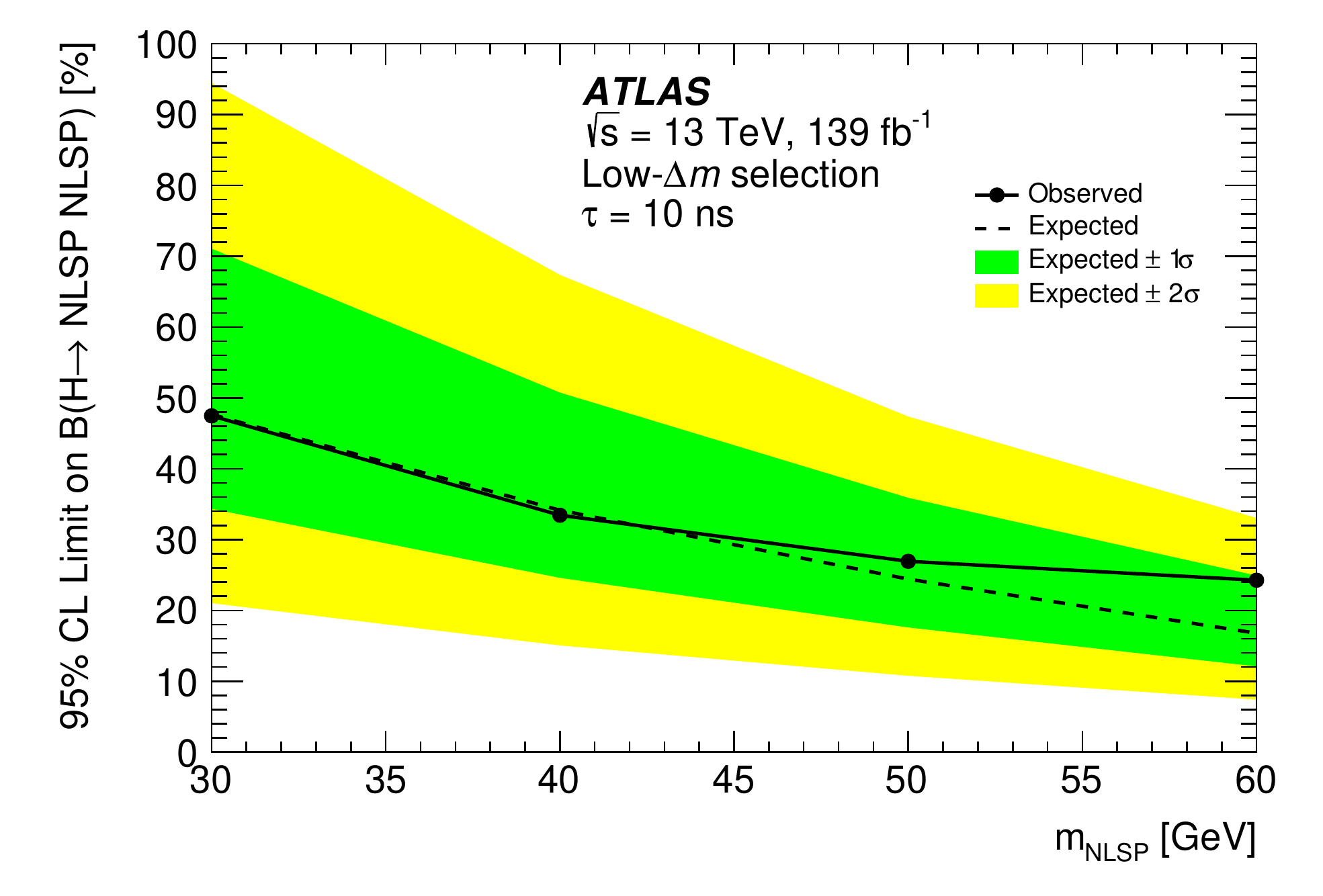}
\caption{ Interpolated 95\% CL limits on $\mathcal{B}$(\Hnlspnlsp) as a function of NLSP mass for signal points in the the \lowdm \ selection. Straight lines are drawn between
the results for the
discrete signal models considered by the \lowdm \ selection,
which were simulated with $\Delta m = 10$~\GeV.
The plot on the left (right) corresponds to the
results with the NLSP lifetime equal to 2~ns (10~ns).
\label{fig:InterpretationMassLow} }
\end{figure*}
 
Examining the results in Figures~\ref{fig:InterpretationMassHigh}
and \ref{fig:InterpretationMassLow}, there is a clear dependence of the achieved
limit on the signal mass splitting, $\Delta m$. The most
stringent constraints are achieved toward
the bottom right corner of Figure~\ref{fig:InterpretationMassHigh},
where 95\% CL upper limits on $\mathcal{B}$(\Hnlspnlsp) of order 1\%
are obtained.  In that region of the signal mass grid, the
mass splitting between the NLSP and LSP is large, giving
rise to higher-energy photons and, therefore, higher photon acceptance and lower backgrounds. Moving from the bottom
right to upper left in Figure~\ref{fig:InterpretationMassHigh},
the limits smoothly become less stringent as the mass
splitting decreases, a trend which continues into the \lowdm \
results in Figure~\ref{fig:InterpretationMassLow}.
However, it should be noted that the 95\% CL upper
limit on $\mathcal{B}$(\Hnlspnlsp) is below 10\% over most of the two-dimensional mass
plane for the \highdm \ case shown in Figure~\ref{fig:InterpretationMassHigh}.
 
The observed BSM limits also depend on the NLSP lifetime.
Comparing the left and right plots of Figures~\ref{fig:InterpretationMassHigh}
and \ref{fig:InterpretationMassLow} shows that, for
a given set of NLSP and LSP mass values, the upper limits
on $\mathcal{B}$(\Hnlspnlsp)
for an NLSP lifetime of 2~ns are tighter than those for a 10~ns NLSP lifetime.
This reflects the fact that the probability of at least one NLSP decaying before reaching the EM calorimeter is higher for low-lifetime signal points, resulting in higher photon
(and therefore signal) acceptance. The lifetime dependence
of the results is further explored by appropriately reweighting
the simulated MC samples, typically generated with fixed NLSP lifetimes
of either 2~ns or 10~ns, in order to produce reweighted
samples corresponding to other lifetime values, ranging from
250~ps to 100~ns.
 
Using this lifetime reweighting method, Figure~\ref{fig:InterpretationLifetime} presents the 95\% CL
limits on $\mathcal{B}$(\Hnlspnlsp) in the two-dimensional plane of
LSP mass versus NLSP lifetime. An interpolation is again used
to provide smooth results throughout the plane, connecting the discrete mass
values that were simulated.
The results are shown separately in four plots, with the
NLSP mass fixed within a given plot, and with the four fixed
NLSP mass values considered being 30, 40, 50, and 60~\GeV.
 
\begin{figure*}[!hp]
\centering
\includegraphics[width=0.48\columnwidth]{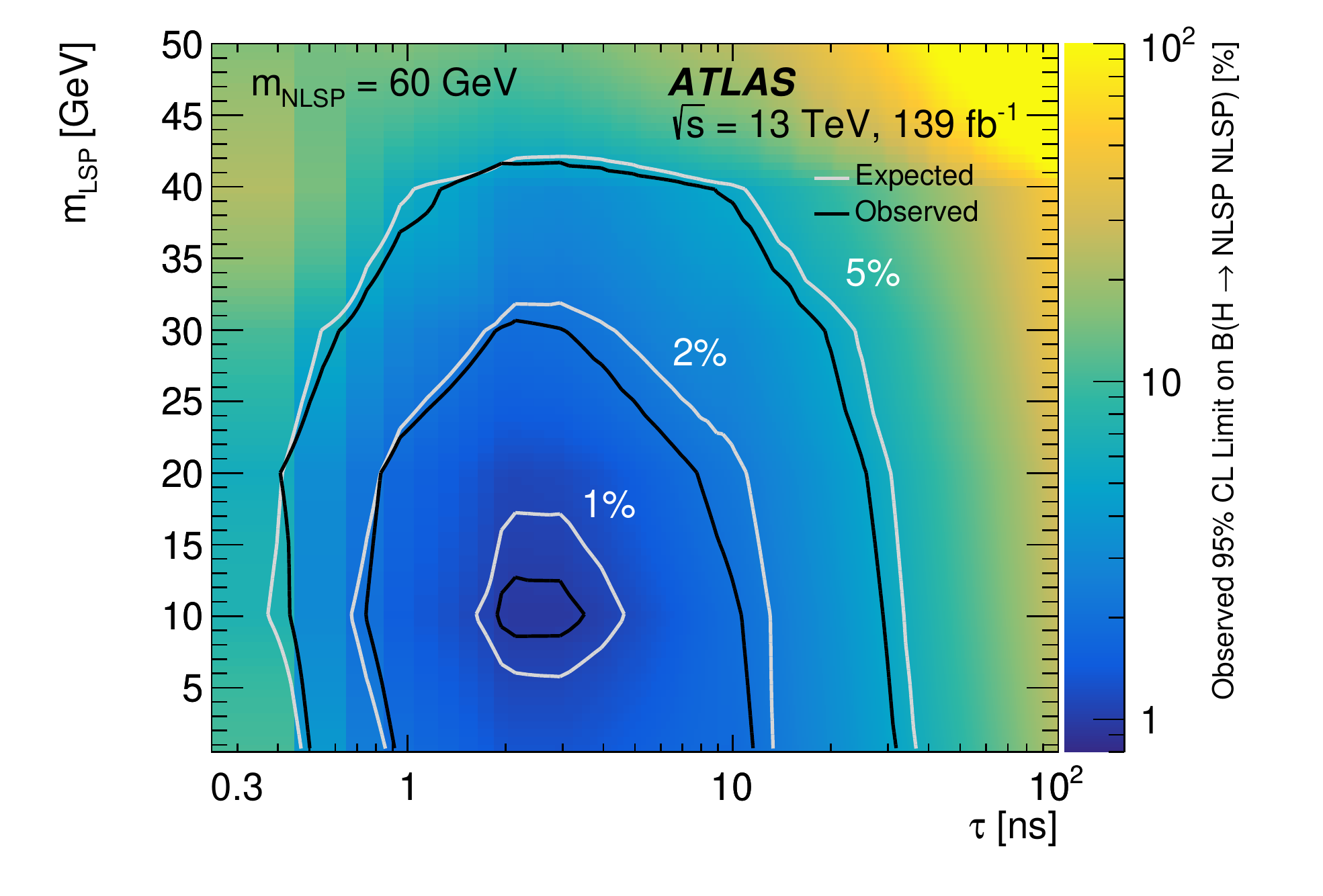}
\includegraphics[width=0.48\columnwidth]{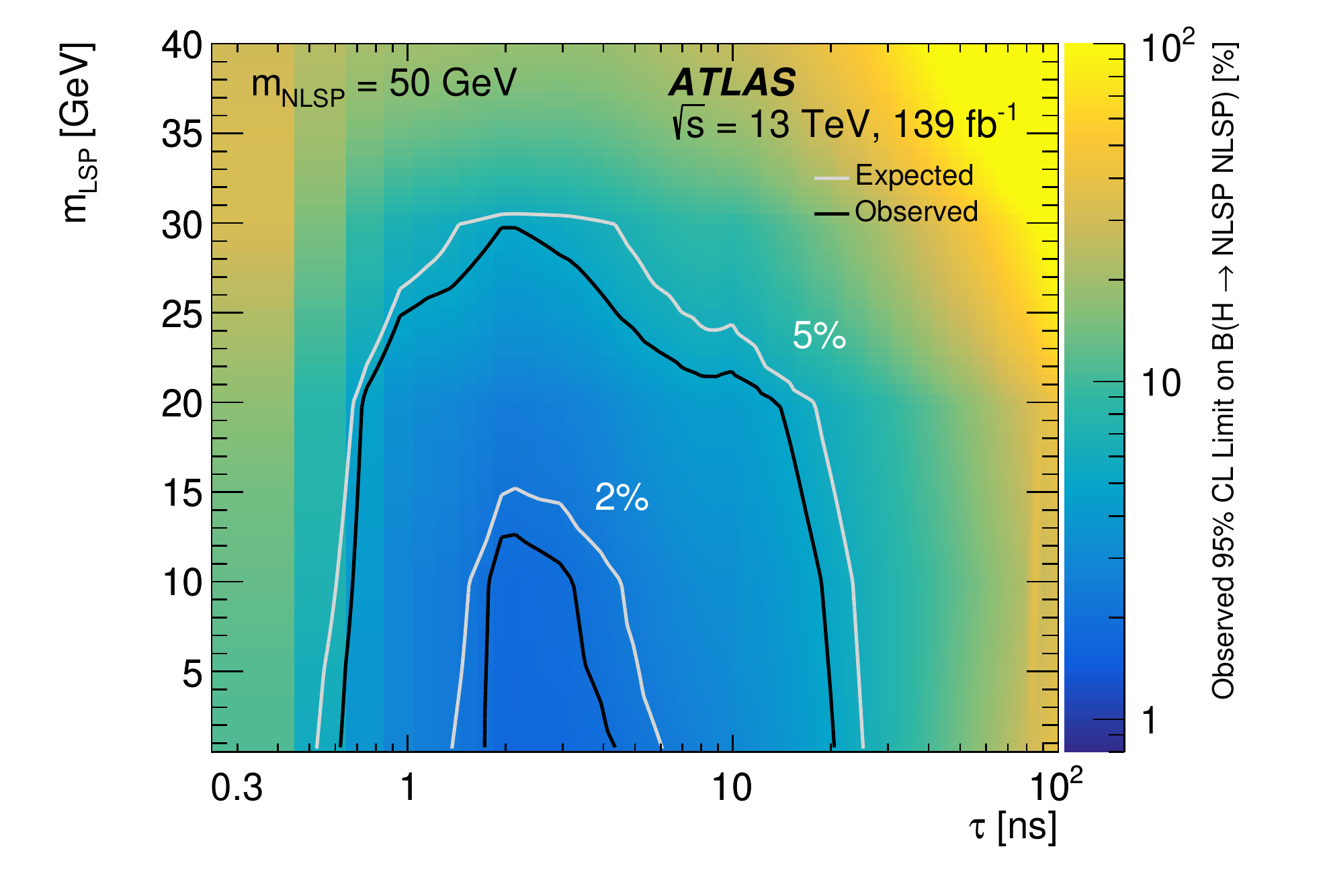}
\includegraphics[width=0.48\columnwidth]{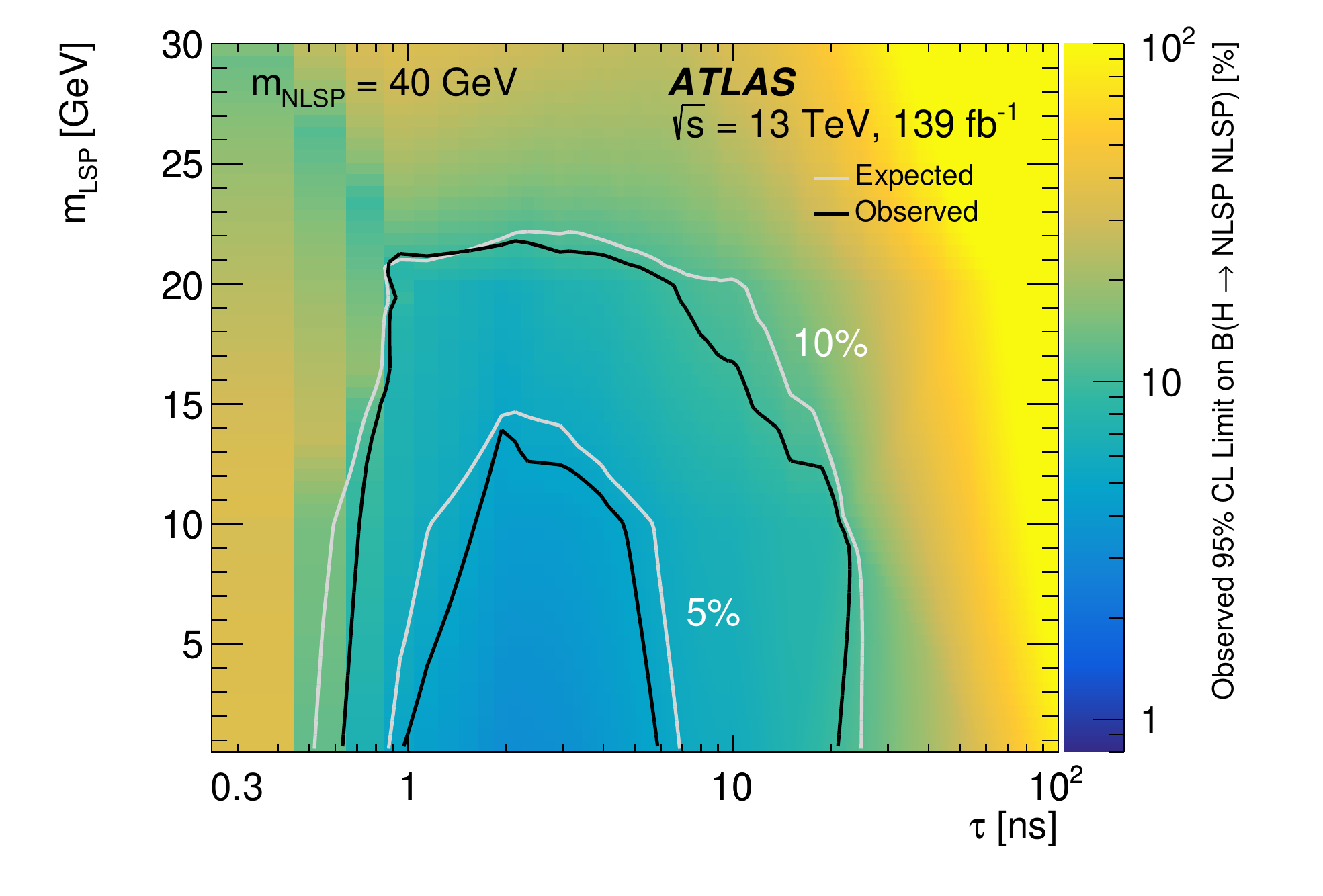}
\includegraphics[width=0.48\columnwidth]{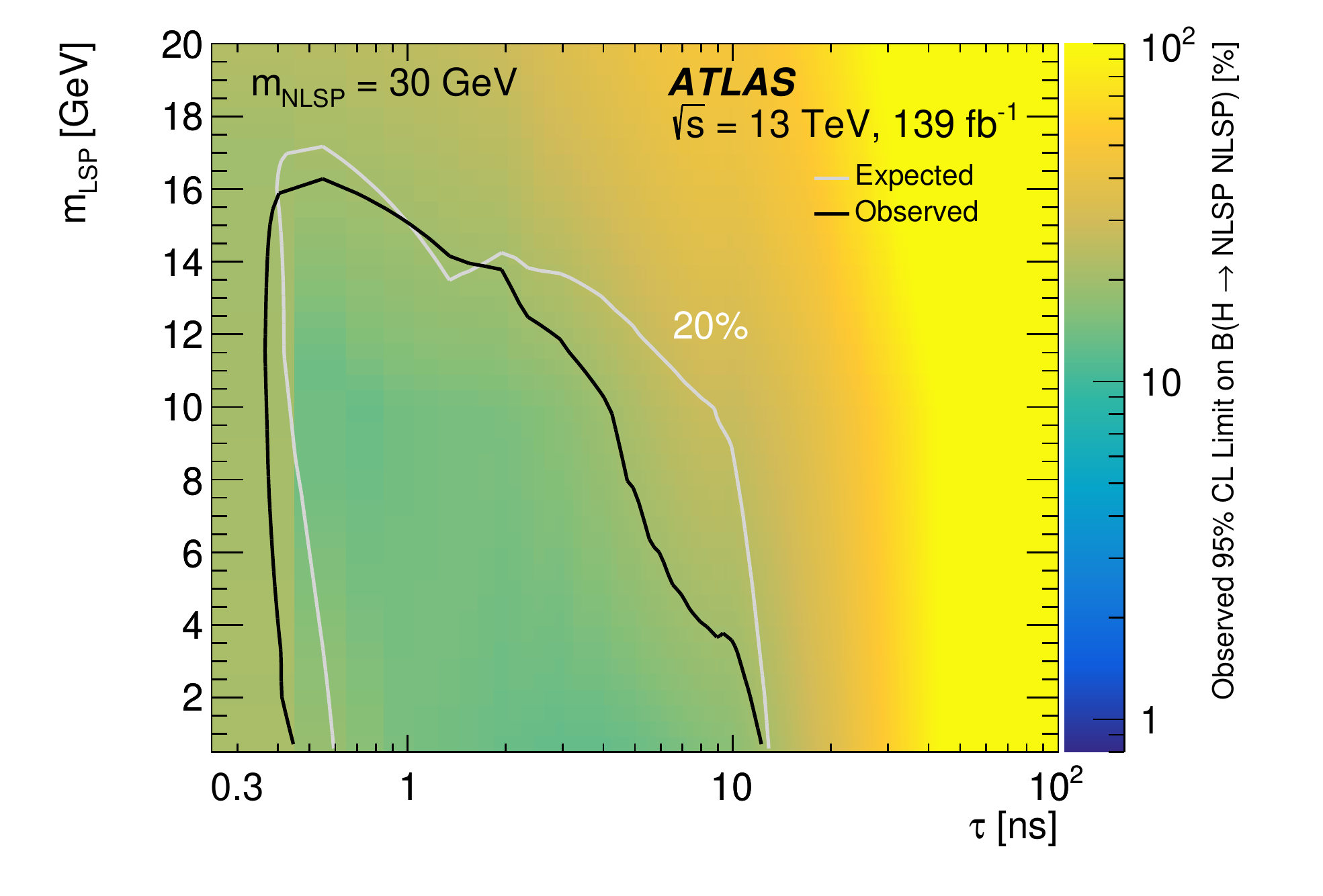}
\caption{Interpolated 95\% CL limits on $\mathcal{B}$(\Hnlspnlsp) in the
plane of LSP mass versus NLSP lifetime. Within each plot,
the NLSP mass is held fixed, with values of 60~\GeV \ (top left), 50~\GeV \ (top right), 40~\GeV \ (bottom left), and 30~\GeV \ (bottom right). A smoothing function is used to continuously
fill the plane, interpolating between the results for the
discrete signal models, which
were simulated on a grid with 10~\GeV \ spacing.
For visualization purposes,
contours are superimposed to
indicate where the expected and observed 95\% CL limits
on $\mathcal{B}$(\Hnlspnlsp) reach values of  1\%, 2\%, 5\%, 10\% or 20\%, depending on the plot.
\label{fig:InterpretationLifetime} }
\end{figure*}
\FloatBarrier
 
The results in Figure~\ref{fig:InterpretationLifetime} demonstrate
that the limits are most stringent for NLSP lifetimes of the
order of a few ns, as expected
given the size of the ID. The limits become weaker for both
shorter and longer lifetimes.  The sensitivity decreases for longer lifetimes because more NLSP decays
occur outside the ID volume and therefore fewer decay photons are
reconstructed in the EM calorimeter, with a resultant loss
of signal acceptance.
For very short lifetimes, the signal
pointing and timing distributions become more difficult
to separate from the prompt background, reducing the
sensitivity of the analysis. These trends can be illustrated
more clearly by examining the evolution of the limits
as a function of NLSP lifetime for fixed NLSP and LSP masses,
as shown in Figure~\ref{fig:InterpretationLimitVsLifetime}.
 
\begin{figure*}[!ph]
\centering
\includegraphics[width=0.48\columnwidth]{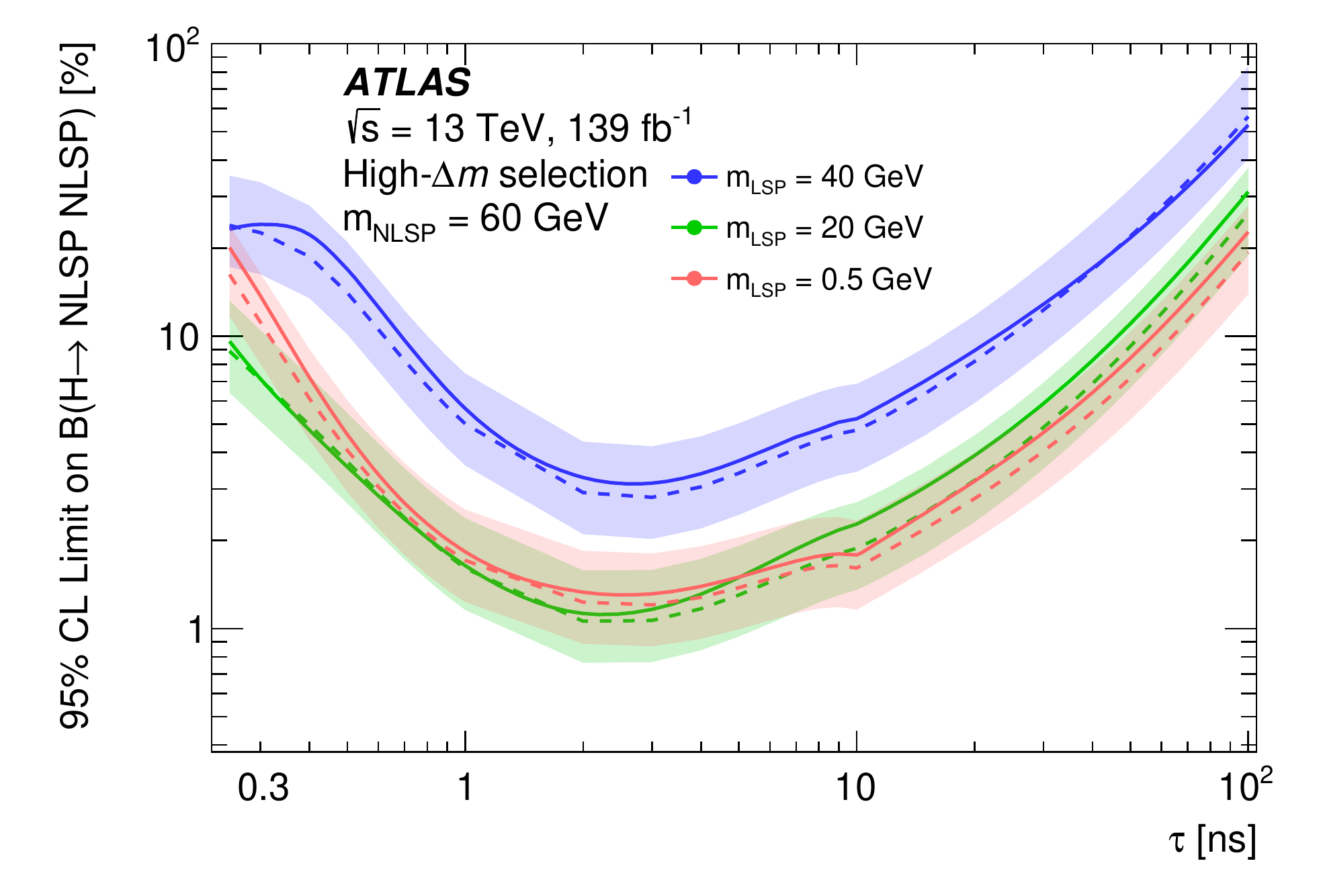}
\includegraphics[width=0.48\columnwidth]{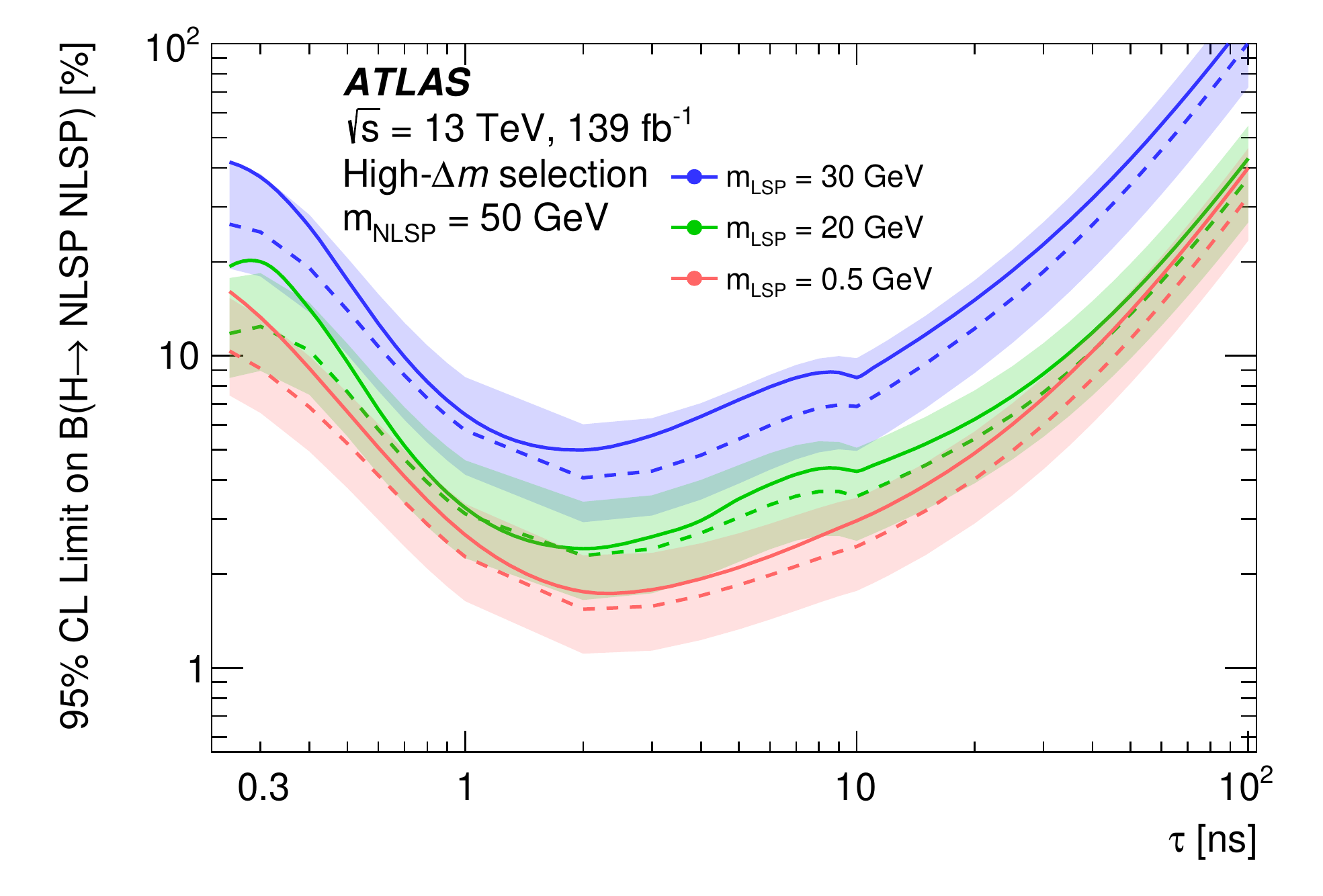}
\includegraphics[width=0.48\columnwidth]{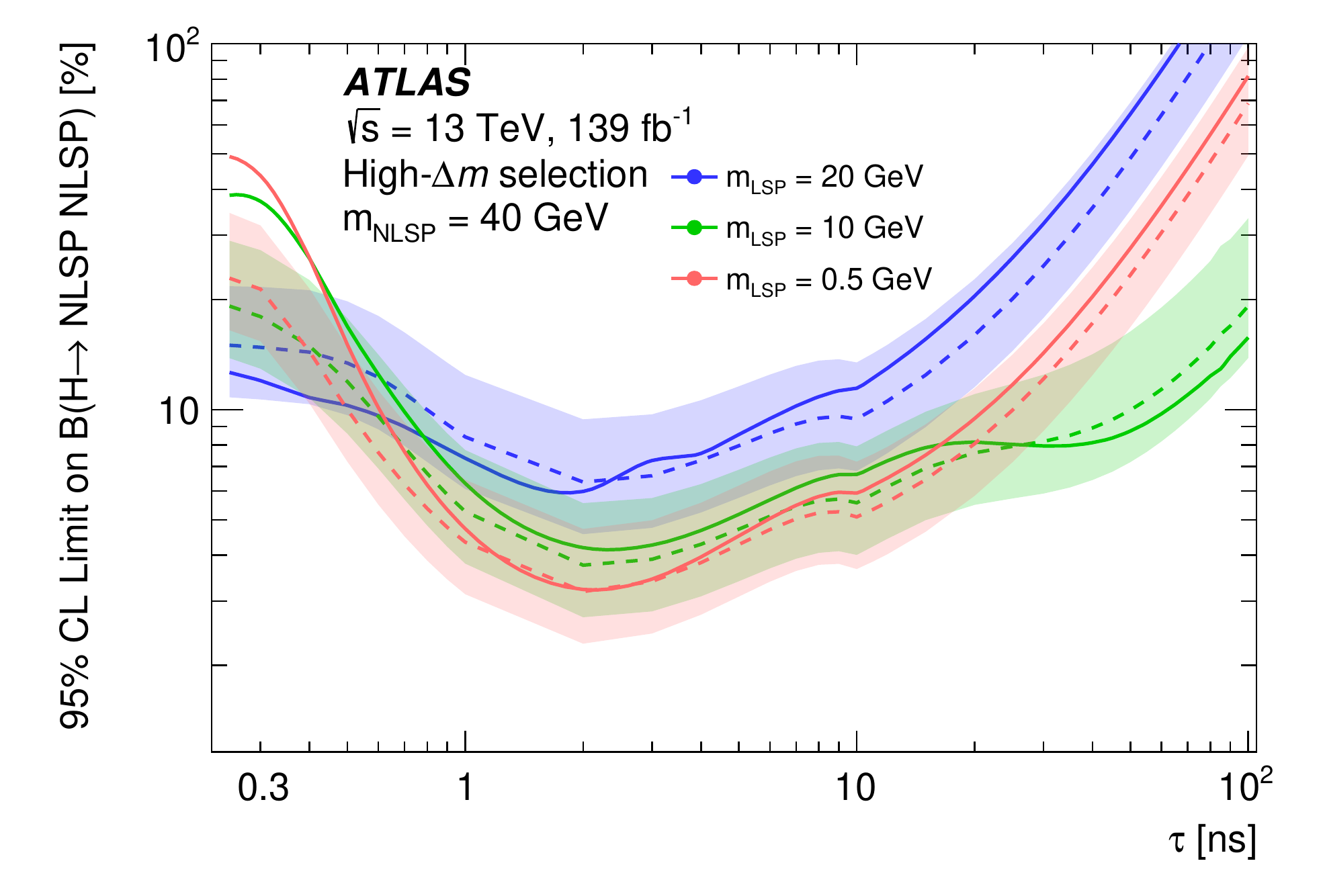}
\includegraphics[width=0.48\columnwidth]{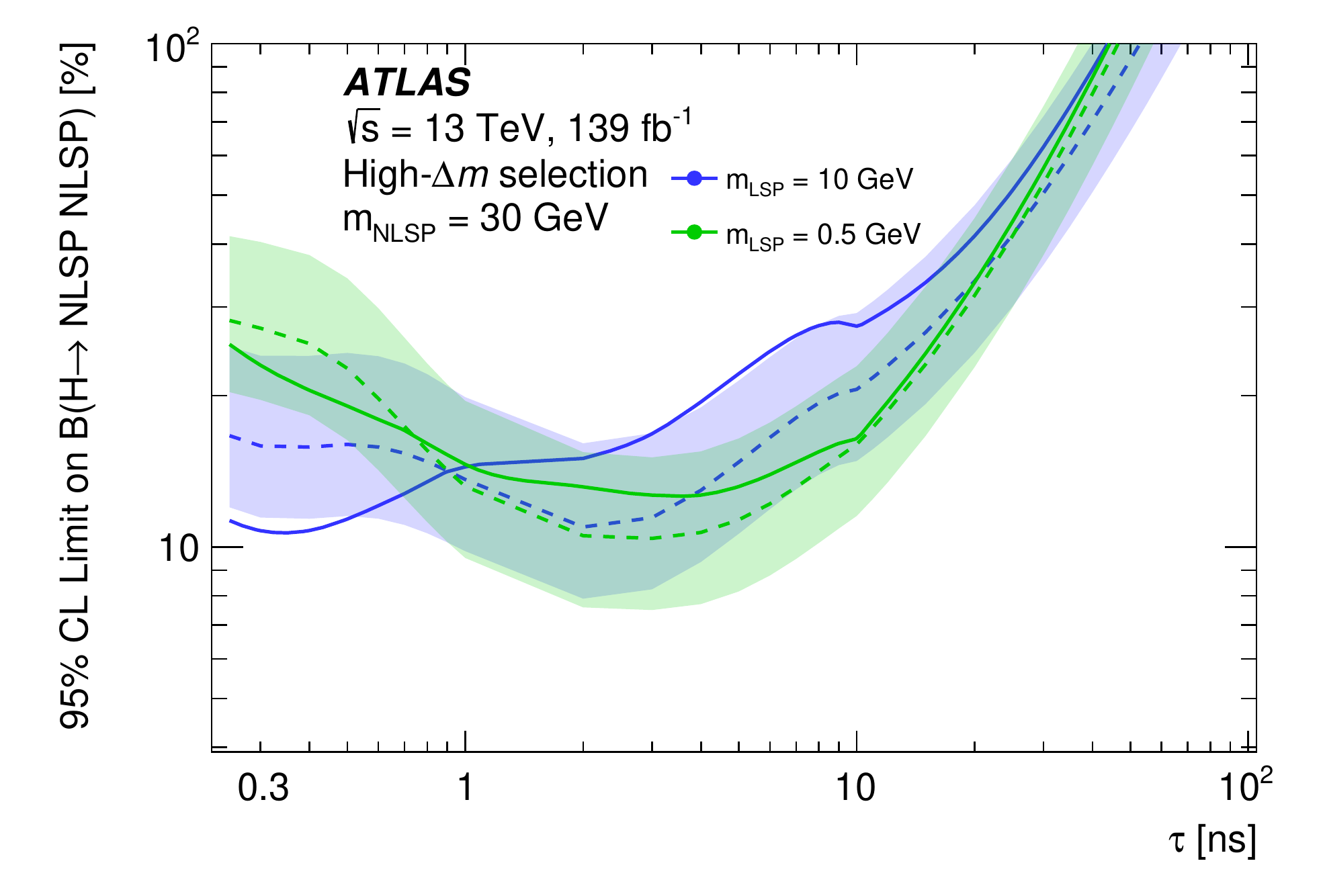}
\caption{The 95\% CL limits on $\mathcal{B}$(\Hnlspnlsp) from the \highdm \ selection as a function of
NLSP lifetime for fixed values of the NLSP and LSP masses. Within each plot,
the NLSP mass is held fixed, with values of 60~\GeV \ (top left), 50~\GeV \ (top right), 40~\GeV \ (bottom left), and 30~\GeV \ (bottom right). Each plot includes multiple curves, each with the
LSP mass fixed, as shown in the legend. The 95\% CL observed limits are shown as solid lines, while the 95\% CL expected limits and their $\pm1\sigma$ uncertainties are shown as the dotted lines and shaded bands, respectively. The crossings of the observed and expected limits are a result of changing relative importance of the $1\gamma$ and $\ge 2\gamma$ channels as a function of signal lifetimes. \label{fig:InterpretationLimitVsLifetime} }
\end{figure*}
\FloatBarrier

In addition to the results reported above for LLPs produced
in Higgs boson decays, a more model-independent test
is performed to provide a limit on the
production of displaced photons with large values of both
pointing and timing, using the selection requirements of
the \highdm \ analysis. The test is performed separately
for
the $1\gamma$ and $\ge2\gamma$ channels, as well as their
combination. The model-independent test region is
defined for each channel by the lower bound for
the single bin with the highest pointing and timing values.
The pointing requirement is thus $\abszdca > 300$~mm for
both channels, while the timing requirement is
$\tg > 1.5$~ns ($1$~ns) for the $1\gamma$ ($\ge2\gamma$) channel.
The same likelihood fit procedure described in Section~\ref{sec:stats} is applied, but eliminating the
model-independent test regions from the fit.
The background prediction and uncertainty in the model-independent test regions are then determined by extrapolating the
results of this background-only fit to the model-independent
region with higher pointing and timing values.
 
The model-independent limits are summarized in Table~\ref{tab:model_independent}. For the two channels combined, four events are observed in the data, whereas
the background prediction is $4.1 \pm 1.7$ events.
Since the observed numbers of events are consistent with the background expectation in all regions, the results are
used to set 95\%~CL upper limits on the visible cross sections of BSM signals in the model-independent test regions. The reconstruction efficiency for signal events, after taking into account the acceptance of the selection, is typically around
5\% for signal samples considered in this analysis.
 
\begin{table*}[hbpt]
\centering
\caption{Expected background event yield and observed event yield in the model-independent test regions, as defined for the
$1\gamma$ and $\ge2\gamma$ channels by the listed
requirements, as well as for the combination of the
two channels. Also shown are the values of $\sigma^{95}_\mathrm{vis}$, the corresponding
95\% CL limit on the visible cross section of BSM signals.}
\label{tab:model_independent}
\begin{tabular}{lcccc}
\hline
Channel & Test region requirements & Expected & Observed & $\sigma^{95}_\mathrm{vis}$ [fb]\\
\midrule
$1\gamma$&$ 1.5 < \tg < 12$ ns, |\zdca| $>$ 300 mm & 3.8~~ $\pm$ 1.6~~ & 4  & 0.042 \\
$\ge 2\gamma$&$ 1.0 < \tg < 12$ ns, |\zdca| $>$ 300 mm & 0.28 $\pm$ 0.04 & 0 & 0.022\\ \hline
$\ge1\gamma$ & & 4.1~~ $\pm$ 1.7~~ & 4 &  0.041 \\
\hline
\end{tabular}
\end{table*}
\FloatBarrier


\section{Conclusion}
\label{sec:conclusion}

 
A search is performed for delayed and nonpointing photons produced from exotic decays of the 125~\GeV \
Higgs boson into a pair of BSM LLPs, such as the NLSP
present in GMSB models.
The data set used was recorded by the ATLAS detector at the LHC and corresponds to an integrated luminosity of 139~\ifb\ of $pp$ collisions at a center-of-mass energy of $\sqrt{s}=13$~\TeV.
 
Using measurements of the trajectories and
arrival times of photons in the ATLAS LAr EM calorimeter to
search for possible displaced photons, no excess is observed
above the expectation from prompt backgrounds,
which are modeled entirely using data control samples.
The results are used to set limits on $\mathcal{B}$(\Hnlspnlsp) for
NLSP lifetimes ranging from 0.25~ns to 100~ns,
assuming that $\mathcal{B}(\mathrm{NLSP} \rightarrow \mathrm{LSP} + \gamma) = 100$\%.
The limits are determined in a two-dimensional grid of NLSP mass values from
30~\GeV\ to 60~\GeV \ and LSP masses from 0.5~\GeV\ up to
$\mNLSP - 10$~\GeV. The most stringent constraints limit
$\mathcal{B}$(\Hnlspnlsp) to less than about 1\% for intermediate NLSP
lifetime values and for the largest NLSP mass
and mass-splitting between the NLSP and LSP.
A model-independent
limit is also set on the production of photons with large values of
displacement and time delay.


\section*{Acknowledgments}


We thank CERN for the very successful operation of the LHC, as well as the
support staff from our institutions without whom ATLAS could not be
operated efficiently.
 
We acknowledge the support of
ANPCyT, Argentina;
YerPhI, Armenia;
ARC, Australia;
BMWFW and FWF, Austria;
ANAS, Azerbaijan;
CNPq and FAPESP, Brazil;
NSERC, NRC and CFI, Canada;
CERN;
ANID, Chile;
CAS, MOST and NSFC, China;
Minciencias, Colombia;
MEYS CR, Czech Republic;
DNRF and DNSRC, Denmark;
IN2P3-CNRS and CEA-DRF/IRFU, France;
SRNSFG, Georgia;
BMBF, HGF and MPG, Germany;
GSRI, Greece;
RGC and Hong Kong SAR, China;
ISF and Benoziyo Center, Israel;
INFN, Italy;
MEXT and JSPS, Japan;
CNRST, Morocco;
NWO, Netherlands;
RCN, Norway;
MEiN, Poland;
FCT, Portugal;
MNE/IFA, Romania;
MESTD, Serbia;
MSSR, Slovakia;
ARRS and MIZ\v{S}, Slovenia;
DSI/NRF, South Africa;
MICINN, Spain;
SRC and Wallenberg Foundation, Sweden;
SERI, SNSF and Cantons of Bern and Geneva, Switzerland;
MOST, Taiwan;
TENMAK, T\"urkiye;
STFC, United Kingdom;
DOE and NSF, United States of America.
In addition, individual groups and members have received support from
BCKDF, CANARIE, Compute Canada and CRC, Canada;
PRIMUS 21/SCI/017 and UNCE SCI/013, Czech Republic;
COST, ERC, ERDF, Horizon 2020 and Marie Sk{\l}odowska-Curie Actions, European Union;
Investissements d'Avenir Labex, Investissements d'Avenir Idex and ANR, France;
DFG and AvH Foundation, Germany;
Herakleitos, Thales and Aristeia programmes co-financed by EU-ESF and the Greek NSRF, Greece;
BSF-NSF and MINERVA, Israel;
Norwegian Financial Mechanism 2014-2021, Norway;
NCN and NAWA, Poland;
La Caixa Banking Foundation, CERCA Programme Generalitat de Catalunya and PROMETEO and GenT Programmes Generalitat Valenciana, Spain;
G\"{o}ran Gustafssons Stiftelse, Sweden;
The Royal Society and Leverhulme Trust, United Kingdom.
 
The crucial computing support from all WLCG partners is acknowledged gratefully, in particular from CERN, the ATLAS Tier-1 facilities at TRIUMF (Canada), NDGF (Denmark, Norway, Sweden), CC-IN2P3 (France), KIT/GridKA (Germany), INFN-CNAF (Italy), NL-T1 (Netherlands), PIC (Spain), ASGC (Taiwan), RAL (UK) and BNL (USA), the Tier-2 facilities worldwide and large non-WLCG resource providers. Major contributors of computing resources are listed in Ref.~\cite{ATL-SOFT-PUB-2021-003}.


\clearpage

\printbibliography

@article{Cowan:2010js,
    author = "Cowan, Glen and Cranmer, Kyle and Gross, Eilam and Vitells, Ofer",
    title = "{Asymptotic formulae for likelihood-based tests of new physics}",
    eprint = "1007.1727",
    archivePrefix = "arXiv",
    primaryClass = "physics.data-an",
    doi = "10.1140/epjc/s10052-011-1554-0",
    journal = "Eur. Phys. J. C",
    volume = "71",
    pages = "1554",
    year = "2011",
    note = "[Erratum: Eur.Phys.J.C 73, 2501 (2013)]"
}

@article{HIGG-2021-23,
    collaboration = "ATLAS Collaboration",
    title = "{A detailed map of Higgs boson interactions by the ATLAS experiment ten years after the discovery}",
    eprint = "2207.00092",
    archivePrefix = "arXiv",
    primaryClass = "hep-ex",
    reportNumber = "CERN-EP-2022-057",
    doi = "10.1038/s41586-022-04893-w",
    journal = "Nature",
    volume = "607",
    number = "7917",
    pages = "52--59",
    year = "2022"
}

@Article{Neveu:1971rx,
	author    = "Neveu, A. and Schwarz, J. H.",
	title     = "{Factorizable dual model of pions}",
	journal   = "Nucl. Phys. B",
	volume    = "31",
	year      = "1971",
	pages     = "86",
	doi       = "10.1016/0550-3213(71)90448-2",
	SLACcitation  = "%%CITATION = NUPHA,B31,86;%%"
}

@Article{Neveu:1971iv,
	author    = "Neveu, A. and Schwarz, J. H.",
	title     = "{Quark model of dual pions}",
	journal   = "Phys. Rev. D",
	volume    = "4",
	year      = "1971",
	pages     = "1109",
	doi       = "10.1103/PhysRevD.4.1109",
	SLACcitation  = "%%CITATION = PHRVA,D4,1109;%%"
}

@Article{Ramond:1971gb,
	author    = "Ramond, Pierre",
	title     = "{Dual theory for free fermions}",
	journal   = "Phys. Rev. D",
	volume    = "3",
	year      = "1971",
	pages     = "2415",
	doi       = "10.1103/PhysRevD.3.2415",
	SLACcitation  = "%%CITATION = PHRVA,D3,2415;%%"
}

@Article{Wess:1973kz,
	author    = "Wess, J. and Zumino, B.",
	title     = "{A lagrangian model invariant under supergauge
	Transformations}",
	journal   = "Phys. Lett. B",
	volume    = "49",
	year      = "1974",
	pages     = "52",
	doi       = "10.1016/0370-2693(74)90578-4",
	SLACcitation  = "%%CITATION = PHLTA,B49,52;%%"
}

@Article{Fayet:1979sa,
	author    = "Fayet, P.",
	title     = "{Relations between the masses of the superpartners of
	leptons and quarks, the goldstino couplings and the neutral
	currents}",
	journal   = "Phys. Lett. B",
	volume    = "84",
	year      = "1979",
	pages     = "416",
	doi       = "10.1016/0370-2693(79)91229-2",
	SLACcitation  = "%%CITATION = PHLTA,B84,416;%%"
}

@Article{Nappi:1982,
	author    = "Nappi, Chiara R. and Ovrut, Burt A.",
	title     = "{Supersymmetric extension of the SU(3)$\times$SU(2)$\times$U(1) model}",
	journal   = "Phys. Lett. B",
	volume    = "113",
	year      = "1982",
	pages     = "175",
	doi       = "10.1016/0370-2693(82)90418-X",
	SLACcitation  = "%%CITATION = PHLTA,B113,175;%%"
}

@Article{Dine:1993yw,
	author    = "M. Dine and A.E. Nelson",
	title     = "Dynamical supersymmetry breaking at low-energies",
	journal   = "Phys. Rev. D",
	volume    = "48",
	year      = "1993",
	pages     = "1277",
	eprint    = "hep-ph/9303230",
	doi            = "10.1103/PhysRevD.48.1277",
	SLACcitation  = "%%CITATION = HEP-PH/9303230;%%"
}

@Article{Dine:1994vc,
	author    = "M. Dine and A.E. Nelson and Y. Shirman",
	title     = "Low-energy dynamical supersymmetry breaking simplified",
	journal   = "Phys. Rev. D",
	volume    = "51",
	year      = "1995",
	pages     = "1362",
	eprint    = "hep-ph/9408384",
	doi            = "10.1103/PhysRevD.51.1362",
	SLACcitation  = "%%CITATION = HEP-PH/9408384;%%"
}

@Article{Dine:1995ag,
	author    = "M. Dine and A.E. Nelson and Y.Nir and Y. Shirman",
	title     = "New tools for low-energy dynamical supersymmetry breaking",
	journal   = "Phys. Rev. D",
	volume    = "53",
	year      = "1996",
	pages     = "2658",
	eprint    = "hep-ph/9507378",
	doi            = "10.1103/PhysRevD.53.2658",
	SLACcitation  = "%%CITATION = HEP-PH/9507378;%%"
}

@article{Abreu:2010zzc,
	author = "Abreu, H. and others",
	title = "{Performance of the electronic readout of the ATLAS liquid argon calorimeters}",
	doi = "10.1088/1748-0221/5/09/P09003",
	journal = "JINST",
	volume = "5",
	pages = "P09003",
	year = "2010"
}

@article{Avoni_2018,
    author = "Avoni, G. and others",
    title = "{The new LUCID-2 detector for luminosity measurement and monitoring in ATLAS}",
    doi = "10.1088/1748-0221/13/07/P07017",
    journal = "JINST",
    volume = "13",
    number = "07",
    pages = "P07017",
    year = "2018"
}

@article{Buchanan:2008zza,
	author         = "Buchanan, N. J. and others",
	title          = "{Design and implementation of the Front End Board for the
	readout of the ATLAS liquid argon calorimeters}",
	journal        = "JINST",
	volume         = "3",
	pages          = "P03004",
	doi            = "10.1088/1748-0221/3/03/P03004",
	year           = "2008",
	SLACcitation   = "%%CITATION = JINST,3,P03004;%%",
}

@article{cleland,
    author = "Cleland, W. E. and Stern, E. G.",
    title = "{Signal processing considerations for liquid ionization calorimeters in a high rate environment}",
    doi = "10.1016/0168-9002(94)91332-3",
    journal = "Nucl. Instrum. Meth. A",
    volume = "338",
    number = "2-3",
    pages = "467--497",
    year = "1994"
}

@ARTICLE{2008NIMPA.597..178A,
    author = "Aharrouche, M. and others",
    title = "{Time resolution of the ATLAS barrel liquid argon electromagnetic calorimeter}",
    doi = "10.1016/j.nima.2008.08.142",
    journal = "Nucl. Instrum. Meth. A",
    volume = "597",
    pages = "178--188",
    year = "2008"
}

@article{Golfand:1971iw,
      author         = {Golfand, Y.A.  and Likhtman, E.P.},
      title          = "{Extension of the Algebra of Poincare Group Generators
                        and Violation of P Invariance}",
      journal        = "JETP Lett.",
      volume         = "13",
      year           = "1971",
      pages          = "323",
      note           = "[Pisma Zh. Eksp. Teor. Fiz. \textbf{13} (1971) 452]",
      SLACcitation   = "%%CITATION = JTPLA,13,323;%%"
}

@article{Volkov:1973ix,
      author         = "Volkov, D.V. and Akulov, V.P.",
      title          = "{Is the neutrino a goldstone particle?}",
      journal        = "Phys. Lett. B",
      volume         = "46",
      year           = "1973",
      pages          = "109",
      doi            = "10.1016/0370-2693(73)90490-5",
      SLACcitation   = "%%CITATION = PHLTA,B46,109;%%"
}

@article{Wess:1974tw,
      author         = "Wess, J. and Zumino, B.",
      title          = "{Supergauge transformations in four dimensions}",
      journal        = "Nucl. Phys. B",
      volume         = "70",
      year           = "1974",
      pages          = "39",
      doi            = "10.1016/0550-3213(74)90355-1",
      SLACcitation   = "%%CITATION = NUPHA,B70,39;%%"
}

@article{Fayet:1976et,
      author         = "Fayet, Pierre",
      title          = "{Supersymmetry and weak, electromagnetic and strong
                        interactions}",
      journal        = "Phys. Lett. B",
      volume         = "64",
      year           = "1976",
      pages          = "159",
      doi            = "10.1016/0370-2693(76)90319-1",
      reportNumber   = "PTENS 76/14",
      SLACcitation   = "%%CITATION = PHLTA,B64,159;%%"
}

@article{Fayet:1977yc,
      author         = "Fayet, Pierre",
      title          = "{Spontaneously broken supersymmetric theories of weak,
                        electromagnetic and strong interactions}",
      journal        = "Phys. Lett. B",
      volume         = "69",
      year           = "1977",
      pages          = "489",
      doi            = "10.1016/0370-2693(77)90852-8",
      reportNumber   = "LPTENS 77/11",
      SLACcitation   = "%%CITATION = PHLTA,B69,489;%%"
}

@article{Farrar:1978xj,
      author         = "Farrar, Glennys R. and Fayet, Pierre",
      title          = "{Phenomenology of the production, decay, and detection of
                        new hadronic states associated with supersymmetry}",
      journal        = "Phys. Lett. B",
      volume         = "76",
      year           = "1978",
      pages          = "575",
      doi            = "10.1016/0370-2693(78)90858-4",
      reportNumber   = "CALT-68-648",
      SLACcitation   = "%%CITATION = PHLTA,B76,575;%%"
}

@article{Dine:1981gu,
      author         = "Dine, Michael and Fischler, Willy",
      title          = "{A phenomenological model of particle physics based on supersymmetry}",
      journal        = "Phys. Lett. B",
      volume         = "110",
      year           = "1982",
      pages          = "227",
      doi            = "10.1016/0370-2693(82)91241-2",
      reportNumber   = "PRINT-82-0029 (IAS,PRINCETON)",
      SLACcitation   = "%%CITATION = PHLTA,B110,227;%%"
}

@article{AlvarezGaume:1981wy,
      author         = "Alvarez-Gaumé, Luis and Claudson, Mark and Wise, Mark B.",
      title          = "{Low-energy supersymmetry}",
      journal        = "Nucl. Phys. B",
      volume         = "207",
      year           = "1982",
      pages          = "96",
      doi            = "10.1016/0550-3213(82)90138-9",
      reportNumber   = "HUTP-81-A063",
      SLACcitation   = "%%CITATION = NUPHA,B207,96;%%"
}

@article{Dimopoulos:1981zb,
      author         = "Dimopoulos, Savas and Georgi, Howard",
      title          = "{Softly broken supersymmetry and SU(5)}",
      journal        = "Nucl. Phys. B",
      volume         = "193",
      year           = "1981",
      pages          = "150",
      doi            = "10.1016/0550-3213(81)90522-8",
      reportNumber   = "HUTP-81/A022",
      SLACcitation   = "%%CITATION = NUPHA,B193,150;%%"
}

@Article{Cacciari:2008gp,
     author    = "Cacciari, Matteo and Salam, Gavin P. and Soyez, Gregory",
     title     = "{The anti-\(k_{t}\) jet clustering algorithm}",
     journal   = "JHEP",
     volume    = "04",
     year      = "2008",
     pages     = "063",
     eprint    = "0802.1189",
     archivePrefix = "arXiv",
     primaryClass  =  "hep-ph",
     doi       = "10.1088/1126-6708/2008/04/063",
     SLACcitation  = "%%CITATION = 0802.1189;%%"
}

@Article{Fastjet,
      author         = "Cacciari, Matteo and Salam, Gavin P. and Soyez, Gregory",
      title          = "{FastJet user manual}",
      journal        = "Eur. Phys. J. C",
      volume         = "72",
      year           = "2012",
      pages          = "1896",
      doi            = "10.1140/epjc/s10052-012-1896-2",
      eprint         = "1111.6097",
      archivePrefix  = "arXiv",
      primaryClass   = "hep-ph",
      reportNumber   = "CERN-PH-TH-2011-297",
      SLACcitation   = "%%CITATION = ARXIV:1111.6097;%%"
}

@article{Ball:2014uwa,
      author         = "Ball, Richard D. and others",
      title          = "{Parton distributions for the LHC run II}",
      collaboration  = "NNPDF",
      journal        = "JHEP",
      volume         = "04",
      year           = "2015",
      pages          = "040",
      doi            = "10.1007/JHEP04(2015)040",
      eprint         = "1410.8849",
      archivePrefix  = "arXiv",
      primaryClass   = "hep-ph",
      reportNumber   = "EDINBURGH-2014-15, IFUM-1034-FT, CERN-PH-TH-2013-253,
                        OUTP-14-11P, CAVENDISH-HEP-14-11",
      SLACcitation   = "%%CITATION = ARXIV:1410.8849;%%"
}

@article{Sjostrand:2014zea,
      author         = "Sj{\"o}strand, Torbj{\"o}rn and Ask, Stefan and Christiansen,
                        Jesper R. and Corke, Richard and Desai, Nishita and Ilten,
                        Philip and Mrenna, Stephen and Prestel, Stefan and
                        Rasmussen, Christine O. and Skands, Peter Z.",
      title          = "{An introduction to PYTHIA 8.2}",
      journal        = "Comput. Phys. Commun.",
      volume         = "191",
      year           = "2015",
      pages          = "159",
      doi            = "10.1016/j.cpc.2015.01.024",
      eprint         = "1410.3012",
      archivePrefix  = "arXiv",
      primaryClass   = "hep-ph",
      reportNumber   = "LU-TP-14-36, MCNET-14-22, CERN-PH-TH-2014-190,
                        FERMILAB-PUB-14-316-CD, DESY-14-178, SLAC-PUB-16122,
                        --FERMILAB-PUB-14-316-CD",
      SLACcitation   = "%%CITATION = ARXIV:1410.3012;%%"
}

@Article{Alwall:2011uj,
      author         = "Alwall, Johan and Herquet, Michel and Maltoni, Fabio and
                        Mattelaer, Olivier and Stelzer, Tim",
      title          = "{MadGraph 5 : Going Beyond}",
      journal        = "JHEP",
      volume         = "06",
      pages          = "128",
      doi            = "10.1007/JHEP06(2011)128",
      year           = "2011",
      eprint         = "1106.0522",
      archivePrefix  = "arXiv",
      primaryClass   = "hep-ph"
}

@article{deFlorian:2016spz,
      author         = "de Florian, D. and others",
      title          = "{Handbook of LHC Higgs Cross Sections: 4. Deciphering the
                        Nature of the Higgs Sector}",
      collaboration  = "LHC Higgs Cross Section Working Group",
      doi            = "10.23731/CYRM-2017-002",
      year           = "2016",
      eprint         = "1610.07922",
      archivePrefix  = "arXiv",
      primaryClass   = "hep-ph",
      reportNumber   = "FERMILAB-FN-1025-T, CERN-2017-002-M",
      SLACcitation   = "%%CITATION = ARXIV:1610.07922;%%"
}

@article{Agostinelli:2002hh,
      author         = "{GEANT4 Collaboration} and Agostinelli, S. and others",
      title          = "{\textsc{Geant4} -- a simulation toolkit}",
      journal        = "Nucl. Instrum. Meth. A",
      volume         = "506",
      year           = "2003",
      pages          = "250",
      doi            = "10.1016/S0168-9002(03)01368-8",
      reportNumber   = "SLAC-PUB-9350, FERMILAB-PUB-03-339",
      SLACcitation   = "%%CITATION = NUIMA,A506,250;%%"
}

@Article{Read:2002hq,
    Author = {Read, Alexander L.},
    Journal = {J. Phys. G},
    Pages = {2693},
    doi = "10.1088/0954-3899/28/10/313",
    Title = "{Presentation of search results: the \(CL_S\) technique}",
    Volume = {28},
    Year = {2002}
}

@Article{PERF-2007-01,
    author         = "{ATLAS Collaboration}",
    title          = "{The ATLAS Experiment at the CERN Large Hadron Collider}",
    journal        = "JINST",
    volume         = "3",
    year           = "2008",
    pages          = "S08003",
    doi            = "10.1088/1748-0221/3/08/S08003",
    primaryClass   = "hep-ex",
}

@Article{SOFT-2010-01,
    author         = "{ATLAS Collaboration}",
    title          = "{The ATLAS Simulation Infrastructure}",
    journal        = "Eur. Phys. J. C",
    volume         = "70",
    year           = "2010",
    pages          = "823",
    doi            = "10.1140/epjc/s10052-010-1429-9",
    eprint         = "1005.4568",
    archivePrefix  = "arXiv",
    primaryClass   = "physics.ins-det",
}

@Article{SUSY-2012-15,
    author         = "{ATLAS Collaboration}",
    title          = "{Search for nonpointing photons in the diphoton and \(E^{\text{miss}}_{\text{T}}\) final state in \(\sqrt{s} = 7\,\text{TeV}\) proton--proton collisions using the ATLAS detector}",
    journal        = "Phys. Rev. D",
    volume         = "88",
    year           = "2013",
    pages          = "012001",
    doi            = "10.1103/PhysRevD.88.012001",
    reportNumber   = "CERN-PH-EP-2013-049",
    eprint         = "1304.6310",
    archivePrefix  = "arXiv",
    primaryClass   = "hep-ex",
}

@Article{DAPR-2013-01,
    author         = "{ATLAS Collaboration}",
    title          = "{Luminosity determination in \(pp\) collisions at \(\sqrt{s} = 8\,\text{TeV}\) using the ATLAS detector at the LHC}",
    journal        = "Eur. Phys. J. C",
    volume         = "76",
    year           = "2016",
    pages          = "653",
    doi            = "10.1140/epjc/s10052-016-4466-1",
    eprint         = "1608.03953",
    archivePrefix  = "arXiv",
    primaryClass   = "hep-ex",
}

@Article{SUSY-2013-17,
    author         = "{ATLAS Collaboration}",
    title          = "{Search for nonpointing and delayed photons in the diphoton and missing transverse momentum final state in \(8\,\text{TeV}\) \(pp\) collisions at the LHC using the ATLAS detector}",
    journal        = "Phys. Rev. D",
    volume         = "90",
    year           = "2014",
    pages          = "112005",
    doi            = "10.1103/PhysRevD.90.112005",
    reportNumber   = "CERN-PH-EP-2014-215",
    eprint         = "1409.5542",
    archivePrefix  = "arXiv",
    primaryClass   = "hep-ex",
}

@Article{PERF-2014-03,
    author         = "{ATLAS Collaboration}",
    title          = "{Performance of pile-up mitigation techniques for jets in \(pp\) collisions at \(\sqrt{s} = 8\,\text{TeV}\) using the ATLAS detector}",
    journal        = "Eur. Phys. J. C",
    volume         = "76",
    year           = "2016",
    pages          = "581",
    doi            = "10.1140/epjc/s10052-016-4395-z",
    reportNumber   = "CERN-PH-EP-2015-206",
    eprint         = "1510.03823",
    archivePrefix  = "arXiv",
    primaryClass   = "hep-ex",
}

@Article{PERF-2014-07,
    author         = "{ATLAS Collaboration}",
    title          = "{Topological cell clustering in the ATLAS calorimeters and its performance in LHC Run~1}",
    journal        = "Eur. Phys. J. C",
    volume         = "77",
    year           = "2017",
    pages          = "490",
    doi            = "10.1140/epjc/s10052-017-5004-5",
    reportNumber   = "CERN-PH-EP-2015-304",
    eprint         = "1603.02934",
    archivePrefix  = "arXiv",
    primaryClass   = "hep-ex",
}

@Article{PERF-2015-09,
    author         = "{ATLAS Collaboration}",
    title          = "{Jet reconstruction and performance using particle flow with the ATLAS Detector}",
    journal        = "Eur. Phys. J. C",
    volume         = "77",
    year           = "2017",
    pages          = "466",
    doi            = "10.1140/epjc/s10052-017-5031-2",
    reportNumber   = "CERN-EP-2017-024",
    eprint         = "1703.10485",
    archivePrefix  = "arXiv",
    primaryClass   = "hep-ex",
}

@Article{PERF-2016-07,
    author         = "{ATLAS Collaboration}",
    title          = "{Performance of missing transverse momentum reconstruction with the ATLAS detector using proton--proton collisions at \(\sqrt{s} = 13\,\text{TeV}\)}",
    journal        = "Eur. Phys. J. C",
    volume         = "78",
    year           = "2018",
    pages          = "903",
    doi            = "10.1140/epjc/s10052-018-6288-9",
    reportNumber   = "CERN-EP-2017-274",
    eprint         = "1802.08168",
    archivePrefix  = "arXiv",
    primaryClass   = "hep-ex",
}

@Article{TRIG-2016-01,
    author         = "{ATLAS Collaboration}",
    title          = "{Performance of the ATLAS trigger system in 2015}",
    journal        = "Eur. Phys. J. C",
    volume         = "77",
    year           = "2017",
    pages          = "317",
    doi            = "10.1140/epjc/s10052-017-4852-3",
    reportNumber   = "CERN-EP-2016-241",
    eprint         = "1611.09661",
    archivePrefix  = "arXiv",
    primaryClass   = "hep-ex",
}

@Article{PERF-2017-02,
    author         = "{ATLAS Collaboration}",
    title          = "{Measurement of the photon identification efficiencies with the ATLAS detector using LHC Run~2 data collected in 2015 and 2016}",
    journal        = "Eur. Phys. J. C",
    volume         = "79",
    year           = "2019",
    pages          = "205",
    doi            = "10.1140/epjc/s10052-019-6650-6",
    reportNumber   = "CERN-EP-2018-216",
    eprint         = "1810.05087",
    archivePrefix  = "arXiv",
    primaryClass   = "hep-ex",
}

@Article{DAPR-2018-01,
    author         = "{ATLAS Collaboration}",
    title          = "{ATLAS data quality operations and performance for 2015--2018 data-taking}",
    journal        = "JINST",
    volume         = "15",
    year           = "2020",
    pages          = "P04003",
    doi            = "10.1088/1748-0221/15/04/P04003",
    reportNumber   = "CERN-EP-2019-207",
    eprint         = "1911.04632",
    archivePrefix  = "arXiv",
    primaryClass   = "physics.ins-det",
}

@Article{EGAM-2018-01,
    author         = "{ATLAS Collaboration}",
    title          = "{Electron and photon performance measurements with the ATLAS detector using the 2015--2017 LHC proton--proton collision data}",
    journal        = "JINST",
    volume         = "14",
    year           = "2019",
    pages          = "P12006",
    doi            = "10.1088/1748-0221/14/12/P12006",
    reportNumber   = "CERN-EP-2019-145",
    eprint         = "1908.00005",
    archivePrefix  = "arXiv",
    primaryClass   = "hep-ex",
}

@Article{MUON-2018-03,
    author         = "{ATLAS Collaboration}",
    title          = "{Muon reconstruction and identification efficiency in ATLAS using the full Run~2 \(pp\) collision data set at \(\sqrt{s} = 13\,\text{TeV}\)}",
    journal        = "Eur. Phys. J. C",
    volume         = "81",
    year           = "2021",
    pages          = "578",
    doi            = "10.1140/epjc/s10052-021-09233-2",
    reportNumber   = "CERN-EP-2020-199",
    eprint         = "2012.00578",
    archivePrefix  = "arXiv",
    primaryClass   = "hep-ex",
}

@Article{PIX-2018-001,
    author         = "Abbott, B. and others",
    title          = "{Production and integration of the ATLAS Insertable B-Layer}",
    journal        = "JINST",
    volume         = "13",
    year           = "2018",
    pages          = "T05008",
    doi            = "10.1088/1748-0221/13/05/T05008",
    eprint         = "1803.00844",
    archivePrefix  = "arXiv",
    primaryClass   = "physics.ins-det",
}

@Article{TRIG-2018-01,
    author         = "{ATLAS Collaboration}",
    title          = "{Performance of the ATLAS muon triggers in Run~2}",
    journal        = "JINST",
    volume         = "15",
    year           = "2020",
    pages          = "P09015",
    doi            = "10.1088/1748-0221/15/09/p09015",
    reportNumber   = "CERN-EP-2020-031",
    eprint         = "2004.13447",
    archivePrefix  = "arXiv",
    primaryClass   = "hep-ex",
}

@Article{TRIG-2018-05,
    author         = "{ATLAS Collaboration}",
    title          = "{Performance of electron and photon triggers in ATLAS during LHC Run~2}",
    journal        = "Eur. Phys. J. C",
    volume         = "80",
    year           = "2020",
    pages          = "47",
    doi            = "10.1140/epjc/s10052-019-7500-2",
    reportNumber   = "CERN-EP-2019-169",
    eprint         = "1909.00761",
    archivePrefix  = "arXiv",
    primaryClass   = "hep-ex",
}

@Article{SUSY-2019-14,
    author         = "{ATLAS Collaboration}",
    title          = "{Search for displaced photons produced in exotic decays of the Higgs boson using \(13\,\text{TeV}\) \(pp\) collisions with the ATLAS detector}",
    year           = "2022",
    reportNumber   = "CERN-EP-2022-096",
    eprint         = "2209.01029",
    archivePrefix  = "arXiv",
    primaryClass   = "hep-ex",
}

@Booklet{ATL-SOFT-PUB-2021-001,
    author         = "{ATLAS Collaboration}",
    title          = "{The ATLAS Collaboration Software and Firmware}",
    howpublished   = "{ATL-SOFT-PUB-2021-001}",
    url            = "https://cds.cern.ch/record/2767187",
    year           = "2021",
}

@Booklet{ATL-SOFT-PUB-2021-003,
    author         = "{ATLAS Collaboration}",
    title          = "{ATLAS Computing Acknowledgements}",
    howpublished   = "{ATL-SOFT-PUB-2021-003}",
    url            = "https://cds.cern.ch/record/2776662",
    year           = "2021",
}

@Report{ATLAS-TDR-19,
    author         = "{ATLAS Collaboration}",
    title          = "{ATLAS Insertable B-Layer: Technical Design Report}",
    type           = "ATLAS-TDR-19; CERN-LHCC-2010-013",
    year           = "2010",
    url            = "https://cds.cern.ch/record/1291633",
    related        = "ATLAS-TDR-19-addm",
    relatedstring  = "Addendum:",
}

@Article{CMS-EXO-19-005,
    author         = "{CMS Collaboration}",
    title          = "{Search for long-lived particles using delayed photons in proton--proton collisions at \(\sqrt{s} = 13\,\text{TeV}\)}",
    journal        = "Phys. Rev. D",
    volume         = "100",
    year           = "2019",
    pages          = "112003",
    doi            = "10.1103/PhysRevD.100.112003",
    reportNumber   = "CERN-EP-2019-185",
    eprint         = "1909.06166",
    archivePrefix  = "arXiv",
    primaryClass   = "hep-ex",
}

@Booklet{ATL-PHYS-PUB-2014-021,
    author         = "{ATLAS Collaboration}",
    title          = "{ATLAS Pythia~8 tunes to \(7~\text{TeV}\) data}",
    howpublished   = "{ATL-PHYS-PUB-2014-021}",
    url            = "https://cds.cern.ch/record/1966419",
    year           = "2014",
}
 
\clearpage
 
\begin{flushleft}
\hypersetup{urlcolor=black}
{\Large The ATLAS Collaboration}

\bigskip

\AtlasOrcid[0000-0002-6665-4934]{G.~Aad}$^\textrm{\scriptsize 101}$,
\AtlasOrcid[0000-0002-5888-2734]{B.~Abbott}$^\textrm{\scriptsize 119}$,
\AtlasOrcid[0000-0002-7248-3203]{D.C.~Abbott}$^\textrm{\scriptsize 102}$,
\AtlasOrcid[0000-0002-1002-1652]{K.~Abeling}$^\textrm{\scriptsize 55}$,
\AtlasOrcid[0000-0002-8496-9294]{S.H.~Abidi}$^\textrm{\scriptsize 29}$,
\AtlasOrcid[0000-0002-9987-2292]{A.~Aboulhorma}$^\textrm{\scriptsize 35e}$,
\AtlasOrcid[0000-0001-5329-6640]{H.~Abramowicz}$^\textrm{\scriptsize 150}$,
\AtlasOrcid[0000-0002-1599-2896]{H.~Abreu}$^\textrm{\scriptsize 149}$,
\AtlasOrcid[0000-0003-0403-3697]{Y.~Abulaiti}$^\textrm{\scriptsize 116}$,
\AtlasOrcid[0000-0003-0762-7204]{A.C.~Abusleme~Hoffman}$^\textrm{\scriptsize 136a}$
% ...
\bigskip
\\

% ...
$^{29}$Physics Department, Brookhaven National Laboratory, Upton NY; United States of America.\\
% ...
$^{35}$$^{(a)}$Facult\'e des Sciences Ain Chock, R\'eseau Universitaire de Physique des Hautes Energies - Universit\'e Hassan II, Casablanca;$^{(b)}$Facult\'{e} des Sciences, Universit\'{e} Ibn-Tofail, K\'{e}nitra;$^{(c)}$Facult\'e des Sciences Semlalia, Universit\'e Cadi Ayyad, LPHEA-Marrakech;$^{(d)}$LPMR, Facult\'e des Sciences, Universit\'e Mohamed Premier, Oujda;$^{(e)}$Facult\'e des sciences, Universit\'e Mohammed V, Rabat;$^{(f)}$Institute of Applied Physics, Mohammed VI Polytechnic University, Ben Guerir; Morocco.\\
% ...
$^{55}$II. Physikalisches Institut, Georg-August-Universit\"{a}t G\"ottingen, G\"ottingen; Germany.\\
% ...
$^{101}$CPPM, Aix-Marseille Universit\'e, CNRS/IN2P3, Marseille; France.\\
$^{102}$Department of Physics, University of Massachusetts, Amherst MA; United States of America.\\
% ...
$^{116}$Department of Physics, New York University, New York NY; United States of America.\\
% ...
$^{119}$Homer L. Dodge Department of Physics and Astronomy, University of Oklahoma, Norman OK; United States of America.\\
% ...
$^{136}$$^{(a)}$Departamento de F\'isica, Pontificia Universidad Cat\'olica de Chile, Santiago;$^{(b)}$Millennium Institute for Subatomic physics at high energy frontier (SAPHIR), Santiago;$^{(c)}$Instituto de Investigaci\'on Multidisciplinario en Ciencia y Tecnolog\'ia, y Departamento de F\'isica, Universidad de La Serena;$^{(d)}$Universidad Andres Bello, Department of Physics, Santiago;$^{(e)}$Instituto de Alta Investigaci\'on, Universidad de Tarapac\'a, Arica;$^{(f)}$Departamento de F\'isica, Universidad T\'ecnica Federico Santa Mar\'ia, Valpara\'iso; Chile.\\
% ...
$^{149}$Department of Physics, Technion, Israel Institute of Technology, Haifa; Israel.\\
$^{150}$Raymond and Beverly Sackler School of Physics and Astronomy, Tel Aviv University, Tel Aviv; Israel.\\
% ...
$^{a}$ Also Affiliated with an institute covered by a cooperation agreement with CERN.\\
$^{b}$ Also at An-Najah National University, Nablus; Palestine.\\
$^{c}$ Also at Borough of Manhattan Community College, City University of New York, New York NY; United States of America.\\
$^{d}$ Also at Bruno Kessler Foundation, Trento; Italy.\\
$^{e}$ Also at Center for High Energy Physics, Peking University; China.\\
$^{f}$ Also at Centro Studi e Ricerche Enrico Fermi; Italy.\\
% ...

\end{flushleft}


\end{document}